\documentclass[11pt]{article}

\usepackage{booktabs}
\usepackage{multirow}
\usepackage{amssymb}
\usepackage{amsmath,amsfonts}
\usepackage{amsthm}
\usepackage{latexsym}
\usepackage{mathrsfs}
\usepackage{color}
\usepackage{algorithm}
\usepackage{algorithmic}
\usepackage{dsfont}
\usepackage[title]{appendix}

\usepackage[colorlinks=true, pdfstartview=FitV, linkcolor=blue, citecolor=blue, urlcolor=blue,pagebackref=false]{hyperref} % hyperlinks
\usepackage{url}            % simple URL typesetting
\usepackage{booktabs}       % professional-quality tables
\usepackage[a4paper]{geometry}
\usepackage{bm}
\usepackage{graphicx}
\usepackage{natbib}
\usepackage{enumerate}
\usepackage{amsopn}
\usepackage{graphicx}
\usepackage{subcaption}

\newcommand{\email}[1]{\protect\href{mailto:#1}{#1}}
\newcommand{\beaurl}{https://apps.bea.gov/iTable/?reqid=19\&step=2\&isuri=1\&categories=survey\&\_gl=1*123qp9l*\_ga*MjAyNzMzMzA5OC4xNzUwMzA3MTc3*\_ga\_J4698JNNFT*czE3NTE0MzI1MTkkbzMkZzEkdDE3NTE0MzI2MDAkajM5JGwwJGgw\#eyJhcHBpZCI6MTksInN0ZXBzIjpbMSwyLDMsM10sImRhdGEiOltbImNhdGVnb3JpZXMiLCJTdXJ2ZXkiXSxbIk5JUEFfVGFibGVfTGlzdCIsIjY1Il0sWyJGaXJzdF9ZZWFyIiwiMTkyOSJdLFsiTGFzdF9ZZWFyIiwiMjAyNCJdLFsiU2NhbGUiLCItOSJdLFsiU2VyaWVzIiwiQSJdXX0=}
\newcommand{\beaurll}{https://apps.bea.gov/iTable/?reqid=19\&step=2\&isuri=1\&categories=survey\&\_gl=1*123qp9l*\_ga*MjAyNzMzMzA5OC4xNzUwMzA3MTc3*\_ga\_J4698JNNFT*czE3NTE0MzI1MTkkbzMkZzEkdDE3NTE0MzI2MDAkajM5JGwwJGgw\#eyJhcHBpZCI6MTksInN0ZXBzIjpbMSwyLDNdLCJkYXRhIjpbWyJjYXRlZ29yaWVzIiwiU3VydmV5Il0sWyJOSVBBX1RhYmxlX0xpc3QiLCI2NCJdXX0=}
\newcommand{\corp}{https://apps.bea.gov/iTable/?reqid=19\&step=2\&isuri=1\&categories=survey\&\_gl=1*123qp9l*\_ga*MjAyNzMzMzA5OC4xNzUwMzA3MTc3*\_ga\_J4698JNNFT*czE3NTE0MzI1MTkkbzMkZzEkdDE3NTE0MzI2MDAkajM5JGwwJGgw\#eyJhcHBpZCI6MTksInN0ZXBzIjpbMSwyLDNdLCJkYXRhIjpbWyJjYXRlZ29yaWVzIiwiU3VydmV5Il0sWyJOSVBBX1RhYmxlX0xpc3QiLCIyNDgiXV19}
\newcommand{\cens}{https://data.census.gov/}
\newtheorem{theorem}{Theorem}
\newtheorem{corollary}{Corollary}
\newtheorem{definition}{Definition}
\newtheorem{proposition}{Proposition}

\newtheorem{assumption}{Assumption}

\newtheorem{Remark}{Remark}

\newcounter{spb}
\newcommand{\no}{\nonumber}
\newcommand{\lf}{\left}
\newcommand{\rh}{\right}
\newcommand{\RNum}[1]{\uppercase\expandafter{\romannumeral #1\relax}}
\newcommand{\rNum}[1]{\lowercase\expandafter{\romannumeral #1\relax}}
\newcommand{\argmax}{\arg\max}

\newcommand{\dr}{\partial}

\renewcommand{\d}{{\mathrm{d}}}

\newcommand{\eps}{\varepsilon}

\newcommand{\ie}{{\it i.e.}}

\def\b1{\bm{1}}

\def\bc{\bm{c}}
\def\bu{\bm{u}}

\def\CA{\mathcal{A}}

\newcommand{\E}{\mathbb{E}}
\newcommand{\expect}{\mathbb{E}}
\renewcommand{\P}{\mathbb{P}}
\newcommand{\R}{\mathbb{R}}

\newcommand{\oV}{{\overline{V}}}
\newcommand{\ov}{{\overline{v}}}

\newcommand{\om}{{\overline{m}}}

\newcommand{\uV}{{\underline{V}}}
\newcommand{\uv}{{\underline{v}}}

\newcommand{\um}{{\underline{m}}}

\newcommand{\SL}{\mathscr{L}}

\renewcommand{\qed}%{\hfill Q.E.D}
{\hfill \mbox{\raggedright \rule{.07in}{.1in}}}

\renewenvironment{proof}{\vspace{1ex}\noindent{\bf Proof}\hspace{0.5em}}
{\qed\vspace{1ex}}
\newenvironment{pfof}[1]{\vspace{1ex}\noindent{\bf Proof of #1}\hspace{0.5em}}
{\qed\vspace{1ex}}

\newcommand{\add}[1]{{\color{red}#1}}

\numberwithin{equation}{section}

\begin{document}
	\title{Dynamic Asset Pricing with $\alpha$-MEU Model\thanks{We thank conference and seminar participants in Hong Kong University of Science and Technology, Shenzhen University, the 2024 ETH-Hong Kong-Imperial Mathematical Finance Workshop, The 8th Asian Quantitative Finance Conference, the 2024 INFORMS Annual Meeting, the 14th AIMS Meeting of AIMS, and the 10th International Symposium on Backward Stochastic Differential Equations. The second author acknowledges the support from the Research Grants Council (RGC) of Hong Kong (GRF 14207620).}}	\author{
		Jiacheng Fan \thanks{Department of Applied Mathematics, The Hong Kong Polytechnic University,
			Kowloon, Hong Kong, China.  (\email{jiacheng.fan@polyu.edu.hk}).}
		\and Xue Dong He\thanks{Department of Systems Engineering and Engineering Management, The Chinese University of Hong Kong, Shatin N.T., Hong Kong SAR, China.(\email{ xdhe@se.cuhk.edu.hk}).}
		\and Ruocheng Wu\thanks{Department of Mathematics, The Chinese University of Hong Kong, Shatin N.T., Hong Kong SAR, China.(\email{rcwu11@link.cuhk.edu.hk}).}
	}
	\date{\today}
	\maketitle
	%\tableofcontents

	\begin{abstract}
		We study a dynamic asset pricing problem in which a representative agent is ambiguous about the aggregate endowment growth rate and trades a risky stock, human capital, and a risk-free asset to maximize her preference value of consumption represented by the $\alpha$-maxmin expected utility model. This preference model is known to be dynamically inconsistent, so we consider intra-personal equilibrium strategies for the representative agent and define the market equilibrium as the one in which the strategy that clears the market is an intra-personal equilibrium. We prove the existence and uniqueness of the market equilibrium and show that the asset prices in the equilibrium are the same as in the case when the agent does not perceive any ambiguity but believes in a particular probabilistic model of the endowment process. We show that with reasonable parameter values, the more ambiguity the agent perceives or the more ambiguity-averse she is, the lower the risk-free rate, the higher the stock price, the higher the stock risk premium, and the lower the stock volatility.
	\end{abstract}

	\section{Introduction}\label{sec:intro}
	The Ellsberg paradox \citep{Ellsberg:1961Ambiguity}, along with a substantial body of empirical evidence, demonstrates that ambiguity, also known as Knightian uncertainty, is a pervasive factor in decision-making under uncertainty. This ambiguity arises not only from the unknown realizations of a given stochastic element but also from individuals' lack of confidence in the underlying probability distribution governing that element. The experimental findings in \citet{Ellsberg:1961Ambiguity} suggest that individuals tend to be ambiguity-averse. In consequence, decision-theoretic models of preferences with ambiguity aversion have been proposed in the literature, and one of the most popular models is the maxmin expected utility (MEU) model proposed by \citet{GilboaSchmeidler1989:MaxminExpectedUtility}. This model characterizes an agent's perceived ambiguity through a set of possible probability measures, often referred to as multiple priors, and the agent's preference value of a payoff is represented by the worst-case expected utility among these priors. Standard investment models in the literature that allow for ambiguity employ the MEU model for agents' preferences and thus assume that agents are completely averse.
	
	It is, however, found in the literature that individuals can be ambiguity-seeking when their perceived competence level is high \citep{HeathTversky:1991Ambiguity,KilkaWeber2001:WhatDetermines,DimmockEtal2016:AmbiguityAversionAndHousehold}
	and when they face uncertain events with low likelihood \citep{DimmockKouwenbergMitchellPeijnenburg2015:EstimatingAmbiguity,DimmockEtal2015:AmbiguityAttitudes,TrautmannVanDeKuilen2015:Ambiguity}. One of the most commonly used preference models that account for mixed ambiguity attitudes is the $\alpha$-maxmin expected utility ($\alpha$-MEU) model \citep{Marinacci2002:ProbabilisticSophistication,GhirardatoEtal2004:DifferentiatingAmbiguity,Olszewski2007:Preferences}.\footnote{Another two commonly used models are Choquet expected utility model \citep{SchmeidlerD:89rd} and the smooth ambiguity model \citep{KlibanoffEtal2005:SmoothModel}.} In this model, the individual's perceived ambiguity is modeled by a set of probability measures and her preference value for a random payoff is represented by a weighted average of the worst-case expected utility and the best-case expected utility of the payoff among these probability measures. The weight of the worst-case expected utility, denoted as $\alpha$, measures the degree of the individual's ambiguity aversion. The $\alpha$-MEU model not only describes individuals' mixed ambiguity attitudes observed in experimental and empirical studies, but also has an axiomatic characterization in the Anscombe-Aumann framework; see \citet{Hartmann2023:StrengthOfPreference}.

	In this paper, we study equilibrium asset pricing in a pure-exchange economy with a representative agent who has $\alpha$-MEU preferences. There are a risk-free asset and two risky assets in the economy. These two risky assets can be interpreted as two broad asset classes, two countries' asset markets in international finance, or a stock versus human capital \citep{CochraneLongstaffStantaClara2008:TwoTrees}. In what follows, we take the last interpretation. The stock and human capital are in positive supply, producing cash flows referred to as dividend and labor income, respectively, and the sum of these two cash flows is the aggregate endowment in the economy. Empirically, the aggregate endowment is much larger than the stock dividend and their ratio is volatile over time; see a discussion and relevant literature in Section \ref{subse:Calibration}. In the asset pricing literature, some works assume that that human capital is not tradeable \citep[see, e.g.,][]{BrennanXia2001:StockPriceVol,LongstaffPiazzesi2004:CorporateEarnings}, while some others assume tradeable human capital \citep[see, e.g.,][]{CochraneLongstaffStantaClara2008:TwoTrees,SantosVeronesi2006:LaborIncome}. We follow the latter works, i.e., assume that the human capital is tradeable. As we will explain in Section \ref{subse:NontradeableLI}, the study of asset pricing with $\alpha$-MEU preferences and non-tradeable human capital is much more challenging, and we leave it for future research.

The agent is ambiguous about the mean growth rate of the aggregate endowment. We follow \citet{ChenEpstein2002:AmbiguityRiskAssetReturns} and \citet{Schroder2011:Investment} to model the agent's perceived ambiguity by a set of probability measures that are absolutely continuous with respect to a benchmark probability measure with the corresponding density generator processes taking values in a symmetric, compact interval $[-\kappa,\kappa]$. See Section \ref{sec:model1} for details. The agent can trade the assets continuously in time to maximize the $\alpha$-MEU value of the discounted sum of consumption utility. It is known in the literature that the $\alpha$-MEU model, when applied to a dynamic decision problem, is dynamically inconsistent: The actual choice upon arriving at a decision node would differ from the planned choice for that node. See for instance, \citet{Schroder2011:Investment} and \citet{BeissnerEtal2020:DynamicallyConsistent}. Consequently, the agent's action in our model is {\em time-inconsistent}: The optimal choice of future consumption and investment planned by the agent today may no longer be optimal upon arriving at the future time and thus will not be implemented by the agent at the future time. To address the issue of time inconsistency, we follow the literature on time-inconsistent decision problems to consider intra-personal equilibrium strategies that can be consistently implemented by the agent. More precisely, the agent is assumed not to be able to commit her future selves to any plan today and thus considers her selves at different times to be different players of a game. An intra-personal equilibrium strategy is an equilibrium of the game and can be consistently implemented by the agent at any time because the agent has no incentive to deviate from this strategy at any time. For detailed discussions of intra-personal strategies, see for instance \citet{Strotz1955:MyopiaInconsistency}, \citet{EkelandLazrak2006:BeingSeriousAboutNonCommitment}, \citet{Bjork2017:TimeInconsistent}, and a recent survey by \citet{HeZhou2021:WhoAreI}.

	We define the market equilibrium as the set of asset prices with which the particular strategy that clears the market is an intra-personal equilibrium strategy of the representative agent. The classic approaches to deriving the market equilibrium cannot be applied to our problem due to the time inconsistency. Indeed, there are two commonly used approaches to deriving Euler equations for asset returns in pure-exchange economies with representative agents whose preferences are dynamically consistent. The first one is to characterize the asset returns by the first-order conditions of the Hamilton-Bellman-Jacobi (HJB) equation for the representative agent's optimal control; see for instance \citet{DuffieEpstein1992:AssetPricingSDU}. This approach does not work for our problem because the HJB equation is only valid for dynamically consistent control problems. The second approach is to characterize the first-order condition for the representative agent's optimal consumption stream in the space of stochastic processes and derive asset returns from this first-order condition; see for instance \citet{ChenEpstein2002:AmbiguityRiskAssetReturns}. This approach does not work for our problem either: Due to dynamic inconsistency, the intra-personal equilibrium strategy is not optimal under the agent's preference value at any time.
	
	Our methodological contribution is to develop a new approach to deriving the equilibrium asset returns: We first derive a sufficient and necessary condition for a strategy to be an intra-personal equilibrium. Then, we characterize asset returns from this condition, which yields a second-order ordinary differential equation (ODE). Because this ODE is not uniformly elliptic, it is mathematically challenging to derive its solution. Our main technical contribution is to prove the existence and uniqueness of the classical solution to this ODE, from which we derive the asset returns. The assets' mean growth rates and volatilities in the market equilibrium are all functions of the market state represented by the dividend-endowment ratio.
	
	We prove that the equilibrium asset returns are the same as those in an economy with a representative agent who does not perceive any ambiguity and believes in a particular probability measure, referred to as the {\em ambiguity-adjusted probability measure}. The density generator process of this probability measure is a constant, which is a weighted average of $\kappa$ and $-\kappa$ with the weight depending not only on the ambiguity aversion degree $\alpha$ but also on other model parameters. This constant, referred to as the {\em ambiguity-implied density generator}, represents the difference in the mean growth rates of the endowment process, per unit of the volatility, in the ambiguity-adjusted probability measure and in the benchmark probability measure, so it measures how pessimistic or optimistic the agent is under the ambiguity-adjusted probability measure. We prove that the ambiguity-implied density generator is strictly decreasing in $\alpha$, showing that the agent is more pessimistic when she is more averse to ambiguity. The dependence of the ambiguity-implied density generator on the agent's perceived ambiguity level $\kappa$ depends on $\alpha$ and the agent's risk aversion degree $\gamma$ and the dependence relation can be nonmonotone. In the case when $\gamma>1$ and $\alpha\ge 1/2$, which is the empirically relevant case, the ambiguity-implied density generator is strictly decreasing in $\kappa$, showing that the agent is more pessimistic when she perceives a higher level of ambiguity.
	
	We then show that the risk-free rate is strictly increasing in the ambiguity-implied density generator. When $\gamma>1$, the stock price-dividend ratio is strictly decreasing, the stock risk premium is strictly decreasing, and the stock volatility is strictly increasing in the ambiguity-implied density generator. In consequence, we conclude that with $\gamma>1$ and $\alpha\ge 1/2$, if the agent becomes more ambiguity averse (i.e., $\alpha$ is larger) or perceives a higher level of ambiguity (i.e., $\kappa$ is larger), the risk-free rate %, consumption propensity,
and the stock volatility become lower, and the stock price-dividend ratio and risk premium become higher.
	
	We also find that the stock price-dividend ratio is strictly decreasing in the dividend-endowment ratio. The stock risk premium and volatility are strictly increasing in the dividend-endowment ratio in the empirically relevant range. These two findings are largely consistent with the implications of some existing asset pricing models with multiple cash flows in the literature \citep{SantosVeronesi2006:LaborIncome,CochraneLongstaffStantaClara2008:TwoTrees}.
	
	Finally, we compute the unconditional moments of the asset returns under the invariant distribution of the dividend-endowment ratio. We calibrate the parameters of the endowment and dividend processes to the consumption data from 1933 to 2022, choose different values of the ambiguity-implied density generator which encodes the agent's ambiguity aversion degree and her perceived ambiguity level, and calibrate the agent's discount rate and risk aversion degree to match the unconditional stock risk premium and price-dividend ratio empirically estimated in the literature. We find that even with a very high ambiguity aversion degree and perceived ambiguity level, the risk aversion degree that is needed to match the empirical estimate of risk premium is still larger than twenty, which is unrealistically high. More importantly, the model-implied stock volatility is significantly lower than the empirical estimate. Therefore, introducing ambiguity in the expected utility framework cannot completely explain the empirical findings of asset returns in the literature.

	The remainder of the paper is organized as follows. In Section \ref{se:Literature}, we review the literature. We then present the model in Section \ref{sec:model1} and the main results in Section  \ref{sec:main_result}. In Section \ref{se:ComparativeStatics}, we conduct comparative statics, and in Section \ref{se:Numerical}, we conduct a numerical study. Finally, we conclude in Section \ref{se:Conclusions}. Appendix \ref{appx:ODE} presents some results of the existence and uniqueness of the solution to an ODE. All proofs are %in Section \ref{appx:proofs} of
in an electronic companion.

	%%%%%%%%%%%%%%%%%%%%%%%%%%%%%%%%%%%%%%%%%%%%%%%%%%%%%%%%%%%%%%%%%%%%%%%%%%%%
	%%%%%%%%%%%%%%%%%%%%%%%%%%%%%%%%%%%%%%%%%%%%%%%%%%%%%%%%%%%%%%%%%%%%%%%%%%%%
	%%%%%%%%%%%%%%%%%%%%%%%%%%%%%%%%%%%%%%%%%%%%%%%%%%%%%%%%%%%%%%%%%%%%%%%%%%%%
	%%%%%%%%%%%%%%%%%%%%%%%%%%%%%%%%%%%%%%%%%%%%%%%%%%%%%%%%%%%%%%%%%%%%%%%%%%%%
	%%%%%%%%%%%%%%%%%%%%%%%%%%%%%%%%%%%%%%%%%%%%%%%%%%%%%%%%%%%%%%%%%%%%%%%%%%%%
	
	\section{Literature}\label{se:Literature}
	
	\subsection{Portfolio Selection and Asset Pricing under the $\alpha$-MEU Preference Model}
	
	\citet{BossaertsEtal2010:AmbiguityInAssetMarkets} study  single-period portfolio selection and asset pricing in a complete market with three states, where the agents have the $\alpha$-MEU preferences. \citet{AnthropelosSchneider2024:OptimalInvestment} consider a single-period portfolio selection model where the agent's preferences are represented by the $\alpha$-MEU model with the ambiguity set quantified as a neighborhood of a reference measure with quadratic $f$-divergence. The authors also study the equilibrium asset price of a single risky asset that clears the market.
	\citet{BeissnerWerner2023:OptimalAllocations} study optimal risk sharing in a static setting for various preference models including the $\alpha$-MEU model.
	\citet{ChateauneufEtal2007:ChoiceUnderUncertainty} consider the $\alpha$-MEU with a specific ambiguity set, which leads to the Choquet expected utility (CEU) with neo-additive capacities. The authors then apply this preference model to single-period portfolio selection and asset pricing. \citet{GhaziSchneiderStrauss2025:MarketAmbiguity} consider a single-period consumption-based asset pricing model for a representative agent whose preferences are represented by CEU with neo-additive capacities.
	
	\citet{Schroder2011:Investment} and \citet{KimKwakChoi2009:InvestmentUnderAmbiguity} study the optimal investment of a firm with the $\alpha$-MEU preferences. At each time, the firm makes an all-or-nothing decision of investing in a project that generates a cash flow stream. If the firm chooses not to invest, the project is gone, so there is no value in waiting. Therefore, although the investment problem of the firm is formulated in a continuous-time setting, the firm's decision-making is static instead of dynamic. In consequence, the issue of dynamic inconsistency does not arise in their problems.
	
	\citet{Zimper2012:AssetPricing} study the Lucas fruit-tree economy with a representative agent following CEU with neo-additive capacities, which can be regarded as an $\alpha$-MEU model with a specific ambiguity set. Similar to us, they apply the preference model to the discounted sum of consumption utility, which yields time inconsistency, and considers a sophisticated representative agent. They show that an optimistic representative agent gives rise to a high risk free rate and lowers equity premium for a risky asset with a high best-case return, and a pessimistic representative agent does the opposite. This result is consistent with the findings in our model.	\citet{GroneckEtal2016:ALifeCycleModel} use a similar version of \citet{Zimper2012:AssetPricing}'s model to study life-cycle investment with retirement. \citet{HeSun2025:DynamicPortfolioSelection} study portfolio selection and asset pricing in a multi-period setting for an agent whose preferences are represented by the CEU model with neo-additive capacities. Given the time inconsistency, the authors consider three types of behavior, depending on whether the agent is aware of time inconsistency and whether she has self-control. One of the types is the same as the intra-personal equilibrium considered in our work. The models in the above three works differ  from our in two aspects: First, we consider the continuous-time setting while they consider the discrete-time setting. Second, we consider the ambiguity set in \citet{ChenEpstein2002:AmbiguityRiskAssetReturns} and \citet{Schroder2011:Investment}, whereas they use the ambiguity set that induces the CEU model with neo-additive capacities.

	In continuous-time settings, the $\alpha$-MEU model has been applied to portfolio selection \citep{LiEtal2019:EquilibriumStrategies,YangLiZengLiu2025:OptimalInvestment}, optimal stopping \citep{HuangYu2021:OptimalStopping}, and optimal insurance design \citep{LiLiXiong2016:AlphaRobust,ZhangLi2021:OptimalReinsurance,GuanLiangSong2024:AStackelberg}, assuming that the agents take intra-personal equilibrium strategies given time inconsistency. We consider asset pricing, which is a completely different problem and requires new methodologies.
	
	%\citet{YuEtal2020:PortfolioSelection}

	An alternative approach to applying $\alpha$-MEU to dynamic decision problems is to use it to evaluate the risk in each period instead of the aggregate risk in multiple periods. This approach does not lead to time inconsistency because $\alpha$-MEU is applied in every single period, and the resulting preference model is known as the recursive $\alpha$-MEU model. \citet{BeissnerEtal2020:DynamicallyConsistent} study the recursive $\alpha$-MEU model in a continuous-time setting and show that the preference value under this model can be represented by a backward stochastic differential equation. The authors then apply the model to consumption-based asset pricing. \citet{Peijnenburg2018:LifeCycleAssetAllocation} studies a life-cycle asset allocation problem for both the MEU model and the recursive $\alpha$-MEU model.	\citet{Saghafian2018:AmbiguousPartiallyObservableMDP} applies the recursive $\alpha$-MEU model to a partially observable Markov decision problem.

	\subsection{Asset Pricing with Multiple Cash Flows}
	\citet{SantosVeronesi2006:LaborIncome} consider asset pricing in a pure-exchange economy with multiple risky assets for a representative agent with expected utility (EU) preferences. They assume a specific model for the aggregate endowment, which is the sum of the dividends of all risky assets, and the dividend-endowment share. The dividend-endowment share process is stationary and always lies between 0 and 1. The model is specified so that the price of each risky asset is a linear function of the asset dividends. Our model of the dividend and endowment processes is different from \citet{SantosVeronesi2006:LaborIncome} except for the case when the endowment process and dividend-endowment share process are uncorrelated; see Assumptions 1 and 2 in \citet{SantosVeronesi2006:LaborIncome}. The dividend model of \citet{SantosVeronesi2006:LaborIncome} is also used by \citet{Menzly2004:UnderstandingPredictability} to study asset pricing with time-varying preferences, by \citet{SantosVeronesi2010:HabitFormation} to study asset pricing with an external habit, and by \citet{Stathopoulos2021:PortfolioHomeBias} in the context of multiple consumption goods.

	\citet{GuasoniWong2020:AssetPrices} study asset pricing in a pure-exchange economy with two risky assets for a representative agent with EU preferences. They model the aggregate endowment process as a geometric Brownian motion, similar to our model. However, they assume that the dividend-endowment ratio process is uncorrelated with the aggregate endowment process, whereas we consider a general correlation structure. \citet{GuasoniPiccirilliWang2025:AssetPricing} consider the same asset pricing problem as \citet{GuasoniWong2020:AssetPrices} but with a general model of the aggregate endowment process and dividend-endowment process. The authors make an assumption that essentially guarantees the existence of the equilibrium asset prices; see Assumption 3 therein. In the present paper, we prove the existence and uniqueness of the equilibrium asset prices by proving the existence and uniqueness of a classical solution to an ODE. This is not trivial and is one of our main contributions. In a specific setting, \citet{GuasoniPiccirilliWang2025:AssetPricing} are able to derive the price-endowment ratio in closed form, which is linear in the dividend-endowment ratio. Our model is different from this specific setting because the price-endowment ratio in our model is not linear in the dividend-endowment ratio except for the case when the endowment and dividend-endowment ratio are uncorrelated.
	
	\citet{CochraneLongstaffStantaClara2008:TwoTrees} consider asset pricing in a pure-exchange economy with two risky assets. The dividends of the assets follow geometric Brownian motions. Assuming a representative agent with EU preferences and a logarithmic utility function, the authors derive the asset prices in closed form. \citet{Martin2013:LucasOrchard} considers a more general setup: They assume a representative agent with EU preferences and constant relative risk aversion and multiple risky assets with dividends following a multi-dimensional exponential L\'evy process. The author then derives formulas of the risky asset prices and risk-free rate from Euler equations. \citet{BhamraCoeurdacierGuibaud2014:ADynamicEquilibriumModel} use the same dividend model as \citet{CochraneLongstaffStantaClara2008:TwoTrees} to study the dynamic equilibrium of imperfectly financial integrated markets. \citet{Sauzet2022:TwoInvestors} also uses the same dividend model as in \citet{CochraneLongstaffStantaClara2008:TwoTrees} to study an asset pricing problem of two heterogeneous investors with recursive utility preferences in a two-tree, two-good environment. \citet{ParlourStantonWalden2011:RevisitingAssetPricingPuzzles} consider a two-Lucas-tree setting, which can be regarded as a special case of \citet{CochraneLongstaffStantaClara2008:TwoTrees} and \citet{Martin2013:LucasOrchard}. In this setting, the first asset is risky and its dividend follows a geometric Brownian motion. The second asset is risk-free and generates a constant dividend. \citet{ParlourStantonWalden2012:FinancialFlexibility} consider a similar model to \citet{ParlourStantonWalden2011:RevisitingAssetPricingPuzzles}, incorporating negative jumps in the dividend of the risk-free asset and a financial intermediary. Our dividend model is different from those in \citet{CochraneLongstaffStantaClara2008:TwoTrees} and \citet{Martin2013:LucasOrchard}: In our model, the outputs of the risky stock and human capital are not exponential L\'evy processes. Moreover, the dividend-endowment ratio is stationary in our model but not stationary in \citet{CochraneLongstaffStantaClara2008:TwoTrees} and \citet{Martin2013:LucasOrchard}.
	
	\citet{Johnson2006:DynamicLiquidity} studies dynamic liquidity in a pure-exchange economy. To derive tractable solutions, the authors consider the dividend models of \citet{SantosVeronesi2006:LaborIncome} and \citet{CochraneLongstaffStantaClara2008:TwoTrees}. \citet{ChenJoslin2012:GeneralizedTransformAnalysis} study a class of transforms for processes with tractable characteristic functions and apply their results to asset pricing with multiple assets. They assume that the representative agent has EU preferences and that the dividend processes are affine. In our model, the dividends of the risky stock and human capital are not affine.

	\citet{LongstaffPiazzesi2004:CorporateEarnings} study asset pricing in a pure-exchange economy with a representative agent maximizing the EU of her consumption. The aggregate endowment consists of the dividend of a financial asset and a nonfinancial income. The authors assume that the aggregate endowment follows a jump-diffusion process with a constant growth rate and the log dividend-endowment ratio follows a square-root jump-diffusion process. \citet{Marfe2016:CorporateFraction} assumes a similar dividend model and considers a representative agent with recursive utility preferences. \citet{BekaertEngstromXing2009:RiskUncertainty} use a similar dividend model, %to \citet{LongstaffPiazzesi2004:CorporateEarnings} and \citet{Marfe2016:CorporateFraction}
albeit in a discrete-time setting, and study the effect of conditional variance of fundamentals and changes in risk aversion on asset prices. In \citet{LongstaffPiazzesi2004:CorporateEarnings} and \citet{Marfe2016:CorporateFraction}, the log aggregate endowment and dividend-endowment ratio processes are affine processes, so the authors can derive asset prices in closed form. However, due to the affine structure, the dividend-endowment ratio can be higher than 1. By contrast, in our model, the dividend-endowment ratio process is not affine and stays between 0 and 1.
	
	\citet{AtmazBasak2022:StockMarket} consider an asset pricing model in a pure-exchange economy with a dividend-paying stock and two stocks paying dividends alternately. The stock dividends are modeled as geometric Brownian motions and the representative agent is assumed to have EU preferences. Our model is different from theirs in that the two risky assets in our model pay dividends continuously.
	%Dividends are observed only when they are distributed. Thus, for non-dividend-paying stocks, the agent does not observe the dividend until it is paid. Instead, she observes some process that signals the value of the dividend to be paid in the future. The aggregate endowment in the economy is {\em different} from the aggregate dividends paid by the stocks and assumed to be a geometric Brownian motion. Thus, the aggregate endowment can be lower than the aggregate dividends in the economy. The representative agent is assumed to be an expected utility maximizer with constant relative risk aversion.
	
	\subsection{Asset Pricing with Time-Inconsistent Preferences}

	There have been a few works on competitive market equilibrium for agents with time-inconsistent behaviour. \citet{LuttmerMariotti2003:SubjectiveDiscounting} describe the equilibrium in a pure-exchange discrete-time economy with time inconsistency caused by non-exponential discounting and provide an analytical, continuous-time approximation. \citet{LuttmerMariotti2006:CompetitiveEquilibrium} study competitive equilibria in a discrete-time economy without uncertainty where the agents have time-varying preferences and are thus time inconsistent. \citet{LuttmerMaritti2007:EfficiencyAndEquilibrium} study efficient allocations and competitive equilibria in a three-period exchange economy with deterministic endowment and consumption and with agents whose discount rate is non-exponential. \citet{Bjork2017:TimeInconsistent} consider a stochastic production economy of Cox-Ingersoll-Ross type with a representative agent who maximizes the expected discounted sum of consumption utility. The utility function is time-varying, leading to time inconsistency. The authors characterize the market equilibrium by the solution to an extended HJB equation. \citet{Khapko2023:AssetPricing} study market competitive equilibrium in a continuous-time endowment economy with a representative agent who maximizes the expected discounted sum of consumption utility with time-varying and state-dependent utility functions. In all the above works, the authors consider intra-personal equilibrium strategies for the agents in their models. Our work differs from those works in that we consider ambiguity and the $\alpha$-MEU preference model.

In a time-inconsistent dynamic decision problem, if the agent is not aware of time inconsistency and revisits the problem at every time, she would implement, at each time, the strategy that is optimal for her at that time. At the next instant, however, she is not able to commit to the planned strategy; instead, she would take another strategy that is optimal at this new instant. Such types of agents are called naive.  \citet{HeringsRohde2006:TimeInconsistent} introduce a notion of competitive equilibrium for naive agents in a deterministic consumption problem.
	\citet{GabrieliSayantan2013:NonExistence} consider two different notions of competitive equilibrium for naive agents, one referred to as perfect foresight competitive equilibrium and the other referred to as temporary competitive equilibrium. The latter notion is the same as the one used in \citet{HeringsRohde2006:TimeInconsistent}.

	\section{Model}\label{sec:model1}
	\subsection{Economy} \label{subsec:model1}
	Consider a pure-exchange economy with a risk-free asset and two risky assets that are traded continuously over time. The risk-free asset is in zero net supply and the two risky assets are in unit supply. The first risky asset is interpreted as a stock and the other risky asset represents human capital. In the period $[t,t+dt)$, the stock distributes a dividend of amount $D_t dt$ and the human capital distributes labor income of amount $L_t dt$. The stock dividend and labor income constitute the aggregate endowment $\bar C_t dt$ in the economy. We assume that the aggregate endowment and the dividend-endowment ratio $\omega_t:=D_t/\bar C_t$ follow
	\begin{align}
		&		d \bar C_t / \bar C_t = \mu_C d t + \sigma_C d B_{1,t}^0,\label{eq:consumption}\\
		& d \omega_t = \mu_{\omega}(\omega_t) d t + \sigma_{\omega} (\omega_t )d B^0_{2,t},\label{eq:omega}
	\end{align}
	where $(B^0_{1,t},B^0_{2,t})_{t \ge 0}$ is a two-dimensional Brownian motion with a constant correlation $\rho$ on a filtered probability space $(\Omega,\mathcal{F},(\mathcal{F}_{t})_{t\ge 0},\P^0)$ satisfying the usual condition, $\mu_C$ and $\sigma_C$ are two constants, and $\mu_{\omega}$ and $\sigma_{\omega}$ are two deterministic functions on $(0,1)$.
	
	\begin{assumption}\label{as:SDE}
		$\mu_\omega$ and $\sigma_\omega$ are twice continuously differentiable on $(0,1)$ with bounded derivatives. Moreover, $\sigma_\omega(x)>0,x\in (0,1)$ and
		\begin{align}
			&\lim_{x\downarrow 0}\mu_\omega(x)>0,\quad \lim_{x\uparrow 1 }\mu_\omega(x)<0,\quad \lim_{x\downarrow 0}\sigma_\omega(x)=\lim_{x\uparrow 1} \sigma_\omega(x)=0.
		\end{align}
	\end{assumption}
	
	\begin{proposition}\label{prop:SDEExistence}
		Let  Assumption \ref{as:SDE} hold. Then, for any initial value $\omega_0\in (0,1)$, \eqref{eq:omega} admits a unique strong solution $(\omega_t)_{t\ge 0}$ taking values in $(0,1)$. Moreover, there exists a unique invariant distribution for $(\omega_t)_{t\ge 0}$.
	\end{proposition}

	A commonly used specification of $\mu_\omega$ and $\sigma_{\omega}$ is
	\begin{align}
		&\mu_{\omega}(x)=\lambda (\overline{\omega} -x),\quad \sigma_{\omega}(x) = \nu x(1-x),\label{eq:QudraticDiffusion}
	\end{align}
	where $\overline{\omega}\in (0,1)$ represents the long-run mean of $\omega_t$, $\lambda>0$ measures the speed of $\omega_t$ reverting to the long-run mean, and $\nu>0$ is a constant. %See for instance \citet{SantosVeronesi2006:LaborIncome}.
	It is straightforward to see that this specification satisfies Assumption \ref{as:SDE}.

	\subsection{Preferences}
	%Consider consumption in continuous time. Let $(\Omega,\mathcal{F},(\mathcal{F}_{t})_{t\ge 0})$ be a filtered probability space.
	A consumption stream $(C_t)_{t\ge 0}$ is represented as an adapted process on $(\Omega,\mathcal{F},(\mathcal{F}_{t})_{t\ge 0})$. The representative agent's objective is to maximize the expected utility of her consumption in an infinite horizon. The agent, however, is ambiguous about the probability model describing the randomness in the market. Following \cite{Schroder2011:Investment} and \cite{ChenEpstein2002:AmbiguityRiskAssetReturns}, we assume that the agent's ambiguity is represented by a set of Radon-Nikodym densities with respect to the benchmark probability model $\P^0$. More precisely, we assume an interval $[-\kappa,\kappa]$ with $\kappa\ge 0$ and define $\Theta$ as the set of adapted processes $(\theta_t)_{t\ge 0} $ such that $\theta_t\in[-\kappa,\kappa]$ for all $t\ge 0$. For each $\theta \in \Theta$, define a new probability measure $\P^\theta$ on $(\Omega,\mathcal{F},(\mathcal{F}_{t})_{t\ge 0})$ by\footnote{For the existence of $\P^\theta$, see \citet[Proposition 7.4, Chapter 1]{KaratzasIShreveS:98momf}.}
	\begin{align}\label{density-generator}
		\frac{d\P^\theta}{d\P^0}\bigg|_{\mathcal{F}_t}:= e^{-\frac{1}{2}\int_0^t\theta_s^2ds  + \int_0^t\theta_s d B^0_{1,s}},\quad t\ge 0.
	\end{align}
$\theta$ is called the {\em density generator} of $\P^\theta$.
	The agent's ambiguity set is defined as
	\begin{align*}
		{\cal P}:=\{\P^\theta:\theta \in \Theta\}.
	\end{align*}
	Thus, the agent believes that every model in ${\cal P}$ is possible. The parameter $\kappa$ measures the {\em level of perceived ambiguity} by the agent.
	%Note that the ambiguity interval $[-\kappa,\kappa]$ is symmetric

	Under $\P^\theta$, $\left(B^\theta_{1,t},B^\theta_{2,t}\right):=\left(B^0_{1,t} - \int_0^t \theta_{s} \d s,B^0_{2,t}\right)$, $t\ge 0$ is a two-dimensional Brownian motion with correlation $\rho$. By \eqref{eq:consumption} and \eqref{eq:omega}, we have
	\begin{align}
		&	d \bar C_t/ \bar C_t = (\mu_C + \theta_{t}\sigma_C) d t + \sigma_C d B^\theta_{1,t},\label{eq:consumption-ambiguity}\\
		&d \omega_t = \mu_{\omega}(\omega_t) d t + \sigma_{\omega} (\omega_t )d B^\theta_{2,t}.\label{eq:omega-ambiguity}
	\end{align}
	Thus, the mean growth rate of $\bar C_t$ changes from $\mu_C$ in the benchmark model $\P^0$ to $\mu_C + \theta_{t}\sigma_C$ in the model $\P^\theta$. Therefore, the density generator $\theta$ measures the difference in the mean growth rates of the aggregate endowment under $\P^\theta$ and $\P^0$, per unit of the endowment volatility.
	The mean growth rate of $\omega_t$ and the volatilities of $\bar C_t$ and $\omega_t$ are the same under different models $\P^\theta$. Therefore, the agent is only ambiguous about the mean growth rate of the aggregate consumption. There are two reasons why we do not consider another source of  uncertainty: First, it is technically challenging to study portfolio selection and asset pricing with ambiguity in volatility and correlation. Second, the mean growth rate of the aggregate consumption is more difficult to estimate than other model parameters. Thus, the agent does not learn the mean growth rate of the aggregate consumption over time. This assumption is also made by \cite{Schroder2011:Investment} and \cite{ChenEpstein2002:AmbiguityRiskAssetReturns}.
	
	We assume that the agent's preferences for consumption are represented by the $\alpha$-MEU model \citep{Marinacci2002:ProbabilisticSophistication,GhirardatoEtal2004:DifferentiatingAmbiguity,Olszewski2007:Preferences}. More precisely, at each time $t$, the agent evaluates a consumption stream $(C_s)_{s\ge t}$ by
	\begin{equation} \label{eq:alphaMEU}
		\alpha \inf_{\theta\in\Theta} \E^{\theta}_t \lf[ \int_t^{\infty} e^{-\phi(s-t)} U(C_s) \d s \rh] + (1 - \alpha) \sup_{\theta\in\Theta} \E^{\theta}_t \lf[ \int_t^{\infty} e^{-\phi(s-t)} U(C_s) \d s \rh],
	\end{equation}
	where $\E^\theta_t$ denotes the expectation under $\P^\theta$, given the information at time $t$, $\alpha\in[0,1]$ is a constant, $\phi>0$ is a discount rate, and
	\begin{equation} \label{eq:utility}
		U(x) =  \begin{cases}
			\frac{x^{1-\gamma} -1}{1-\gamma} & ~\mbox{if}~ \gamma \neq 1 \\
			\log (x) & ~\mbox{if}~ \gamma = 1
		\end{cases},
	\end{equation}
	is a utility function with the parameter $\gamma>0$ representing the {\em relative risk aversion degree}. In this preference model, the agent is concerned about both the worst-case probabilistic model and the best-case probabilistic model, and the parameter $\alpha$ measures the {\em ambiguity aversion degree}. When $\alpha=1$, the $\alpha$-MEU model simply becomes the MEU model \citep{GilboaSchmeidler1989:MaxminExpectedUtility}.

	\subsection{Definition of Market Equilibrium}
	We aim to find the stock price and the risk-free rate in market equilibrium, i.e., when the assets' supply is equal to their demand. We follow \citet{CochraneLongstaffStantaClara2008:TwoTrees} and \citet{SantosVeronesi2006:LaborIncome} to assume that human capital is also tradeable, whose price represents the value of the labor income in the future. Therefore, we also need to determine the human capital price in market equilibrium.

	The agent's demand for the assets is determined by the consumption and investment strategy that maximizes her preference value. It is known that the $\alpha$-MEU model is not dynamically consistent; see for instance \citet{Schroder2011:Investment} and \citet{BeissnerEtal2020:DynamicallyConsistent}. In consequence, a strategy that maximizes the preference value at current time $t$ does not necessarily maximize the preference value at future time $s>t$ and thus cannot be consistently implemented. We follow the literature on time-inconsistent problems, e.g., \citet{Strotz1955:MyopiaInconsistency}, \citet{EkelandLazrak2006:BeingSeriousAboutNonCommitment}, and \citet{Bjork2017:TimeInconsistent}, to consider the intra-personal equilibrium strategy: At each time, the agent correctly anticipates her actions in the future and takes action at the current time accordingly to maximize her preference value at the current time.

	The definition of intra-personal equilibrium relies on the strategic actions of the agent at each time. The agent's actions consist of the dollar amount invested in the stock, the dollar amount invested in the human capital, and the dollar amount consumed. At each time, the agent's actions are contingent on her wealth and a market state that determines the asset returns in the next period. Denote by $S_t$ and $H_t$ the prices of the stock and human capital, respectively, at time $t$. %We further define $B^0_t:=(B^0_{1,t},B^0_{2,t})^\top$.
	In market equilibrium, we expect that the market state is the dividend-endowment ratio $\omega_t$. Thus, we expect
	\begin{align}
		&(dS_t+D_tdt)/S_t = \mu_S(\omega_t)dt + \sigma_S(\omega_t)dB^0_t,\label{eq:StockPrice}\\
		&(dH_t+L_tdt)/H_t = \mu_H(\omega_t)dt + \sigma_H(\omega_t)dB^0_t,\label{eq:HumanCapitalPrice}
	\end{align}
	where $B^0_t:=(B^0_{1,t},B^0_{2,t})^\top$, and $\mu_i:(0,1)\rightarrow \R$ and $\sigma_i:(0,1)\rightarrow \R^2$, $i\in \{S,H\}$ are deterministic functions to be determined in market equilibrium. Similarly, we expect that the risk-free net return in the period $[t,t+dt)$ is $r_f(\omega_t)dt $ for some function $r_f$ to be determined in market equilibrium. Given the preference model \eqref{eq:alphaMEU} with the utility function \eqref{eq:utility}, the dollar amount of the agent's investment in the stock and human capital and the agent's consumption amount are proportional to her wealth. Thus, we simply represent the agent's action by the percentages of her wealth consumed, invested in the stock, and invested in the human capital, and these percentages are independent of the agent's wealth but dependent on the state variable $\omega_t$. A {\em strategy} is a collection of mappings $\bc$, $\bu_S$, and $\bu_H$ from $(0,1)$ to $\R$ such that $\bc(\omega)>0$ for all $\omega$. At each time $t$, the agent consumes $\bc(\omega_t)$ percentage of her wealth and invests $\bu_S(\omega_t)$ and $\bu_H(\omega_t)$ percentages of wealth in the stock and human capital, respectively. Denote $\bu=(\bu_S,\bu_H)$.
	
	Given the asset prices in \eqref{eq:StockPrice} and \eqref{eq:HumanCapitalPrice}, the risk-free return rate $r_f$, and a strategy $(\bc,\bu)$, the agent's wealth process $(X^{\bc, \bu}_t)_{t\ge 0}$ is
	\begin{align}
		&d X^{\bc, \bu}_t / X^{\bc, \bu}_t =
		\lf[- \bc(\omega_t)+  r_f(\omega_t) (1-\bu_S(\omega_t) - \bu_H(\omega_t) ) \rh] d t \no\\
		& + \bu_S(\omega_t) \lf( \frac{d S_t + D_td t}{S_t} \rh) +\bu_H(\omega_t) \lf( \frac{d H_t + L_td t}{H_t} \rh)dt.
		\label{eq:wealth-labor-income}
	\end{align}
	%Denote by $C^{\bc,\bu}$ the corresponding consumption process, i.e.,
	%\begin{align*}
	%  C^{\bc,\bu}_t:=\bc(\omega_t)X^{\bc, \bu}_t,\quad t\ge 0.
	%\end{align*}
	Then, at time $t$, given a wealth level $x$ and market state $\omega$, the agent's preference value is
	\begin{equation}  \label{eq:alphaMEU-XY}
		V^{\bc, \bu}_t(x,\omega) = \alpha \uV^{\bc,\bu}_t (x,\omega) + (1-\alpha) \oV^{\bc,\bu}_t (x,\omega),
	\end{equation}
	where
	\begin{align}
		\uV^{\bc,\bu}_t (x,\omega) & := \inf_{\theta\in\Theta} \E^{\theta}_t \lf[ \int_t^{\infty} e^{-\phi(s-t)} U(\bc(\omega_s) X^{\bc,\bu}_s) \d s \mid X^{\bc,\bu}_t=x,\omega_t=\omega\rh], \label{eq:uV}\\
		\oV^{\bc,\bu}_t (x,\omega) & := \sup_{\theta\in\Theta} \E^{\theta}_t \lf[ \int_t^{\infty} e^{-\phi(s-t)} U(\bc(\omega_s) X^{\bc,\bu}_s) \d s \mid X^{\bc,\bu}_t=x,\omega_t=\omega\rh]. \label{eq:oV}
	\end{align}

	Following the literature on time-inconsistent problems, e.g., \citet{Strotz1955:MyopiaInconsistency}, \citet{EkelandLazrak2006:BeingSeriousAboutNonCommitment}, and \citet{Bjork2017:TimeInconsistent}, we define intra-personal equilibrium as follows:
	\begin{definition}[Intra-Personal Equilibrium]\label{de:IntraPersonalEqui}
		A given strategy $(\hat{\bc}, \hat{\bu} )$ is an {\em intra-personal equilibrium} if for any time $t$, wealth level $x$, and market state $\omega$,
		\begin{equation}  \label{eq:equilibrium-def}
			\limsup_{\eps\downarrow 0} \frac{ V^{\bc_{t,\eps,c},\bu_{t,\eps,u}}_t(x,\omega) - V^{\hat{\bc},\hat{\bu}}_t(x,\omega)}{\eps} \le 0
		\end{equation}
		for every constant $c>0$ and $u\in\R^2$, where
		\begin{align}
			\bc_{t,\eps,c}(s,y):=\begin{cases}
				c, & s\in [t,t+\eps) ,  \\
				\hat{\bc}(y), & s \not\in[t,t+\eps),\quad
			\end{cases}
			\quad \bu_{t,\eps,u}(s,y) := \begin{cases}
				u, & s\in [t,t+\eps) ,  \\
				\hat{\bu}(y), & s \not\in[t,t+\eps).
			\end{cases}\label{eq:PertubatedStrategy}
		\end{align}
	\end{definition}
	Suppose that at each time $t$, the agent can control herself only in an infinitesimally small period $[t,t+\epsilon)$. Then, any action $(c,u)$ chosen by the agent can be implemented only in $[t,t+\epsilon)$. If the agent believes that her future selves will take the strategy $(\hat{\bc}, \hat{\bu} )$, then the strategy taken by the agent from the current time $t$ throughout the entire time horizon is $(\bc_{t,\eps,c},\bu_{t,\eps,u})$ as defined in \eqref{eq:PertubatedStrategy}. Condition \eqref{eq:equilibrium-def} stipulates that at any time $t$, taking any action $(c,u)$, which can only be implemented in an infinitesimally small period $[t,t+\epsilon)$, cannot lead to a strictly larger preference value compared to following the given strategy $(\hat{\bc}, \hat{\bu} )$. As a result, the agent will follow $(\hat{\bc}, \hat{\bu} )$ throughout the entire horizon.

	Next, we define the market equilibrium. Given the aggregate endowment \eqref{eq:consumption}, the stock and human capital prices are completely determined by their {\em price-endowment ratios} $S_t/\bar C_t$ and $H_t/\bar C_t$. Thus, to find the prices of the stock and human capital in market equilibrium, we only need to identify the price-endowment ratios, which are again assumed to be functions of the market state and denoted as $\varphi_S(\omega_t)$ and $\varphi_H(\omega_t)$, respectively. Then, using It\^o's lemma and straightforward calculation, we can represent the drift and volatility of the asset prices in \eqref{eq:StockPrice} and \eqref{eq:HumanCapitalPrice} as
	\begin{align}
		&\mu_S(\omega)=\mu_C+\frac{1}{\varphi_S(\omega)}\left[\omega + \big(
		\mu_{\omega}(\omega) + \rho\sigma_C\sigma_{\omega}(\omega)
		\big)\varphi_S'(\omega)   + \frac{1}{2}\sigma_{\omega}^2 (\omega) \varphi_S''(\omega)\right],\label{eq:StockDrift}\\
		& \sigma_S(\omega)=\begin{pmatrix}
			\sigma_{S,1}(\omega),\sigma_{S,2}(\omega)
		\end{pmatrix}=\begin{pmatrix}
			\sigma_C, \sigma_{\omega} (\omega ) \varphi_S'(\omega)/\varphi_S(\omega)
		\end{pmatrix},\label{eq:StockVol}\\
		& \mu_H(\omega) = \mu_C + \frac{1}{\varphi_H(\omega)}\left[1-\omega + \big(\mu_{\omega}(\omega) + \rho\sigma_C\sigma_{\omega}(\omega)
		\big)\varphi_H'(\omega)+ \frac{1}{2}\sigma_{\omega}^2 (\omega) \varphi_H''(\omega)\right],\label{eq:HCDrift}\\
		& \sigma_H(\omega)= \begin{pmatrix}
			\sigma_{H,1}(\omega),\sigma_{H,2}(\omega)
		\end{pmatrix}=\begin{pmatrix}
			\sigma_C, \sigma_{\omega} (\omega)\varphi_H'(\omega)/\varphi_H(\omega)
		\end{pmatrix}.\label{eq:HCVol}
	\end{align}
	
	In market equilibrium, the agent's wealth $X_t$ is equal to the total value of the stock and human capital; namely,
	\begin{align}
		X_t = S_t+H_t = \big(\varphi_S(\omega_t)+\varphi_H(\omega_t)\big)\bar C_t. \label{eq:WealthinEquilibrium}
	\end{align}
	In market equilibrium, the agent's consumption is equal to the aggregate endowment $\bar C_t$. In consequence, we derive from \eqref{eq:WealthinEquilibrium} that the agent's percentage of wealth consumed should be
	\begin{align*}
		\frac{\bar C_t}{X_t} = \frac{1}{\varphi_S(\omega_t)+\varphi_H(\omega_t)}.
	\end{align*}
	In market equilibrium, the agent's dollar amount invested in the stock should be equal to the stock price $S_t$. Then, given \eqref{eq:WealthinEquilibrium}, the percentage of wealth invested in the stock should be
	\begin{align*}
		\frac{S_t}{S_t+H_t} = \frac{\varphi_S(\omega_t)}{\varphi_S(\omega_t)+\varphi_H(\omega_t)}.
	\end{align*}
	Similarly, the percentage of wealth invested in the human capital should be $ \varphi_H(\omega_t)/\big(\varphi_S(\omega_t)+\varphi_H(\omega_t)\big)$. This leads to the following definition of market equilibrium:
	\begin{definition}[Market Equilibrium]\label{de:MarketEqui}
		A triplet of functions $r_f$, $\varphi_S$, and $\varphi_H$ on $(0,1)$, which represent the risk-free rate and the price-endowment ratios of the stock and human capital, respectively, constitute a {\em market equilibrium} if
		\begin{align}\label{eq:IntraPersonalEquiInMarketEquil} (\hat{\bc}(\omega_t),\hat{\bu}_S(\omega_t),\hat{\bu}_H(\omega_t)):=\left(\frac{1}{\varphi_S(\omega_t)+\varphi_H(\omega_t)},\frac{\varphi_S(\omega_t)}{\varphi_S(\omega_t)+\varphi_H(\omega_t)},\frac{\varphi_H(\omega_t)}{\varphi_S(\omega_t)+\varphi_H(\omega_t)}\right)
		\end{align}
		is an intra-personal equilibrium strategy of the representative agent.
	\end{definition}

\subsection{Non-tradeable Labor Income}\label{subse:NontradeableLI}
In what follows, we explain why it is difficult to study asset pricing when the labor income is not tradeable. In this case, the agent's wealth process takes the form
\begin{align}
  d X_t / X_t &= \lf[ - c_t + r_{f,t} (1-u_{S,t} ) \rh] d t  + u_{S,t} \lf( \frac{d S_t + D_td t}{S_t} \rh) +  (L_t/X_t)dt,\label{eq:WealthEqNonTrade}
\end{align}
where $r_{f,t}dt $ and $(d S_t + D_td t)/S_t$ are the net returns of the risk-free asset and stock in the period $[t,t+dt)$ to be solved in the equilibrium, $c_t$ is a stochastic process representing the percentage of wealth consumed, $u_{S,t}$ is another process representing the percentage of wealth invested in the stock, and $X_t$ represents the agent's wealth, i.e., the value of her investment in the risk-free asset and stock plus the labor income earned at time $t$. Note that the agent is a price taker, so she takes the asset returns as exogenously given. Suppose that both $r_{f,t}dt $ and $(d S_t + D_td t)/S_t$ are determined by an exogenous market variable $Z_t$ (which can be of high dimension), and, without loss of generality, assume that $L_t/\bar C_t$ is also determined by $Z_t$.

To study intra-personal equilibrium strategies in the presence of time inconsistency, we need to focus on so-called feedback strategies, in which the agent's action at each time $t$ is a function of $t$ and the controlled path up to time $t$. See a detailed discussion in Section 4 of \citet{HeZhou2021:WhoAreI}. The wealth equation \eqref{eq:WealthEqNonTrade} implies that the agent's action $(c_t,u_{S,t})$ depends not only on the exogenous market variable $Z_t$ but also on the agent's income-wealth ratio $L_t/X_t$, which is endogenous in the agent's portfolio selection problem. Thus, we can represent the agent's trading strategy by some functions $\bc$ and $\bu_S$ of $(z,y)$ so that the percentages of wealth consumed and invested in the stock at time $t$ are $\bc(z,y)$ and $\bu_S(z,y)$, respectively, when $(Z_t,L_t/X_t)=(z,y)$.

In market equilibrium, the strategy $(\bc,\bu_S)$ chosen by the agent should clear the market. In other words,
\begin{align}
  &\bc(Z_t,L_t/X_t^{\bc,\bu_S}) = \bar C_t/X_t^{\bc,\bu_S} = \bar C_t/S_t,\label{eq:EquiCondNTConsum}\\
  &\bu_S(Z_t,L_t/X_t^{\bc,\bu_S}) = 1,\label{eq:EquiCondNTPort}
\end{align}
where $X_t^{\bc,\bu_S}$ stands for wealth process under the strategy $(\bc,\bu_S)$, and the second equality in \eqref{eq:EquiCondNTConsum} is the case because $X_t^{\bc,\bu_S}=S_t$ in the market equilibrium. The stock price-endowment ratio $S_t/\bar C_t$ should be independent of the agent's income-wealth ratio and is expected to be a function of the market $Z_t$. Therefore, the right-hand side of both \eqref{eq:EquiCondNTConsum} and \eqref{eq:EquiCondNTPort} does not depend on the agent's income-wealth ratio, while their left-hand side depends on this ratio. Therefore, we cannot expect equations \eqref{eq:EquiCondNTConsum} and \eqref{eq:EquiCondNTPort} to hold for {\em any} possible level of income-wealth ratio; we can only demand these equations to hold along the path of the income-wealth ratio $L_t/X_t^{\bc,\bu_S}$ under the strategy $(\bc,\bu_S)$ that clears the market. Note that $X_t^{\bc,\bu_S}=S_t$ in the market equilibrium and let $S_t/\bar C_t$ and $L_t/\bar C_t$ be $\varphi_S(Z_t)$ and $\ell(Z_t)$, respectively, for some functions $\varphi_S$ and $\ell$. Then, along the path of the income-wealth ratio $L_t/X_t^{\bc,\bu_S}$, equations \eqref{eq:EquiCondNTConsum} and \eqref{eq:EquiCondNTPort} reduce to
\begin{align}
  &\bc\big(Z_t,\ell(Z_t)/\varphi_S(Z_t)\big) = 1/\varphi_S(Z_t),\label{eq:EquiCondNTConsum2}\\
  &\bu_S\big(Z_t,\ell(Z_t)/\varphi_S(Z_t)\big)=1,\label{eq:EquiCondNTPort2}
\end{align}
which need to hold for any possible value of $Z_t$. To derive the market equilibrium, we need to find (i) the risk-free return rate as a function of $Z_t$, denoted as $r_f(Z_t)$, (ii) the stock price-endowment ratio $\varphi_S(Z_t)$, and (iii) the agent's strategy $(\bc,\bu_S)$ as a function of all possible values $(z,y)$ that the market state and the agent's income-wealth ratio can take, such that (a) $(\bc,\bu_S)$ is an intra-personal equilibrium strategy for the agent and (b) the clearing conditions \eqref{eq:EquiCondNTConsum2} and \eqref{eq:EquiCondNTPort2} hold.

Compared to the case of tradeable labor income, there are two genuine difficulties when the labor income is not tradeable. First, \eqref{eq:EquiCondNTConsum2} and \eqref{eq:EquiCondNTPort2} do not provide the information of $\bc(z,y)$ and $\bu_S(z,y)$ for all values of $(z,y)$. In consequence, $(\bc,\bu_S)$ is part of the solution when we solve the market equilibrium. By contrast, in the case of tradeable labor income, the agent's intra-personal equilibrium strategy is completely known in market equilibrium. Second, with non-tradeable labor income, the agent's action depends on her income-wealth ratio, while this ratio is irrelevant in the case of tradeable labor income. This additional state variable makes it more challenging to solve the agent's intra-personal equilibrium and market clearing conditions \eqref{eq:EquiCondNTConsum2} and \eqref{eq:EquiCondNTPort2}. We therefore decide to focus on the case of tradeable labor income in the present paper, and leave the case of non-tradeable labor income for future research.

	\section{Main Result} \label{sec:main_result}
	In this section, we derive the asset prices in market equilibrium. We impose the following growth condition, which is standard in the asset pricing literature:
	\begin{assumption}\label{as:GrowthCondition}
		\begin{align*}
			\delta_+:=\phi-(1-\gamma)\left(\mu_C +\kappa\sigma_C - \frac{\gamma}{2}\sigma_C^2\right)>0,\\
			\delta_-:=\phi-(1-\gamma)\left(\mu_C -\kappa \sigma_C - \frac{\gamma}{2}\sigma_C^2\right)>0.
		\end{align*}
	\end{assumption}
	Note that $\mu_C \pm \kappa \sigma_C$ stands for the mean growth rate of the aggregate endowment process under the most optimistic model and most pessimistic model, respectively, in the ambiguity set. Thus, $\delta_\pm$ stand for the percentages of wealth consumed by the representative agent, i.e., her consumption propensities, in market equilibrium under these two models. %and optimistic belief of the aggregate endowment process. See for instance \citet{Merton:1973ICAPM}. Similarly, $\delta_-$ stands for the consumption propensity of the representative agent with expected utility preferences and pessimistic belief of the aggregate endowment process.

	We first characterize the intra-personal equilibrium strategy given the asset prices, which amounts to computing the limit in \eqref{eq:equilibrium-def}.
	
	\begin{proposition} \label{prop:limit_labor}
		Let Assumptions \ref{as:SDE} and \ref{as:GrowthCondition} hold. Suppose that $r_f$, $\mu_S$, $\mu_H$, $\sigma_S$, and $\sigma_H$ are continuous and bounded on $(0,1)$. Let $(\hat{\bc}, \hat{\bu} )$ be a given strategy such that $\hat{\bc}$ and $\hat{\bu}$ are bounded on $(0,1)$, that
\begin{align}
  \lim_{T\rightarrow \infty}\sup_{\theta\in\Theta}\E_t^\theta\left[\int_T^\infty e^{-\phi(s-t)}\left|U(\bc_{t,\eps,c}(\omega_s) X^{\bc_{t,\eps,c},\bu_{t,\eps,u}}_s)\right|ds|X^{\bc_{t,\eps,c},\bu_{t,\eps,u}}_t=x,\omega_t=\omega\right]=0\label{eq:ConvergentTail}
\end{align}
for any $x>0$, $\omega\in (0,1)$, sufficiently small $\eps\ge 0$, $c>0$, and $u=(u_S,u_H)\in \R^2$, and that $\hat{\bu}$ are bounded on $(0,1)$ and that $\oV_t^{\hat{\bc},\hat{\bu}}$ and $\uV_t^{\hat{\bc},\hat{\bu}}$ are twice continuously differentiable functions of $(x,\omega)$ on $(0,\infty)\times (0,1)$ with the derivatives being polynomial in $x\in (0,\infty)$, uniformly in $\omega\in (0,1)$ (i.e., being bounded by $K(1+x^{-a}+x^a)$ for some constants $K>0$ and $a>0$). Then, for any $x>0$, $\omega\in (0,1)$, $c>0$, and $u=(u_S,u_H)\in \R^2$,
		\begin{align}
			&\lim_{\eps\downarrow 0} \frac{ \uV^{\bc_{t,\eps,c},\bu_{t,\eps,u}}_t(x,\omega) - \uV^{\hat{\bc},\hat{\bu}}_t(x,\omega)}{\eps} = U(cx) + \inf_{\bar \theta\in[-\kappa,\kappa] } \CA_{\bar \theta}^{c,u} \uV_{t}^{\hat{\bc}, \hat{\bu}}(x,\omega),\label{eq:PertubatedValueRateWC}\\
			&\lim_{\eps\downarrow 0} \frac{ \oV^{\bc_{t,\eps,c},\bu_{t,\eps,u}}_t(x,\omega) - \oV^{\hat{\bc},\hat{\bu}}_t(x,\omega)}{\eps} = U(cx) + \sup_{\bar \theta\in[-\kappa,\kappa] } \CA_{\bar \theta}^{c,u} \oV^{\hat{\bc}, \hat{\bu}}_t(x,\omega),\label{eq:PertubatedValueRateBC}
		\end{align}
		where the operator $\CA_{\bar \theta}^{c,u}$ is defined to be
		\begin{align*}
			&\CA_{\bar \theta}^{c,u} g(x,\omega):= -\phi g(x,\omega)
			+ \Big[-c + \lf(1-u_S-u_H\rh)r_f(\omega) + u_S\big(\mu_S(\omega) +\bar\theta \sigma_{S,1}(\omega)  \big)\\
			&\qquad + u_H\big(\mu_H(\omega)+\bar\theta \sigma_{H,1}(\omega)
			\big)\Big]xg_x(x,\omega) +
			\mu_{\omega}(\omega) g_\omega(x,\omega)  + \frac{1}{2} \sigma_{\omega}^2 (\omega) g_{\omega \omega}(x,\omega) \\
			&\qquad+ \frac{1}{2}\Big[\big(u_S\sigma_{S,1}(\omega)+u_H\sigma_{H,1}(\omega)\big)^2 + \big(u_S\sigma_{S,2}(\omega)+u_H\sigma_{H,2}(\omega)\big)^2\\
			& \qquad + 2 \rho\big(u_S\sigma_{S,1}(\omega)+u_H\sigma_{H,1}(\omega)\big)\big(u_S\sigma_{S,2}(\omega)+u_H\sigma_{H,2}(\omega)\big) \Big]x^2g_{xx}(x,\omega)\\
			&\quad + \Big[\rho \big(u_S\sigma_{S,1}(\omega)+u_H\sigma_{H,1}(\omega)\big)+ \big(u_S\sigma_{S,2}(\omega)+u_H\sigma_{H,2}(\omega)\big)  \Big]\sigma_{\omega}(\omega) xg_{x\omega}(x,\omega).
		\end{align*}
		%		and $\mu_S$, $(\sigma_{S,1},\sigma_{S,2})$, $\mu_H$, and $(\sigma_{H,1},\sigma_{H,2})$ are given as in \eqref{eq:StockDrift}, \eqref{eq:StockVol}, \eqref{eq:HCDrift}, and \eqref{eq:HCVol}, respectively. Consequently,
		In consequence,
		\begin{align}
			&\Gamma^{\hat{\bc}, \hat{\bu}}(x,\omega;c,u_S,u_H):=\lim_{\eps\downarrow 0} \frac{ V^{\bc_{t,\eps,c},\bu_{t,\eps,u}}(x,\omega) - V^{\hat{\bc},\hat{\bu}}(x,\omega)}{\eps}\notag\\
			&= U(cx) + \alpha \inf_{\bar\theta\in[-\kappa,\kappa] } \CA_{\bar \theta}^{c,u} \uV^{\hat{\bc}, \hat{\bu}}(x,\omega) +(1-\alpha)\sup_{\bar \theta\in[-\kappa,\kappa] } \CA_{\bar\theta}^{c,u} \oV^{\hat{\bc}, \hat{\bu}}(x,\omega).\label{eq:Gamma}
		\end{align}
	\end{proposition}

The transversality condition \eqref{eq:ConvergentTail} ensures that the $\alpha$-MEU value of the consumption stream under $(\bc_{t,\eps,c},\bu_{t,\eps,u})$ is well defined and finite.Given Proposition \ref{prop:limit_labor} and Definition \ref{de:MarketEqui}, to derive the market equilibrium, we only need to find $(r_f,\varphi_S,\varphi_H)$ such that for $(\hat{\bc},\hat{\bu}_S,\hat{\bu}_H)$ as defined in \eqref{eq:IntraPersonalEquiInMarketEquil},
	\begin{align}
		(\hat{\bc}(\omega),\hat{\bu}_S(\omega),\hat{\bu}_H(\omega))\in \underset{(c,u_S,u_H)}{\mathrm{argmax}} \, \Gamma^{\hat{\bc}, \hat{\bu}}(x,\omega;c,u_S,u_H)\label{eq:IntraPersonalEquiOptimality}
	\end{align}
	for all $x>0$ and $\omega\in (0,1)$. It is thus crucial to compute $\Gamma^{\hat{\bc}, \hat{\bu}}(x,\omega;c,u_S,u_H)$, which amounts to computing $\CA_{\bar \theta}^{c,u} \uV^{\hat{\bc}, \hat{\bu}}$ and $\CA_{\bar \theta}^{c,u} \oV^{\hat{\bc}, \hat{\bu}}$.
	
	By the wealth equation \eqref{eq:wealth-labor-income} and the utility function \eqref{eq:utility}, it is straightforward to see that $\uV^{\hat{\bc}, \hat{\bu}}(x,\omega)$ and $\oV^{\hat{\bc}, \hat{\bu}}(x,\omega)$ are homogeneous in $x$; i.e.,
	\begin{align}\label{eq:HomogeneityV}
		\uV^{\hat{\bc}, \hat{\bu}}(x,\omega) = x^{1-\gamma}\uV^{\hat{\bc}, \hat{\bu}}(1,\omega) + U(x)/\phi,\quad \oV^{\hat{\bc}, \hat{\bu}}(x,\omega) = x^{1-\gamma}\oV^{\hat{\bc}, \hat{\bu}}(1,\omega) + U(x)/\phi.
	\end{align}
	Given \eqref{eq:HomogeneityV}, to compute $\Gamma^{\hat{\bc}, \hat{\bu}}$, we need to know $\uV^{\hat{\bc}, \hat{\bu}}(1,\omega)$ and $\oV^{\hat{\bc}, \hat{\bu}}(1,\omega)$. In general, these two functions are not in closed form and depend on $(r_f,\varphi_S,\varphi_H)$, the asset prices that we want to derive in market equilibrium. This makes it difficult to derive the market equilibrium from \eqref{eq:IntraPersonalEquiOptimality}. Fortunately, for $(\hat{\bc},\hat{\bu}_S,\hat{\bu}_H)$ in market equilibrium, i.e., in \eqref{eq:IntraPersonalEquiInMarketEquil}, we conjecture that $\uV^{\hat{\bc}, \hat{\bu}}(1,\omega)$ and $\oV^{\hat{\bc}, \hat{\bu}}(1,\omega)$ take a simple form, as discussed in the following.
	
	Recall that the aggregate endowment process is a geometric Brownian motion under the benchmark probability measure. It is thus reasonable for us to conjecture that the total price of the stock and human capital is a constant multiple of the aggregate endowment, i.e., $S_t + H_t = \bar C_t/\delta$ for some positive constant $\delta$. In other words,
	\begin{align}
		\varphi_S(\omega)+\varphi_H(\omega) = 1/\delta,\quad \forall \omega \in (0,1). \label{eq:ConstantTotalPriceConsumptionRatio}
	\end{align}
	The constant $\delta$ is to be determined in market equilibrium. Then, for the strategy $(\hat{\bc},\hat{\bu})$ as given by \eqref{eq:IntraPersonalEquiInMarketEquil}, we have
	\begin{align}\label{eq:ConsumptioninEquilibriumConstant}
		\hat{\bc}(\omega)= \delta,\quad \omega \in (0,1).
	\end{align}
Moreover, we can derive from \eqref{eq:wealth-labor-income} that
\begin{align}
 & d X^{\hat{\bc},\hat{\bu}}_t / X^{\hat{\bc},\hat{\bu}}_t = -\delta d t  + \delta\left[\varphi_S(\omega_t) \lf( \frac{d S_t + D_td t}{S_t} \rh) + \varphi_H(\omega_t) \lf( \frac{d H_t + L_td t}{H_t} \rh)\right]\notag\\
 & = -\delta d t  + \delta \left[ \bar C_t^{-1}d (S_t+H_t) + \bar C_t^{-1}(D_t+H_t)d t\right]\notag\\
 & = d\bar C_t/\bar C_t=\mu_Cdt +  \sigma_C dB_{1,t}^0,\label{eq:WealthInEquilibriumGM}
\end{align}
where the second equality is the case because $\varphi_S(\omega_t) = S_t/\bar C_t$ and $\varphi_H(\omega_t) = H_t/\bar C_t$, and the third equality is the case because $D_t+H_t=\bar C_t$ and $S_t+H_t=\big(\varphi_S(\omega_t)+\varphi_H(\omega_t)\big)\bar C_t = \bar C_t/\delta$.

%Because in market equilibrium, the agent's wealth is equal to the sum of the stock price and human capital price, we have
%	\begin{align*}
%		X^{\hat{\bc}, \hat{\bu}}_t =S_t + H_t= \bar C_t/\delta.
%	\end{align*}
%	As a result, we have
%	\begin{align}\label{eq:WealthInEquilibriumGM}
%		dX^{\hat{\bc}, \hat{\bu}}_t = \mu_C X^{\hat{\bc}, \hat{\bu}}_tdt + \sigma_C X^{\hat{\bc}, \hat{\bu}}_t dB_{1,t}^0.
%	\end{align}
%	Moreover, by \eqref{eq:IntraPersonalEquiInMarketEquil},
%	\begin{align}\label{eq:ConsumptioninEquilibriumConstant}
%		\hat{\bc}(\omega)= \delta,\quad \omega \in (0,1).
%	\end{align}
	As a result, we can compute $\uV^{\hat{\bc}, \hat{\bu}}(1,\omega)$ and $\oV^{\hat{\bc}, \hat{\bu}}(1,\omega)$ in closed form.

	\begin{proposition} \label{prop:ov_uv_detailformula}
		Let Assumptions \ref{as:SDE} and \ref{as:GrowthCondition} hold. Consider $(\hat{\bc},\hat{\bu}_S,\hat{\bu}_H)$ with consumption propensity \eqref{eq:ConsumptioninEquilibriumConstant} and associated wealth process \eqref{eq:WealthInEquilibriumGM}. Suppose that $r_f$, $\mu_S$, $\mu_H$, $\sigma_S$, and $\sigma_H$ are continuous and bounded on $(0,1)$. Then,
		\begin{align}
			\uV^{\hat{\bc}, \hat{\bu}}(x,\omega) &= \left(\delta^{1-\gamma}\uv + U(\delta)/\phi\right)x^{1-\gamma}+ U(x)/\phi,\label{eq:uvGMFormula}\\
			\oV^{\hat{\bc}, \hat{\bu}}(x,\omega) &= \left(\delta^{1-\gamma}\ov + U(\delta)/\phi\right)x^{1-\gamma}+ U(x)/\phi,\label{eq:ovGMFormula}
		\end{align}
		where
		\begin{align}
			\uv:&= \frac{\left(\phi-(1-\gamma)\left(\mu_C -\kappa\sigma_C - \frac{\gamma}{2}\sigma_C^2\right)\right)^{-1}-\phi^{-1}}{1-\gamma},\label{eq:uvconstant}\\
			\ov:&=\frac{\left(\phi-(1-\gamma)\left(\mu_C +\kappa\sigma_C - \frac{\gamma}{2}\sigma_C^2\right)\right)^{-1}-\phi^{-1}}{1-\gamma}\label{eq:ovconstant}
		\end{align}
		with the values of $\uv$ and $\ov$ in the case when $\gamma=1$ defined as the limits of \eqref{eq:uvconstant} and \eqref{eq:ovconstant} as $\gamma\rightarrow 1$. Consequently,
		\begin{align}
		&	\CA_{\bar \theta}^{c,u} \uV^{\hat{\bc}, \hat{\bu}}(x,\omega) = - \left(\phi \delta^{1-\gamma}\uv + U(\delta)\right)x^{1-\gamma}-U(x)
			\notag\\
			& + \Big[-c + \lf(1-u_S-u_H\rh)r_f(\omega) + u_S\mu_S(\omega) + u_H\mu_H(\omega)\notag\\
			&  +\bar\theta (u_S+u_H)\sigma_C\Big]\left((1-\gamma)\uv + \phi^{-1}\right)\delta^{1-\gamma} x^{1-\gamma} \notag  \\
			& - \frac{1}{2}\Big[\big((u_S+u_H)\sigma_C\big)^2 + \big(u_S\sigma_{S,2}(\omega)+u_H\sigma_{H,2}(\omega)\big)^2 \notag\\
			&  + 2 \rho(u_S+u_H)\sigma_C\big(u_S\sigma_{S,2}(\omega)+u_H\sigma_{H,2}(\omega)\big) \Big] \notag\\
			& \times \gamma \left((1-\gamma)\uv + \phi^{-1}\right)\delta^{1-\gamma}x^{1-\gamma}\label{eq:GeneratorOnContinuitionValueLow}
		\end{align}
		and $\CA_{\bar \theta}^{c,u} \oV^{\hat{\bc}, \hat{\bu}}(x,\omega)$ is obtained by replacing $\uv$ in the above with $\ov$. Moreover,for any $u_S$ and $u_H$,
		\begin{align}
			&\Gamma^{\hat{\bc}, \hat{\bu}}(x,\omega;c,u_S,u_H)  = x^{1-\gamma}\bigg\{U(c)- \left[\phi \delta^{1-\gamma}\big(\alpha\uv + (1-\alpha)\ov\big) + U(\delta)\right]
			\notag\\
			& + \Big[-c + \lf(1-u_S-u_H\rh)r_f(\omega) + u_S\mu_S(\omega) + u_H\mu_H(\omega)\Big]\left((1-\gamma)\big(\alpha\uv + (1-\alpha)\ov\big) + \phi^{-1}\right)\delta^{1-\gamma}\notag\\
			&  +|u_S+u_H|\sigma_C\left[- \alpha\kappa \big((1-\gamma)\uv + \phi^{-1}\big) + (1-\alpha) \kappa \big((1-\gamma)\ov + \phi^{-1}\big)\right] \delta^{1-\gamma} \notag  \\
			& - \frac{1}{2}\Big[\big((u_S+u_H)\sigma_C\big)^2 + \big(u_S\sigma_{S,2}(\omega)+u_H\sigma_{H,2}(\omega)\big)^2 + 2 \rho(u_S+u_H)\sigma_C\big(u_S\sigma_{S,2}(\omega)+u_H\sigma_{H,2}(\omega)\big) \Big] \notag\\
			& \times \gamma \left((1-\gamma)\big(\alpha\uv + (1-\alpha)\ov\big) + \phi^{-1}\right)\delta^{1-\gamma}\bigg\}.\label{eq:GammaClosedForm}
		\end{align}
	\end{proposition}
	
The above proposition presents the closed-form of $\Gamma^{\hat{\bc}, \hat{\bu}}(x,\omega;c,u_S,u_H)$, the rate of change in the $\alpha$-MEU preference value when the agent deviates from $(\hat{\bc}, \hat{\bu})$ by taking an alternative strategy $(c,u_S,u_H)$ in an infinitesimally small period. We can observe from \eqref{eq:GammaClosedForm} that $\Gamma^{\hat{\bc}, \hat{\bu}}(x,\omega;c,u_S,u_H)$ is concave in $(c,u_S,u_H)$ in the region with $u_S+u_H\ge 0$ and in the region with $u_S+u_H<0$. However, when $- \alpha\kappa \big((1-\gamma)\uv + \phi^{-1}\big) + (1-\alpha) \kappa \big((1-\gamma)\ov + \phi^{-1}\big)>0$, $\Gamma^{\hat{\bc}, \hat{\bu}}(x,\omega;c,u_S,u_H)$ is {\em not} globally concave in the entire region of $(u_S,u_H)$. Note that the above inequality holds when $\alpha$ is small, i.e., when the agent is ambiguity seeking.
 Similar observations have been made in portfolio selection with $\alpha$-MEU preferences in single-period settings; see \citet[Section 1.3.1]{BossaertsEtal2010:AmbiguityInAssetMarkets} and \citet[Figure 3]{AnthropelosSchneider2024:OptimalInvestment}.

	Now, we are ready to present the main result.
	\begin{theorem} \label{thm:main}
%		Assume that the agent is not allowed to invest more than her wealth in the risk-free asset.
Let Assumptions \ref{as:SDE} and \ref{as:GrowthCondition} hold, and assume
		\begin{align}
			\min(\delta_+,\delta_-) >\max\Big\{&\sup_{x\in (0,1)}\left[\mu'_\omega(x)+(1-\gamma)\rho\sigma_C\sigma_\omega'(x)\right],\notag\\
			&\sup_{x\in (0,1)}\left[2\big(\mu'_\omega(x)+(1-\gamma)\rho\sigma_C\sigma_\omega'(x)\big)+(\sigma_\omega')^2(x)\right]\Big\}.\label{eq:GrowthCondition2}
		\end{align}
Assume either (i) the agent is not allowed to invest more than her wealth in the risk-free asset or (ii) $- \alpha\kappa \big((1-\gamma)\uv + \phi^{-1}\big) + (1-\alpha) \kappa \big((1-\gamma)\ov + \phi^{-1}\big)\le 0$.
		Then, there exists a unique market equilibrium in the class of triplets $(r_f,\varphi_S,\varphi_H)$ such that $r_f$ is continuous and bounded, $\varphi_i,i\in\{S,H\}$ are twice continuously differentiable with bounded derivatives, $\varphi_i,i\in\{S,H\}$ are bounded from below by a positive constant, and $\varphi_S+\varphi_H$ is constant on $(0,1)$. Moreover, in the equilibrium,
		\begin{align}
			r_f %= \mu_C - \gamma\sigma_C^2 + \phi-(1-\gamma)\left(\mu_C - \frac{\gamma}{2}\sigma_C^2\right) + \gamma \theta^*\sigma_C\\
			= \phi+ \gamma\mu_C -\frac{\gamma(1+\gamma)}{2}\sigma_C^2+ \gamma \theta^*\sigma_C,\label{eq:MarketEquiRiskFree}
		\end{align}
		where
		\begin{align}
			\theta^*:&= \frac{-\alpha \kappa\delta_-^{-1} + (1-\alpha)\kappa\delta_+^{-1}}{\alpha\delta_-^{-1} +(1-\alpha)\delta_+^{-1}},\label{eq:MarketEquiModel}
		\end{align}
		$\varphi_S$ is the unique solution to the following ODE:
		\begin{align}\label{eq:ODEPriceEndwomentRatio}
			\frac{1}{2}\sigma_{\omega}^2(\omega) \varphi_S''(\omega)
			+ \big[\mu_{\omega}(\omega) + (1-\gamma)\rho\sigma_C\sigma_{\omega}(\omega) \big] \varphi_S'(\omega)
			-\delta \varphi_S(\omega) + \omega =0,\; \omega \in (0,1),
		\end{align}
		and $\varphi_H = 1/\delta-\varphi_S$ and is the unique solution to the following ODE:
		\begin{align}\label{eq:ODEPriceEndwomentRatioHC}
			\frac{1}{2}\sigma_{\omega}^2(\omega) \varphi_H''(\omega)
			+ \big[\mu_{\omega}(\omega) + (1-\gamma)\rho\sigma_C\sigma_{\omega}(\omega) \big] \varphi_H'(\omega)
			-\delta \varphi_S(\omega) + 1-\omega =0,\; \omega \in (0,1),
		\end{align}
		where
		\begin{align}
			\delta &:= \phi-(1-\gamma)\left(\mu_C +\theta^*\sigma_C - \frac{\gamma}{2}\sigma_C^2\right)\label{eq:MarketEquiG}
		\end{align}
		stands for the representative agent's consumption propensity, i.e., her percentage of wealth consumed, in the equilibrium.
	\end{theorem}

The market is in equilibrium if and only if the strategy that clears the market, i.e., the strategy $(\hat{\bc},\hat{\bu}_S,\hat{\bu}_H)$ as defined in \eqref{eq:IntraPersonalEquiInMarketEquil} is an intra-personal strategy. As discussed following Proposition \ref{prop:limit_labor}, $\big(\hat{\bc}(\omega),\hat{\bu}_S(\omega),\hat{\bu}_H(\omega)\big)$  must be the optimizer of $\Gamma^{\hat{\bc}, \hat{\bu}}(x,\omega;c,u_S,u_H)$ in $(c,u_S,u_H)$. The first-order condition of the optimality then implies a set of equations from which we can solve the risk-free rate as in \eqref{eq:MarketEquiRiskFree} and derive that the asset price-endowment ratios satisfy the ODEs \eqref{eq:ODEPriceEndwomentRatio} and \eqref{eq:ODEPriceEndwomentRatioHC}. The existence and uniqueness of classical solutions to these ODEs are nontrivial to prove because $\sigma_\omega(\omega)$ degenerates to zero when $\omega$ goes to 0 and 1 and thus these two ODEs are not uniformly elliptic. One of our methodological contributions is to prove the existence and uniqueness of the solutions to these ODEs; see Appendix \ref{appx:ODE}. Condition \eqref{eq:GrowthCondition2} is a technical assumption that is needed for the proof, and it is satisfied by reasonable parameter values; see Section \ref{se:Numerical}.

As discussed following Proposition \ref{prop:ov_uv_detailformula}, $\Gamma^{\hat{\bc}, \hat{\bu}}(x,\omega;c,u_S,u_H)$ in $(c,u_S,u_H)$ is not globally concave in $(c,u_S,u_H)$ in general. Thus, with the risk-free rate and assert price-endowment ratios as in \eqref{eq:MarketEquiRiskFree}, \eqref{eq:ODEPriceEndwomentRatio}, and \eqref{eq:ODEPriceEndwomentRatioHC}, which are derived from the first-order condition of the optimality of $\big(\hat{\bc}(\omega),\hat{\bu}_S(\omega),\hat{\bu}_H(\omega)\big)$, it is possible that $\big(\hat{\bc}(\omega),\hat{\bu}_S(\omega),\hat{\bu}_H(\omega)\big)$ is not globally optimal. In this case, the strategy that clears the market is not an intra-personal equilibrium strategy, so the market equilibrium does not exist. In Theorem \ref{thm:main}, we assume either that the agent is not allowed to invest more than her wealth in the risk-free asset or that $- \alpha\kappa \big((1-\gamma)\uv + \phi^{-1}\big) + (1-\alpha) \kappa \big((1-\gamma)\ov + \phi^{-1}\big)\le 0$. Under either assumption, $\Gamma^{\hat{\bc}, \hat{\bu}}(x,\omega;c,u_S,u_H)$ is globally concave in the feasible region of $(c,u_S,u_H)$; consequently, the local optimality of $\big(\hat{\bc}(\omega),\hat{\bu}_S(\omega),\hat{\bu}_H(\omega)\big)$ implies global optimality and the market equilibrium exists.

It is observed in the literature that certain non-expected utility preferences are not globally concave and the market equilibrium in a representative-agent economy with these preferences may not exist. See for instance \citet[Section 1.3.1]{BossaertsEtal2010:AmbiguityInAssetMarkets} and \citet[Section 5.2.4]{HeSun2025:DynamicPortfolioSelection}.

\iffalse
	\textcolor{red}{
\begin{Remark}\label{rmk:constraint}
	The constraint that the agent cannot invest more than her wealth in the risk-free asset (equivalently, $u_S + u_H \geq 0$) is a technical condition that ensures the strict concavity of $\Gamma^{\hat{\bc}, \hat{\bu}}(x,\omega;c,u_S,u_H)$ over the admissible strategy space. In time-consistent problems, the value function is typically globally concave, and first-order conditions characterize the global optimum. In our time-inconsistent setting, however, there is no globally concave value function. The intra-personal equilibrium is characterized by checking that the agent has no incentive to deviate at
	each instant, which requires $\Gamma^{\hat{\bc}, \hat{\bu}}$ to be strictly concave over a convex region. The constraint $u_S + u_H \geq 0$ defines this region and ensures that the first-order conditions are necessary and sufficient. Importantly, the market-clearing strategy satisfies $\hat{\bu}_S(\omega) + \hat{\bu}_H(\omega) = 1 > 0$, so the constraint never binds in equilibrium. It serves purely as a technical device for our proof methodology, unlike exogenously imposed portfolio constraints in the literature that bind in equilibrium and affect asset prices.
\end{Remark}
}

\fi
	
	\begin{proposition}\label{prop:SDF}
		Let the assumptions in Theorem \ref{thm:main} hold. Recall $\theta^*$ as defined in \eqref{eq:MarketEquiModel}.
		Then, the asset prices in the market equilibrium are the same as in an economy with a representative agent who is not ambiguous about the mean growth rate of the aggregate endowment and believes in the model $\P^{\theta^*}$. More precisely,
		\begin{align}
			&r_f = \lim_{s\downarrow t}\frac{1}{s-t}\left(\expect^{\theta^*}_t\left[e^{-\phi (s-t)}\left(\frac{\bar C_{s}}{\bar C_t} \right)^{-\gamma}\right]-1\right),\label{eq:equilibriumRiskFreeEQ}\\
			&S_t = \expect^{\theta^*}_t\left[\int_t^{\infty}e^{-\phi(s-t)}\left(\frac{\bar C_{s}}{\bar C_t} \right)^{-\gamma}D_sds\right],\label{eq:equilibriumStockEQ}\\
			& H_t = \expect^{\theta^*}_t\left[\int_t^{\infty}e^{-\phi(s-t)}\left(\frac{\bar C_{s}}{\bar C_t} \right)^{-\gamma}(\bar C_s -D_s)ds\right].\label{eq:equilibriumHCEQ}
		\end{align}
	\end{proposition}
	
	Proposition \ref{prop:SDF} shows that the equilibrium asset prices in our model with a representative agent who perceives ambiguity are the same as in an economy where the representative agent is not ambiguous about the endowment process and believes in the probabilistic model $\P^{\theta^*}$. Note that when the agent does not perceive ambiguity, her preferences become the classical expected utility. In consequence, it is known in the literature that the stochastic discount factor is the discounted marginal utility of the aggregate endowment, $e^{-\phi t}\bar C_t^{-\gamma}$. This explains \eqref{eq:equilibriumRiskFreeEQ}, \eqref{eq:equilibriumStockEQ}, and \eqref{eq:equilibriumHCEQ}. We call $\P^{\theta^*}$ the {\em ambiguity-adjusted probability measure}. Under this measure, the mean growth rate and volatility of the endowment process are $\mu_C+\sigma_C\theta^*$ and $\sigma_C$, respectively. Therefore, $\theta^*$ represents the difference in the mean growth rates of the endowment under the ambiguity-adjusted probability measure and under the benchmark probability measure, per unit of the volatility. We call $\theta^*$ the {\em ambiguity-implied density generator} and the smaller $\theta^*$ is, the more pessimistic the agent is under the ambiguity-adjusted probability measure.

	\section{Comparative Statics}\label{se:ComparativeStatics}
	
	By Theorem \ref{thm:main} and Proposition \ref{prop:SDF}, the agent's perceived degree of ambiguity $\kappa$ and her ambiguity aversion degree $\alpha$ affect the equilibrium asset prices through the ambiguity-implied density generator $\theta^*$. In the following, we first investigate the dependence of $\theta^*$ on $\alpha$ and $\kappa$.

	\begin{proposition}\label{prop:AmImMarketPriceRisk}
		Let Assumption \ref{as:GrowthCondition} hold and recall $\theta^*$ as defined in \eqref{eq:MarketEquiModel}.
		\begin{enumerate}
			\item[(i)] $\theta^*=0$ when $\kappa=0$.
			\item[(ii)] Fixing $\kappa>0$, $\theta^*$ is strictly decreasing in $\alpha \in [0,1]$.
			\item[(iii)] Suppose $\gamma=1$. Then, $\theta^*$ is strictly increasing in $\kappa$ (constant in $\kappa$ and strictly decreasing in $\kappa$, respectively) when $\alpha\in [0,1/2)$ (when $\alpha=1/2$ and when $\alpha\in (1/2,1]$, respectively).
			\item[(iv)] Suppose $\gamma>1$. Then, $\theta^*$ is strictly decreasing in $\kappa$ when $\alpha\in[1/2,1]$, strictly increasing in $\kappa$ when $\alpha=0$, and first strictly increasing and then strictly decreasing in $\kappa$ when $\alpha\in (0,1/2)$.
			\item[(v)] Suppose $\gamma<1$. Then, $\theta^*$ is strictly increasing in $\kappa$ when $\alpha\in[0,1/2]$ and first strictly decreasing and then strictly increasing in $\kappa$ when $\alpha\in (1/2,1]$.
		\end{enumerate}
	\end{proposition}
	
	Proposition \ref{prop:AmImMarketPriceRisk}-(ii) shows that the more ambiguity-averse the agent is, the more pessimistic she is under the ambiguity-adjusted probability measure. On the other hand, the dependence of $\theta^*$ on the perceived ambiguity level $\kappa$ depends on the risk aversion degree $\gamma$ and ambiguity aversion degree $\alpha$. First, when $\gamma>1$ and $\alpha\in (0,1/2)$ and when $\gamma\in(0,1)$ and $\alpha\in (1/2,1]$, $\theta^*$ is not monotone in $\kappa$. Second, for a low level of perceived ambiguity $\kappa$, $\theta^*$ is strictly decreasing in $\kappa$ when the agent is ambiguity averse (i.e., $\alpha\in (1/2,1]$) and is strictly increasing in $\kappa$ when the agent is ambiguity seeking (i.e., $\alpha\in [0,1/2)$). Third, when the agent is ambiguity neutral (i.e., $\alpha=1/2$), $\theta^*$ can still be dependent on $\kappa$ and its monotonicity depends on the value of $\gamma$. %This is different from the recursive $\alpha$-MEU model, where the
	
	Most experimental and empirical estimates of individuals' relative risk aversion degree lie in the range between 1 and 10; see for instance \citet{BlissPanigirtzoglou2004:OptionImpliedRiskAversion}. \citet{DimmockKouwenbergMitchellPeijnenburg2015:EstimatingAmbiguity} and \citet{DimmockEtal2015:AmbiguityAttitudes} use experimental data to estimate the ambiguity aversion degree $\alpha$ in the $\alpha$-MEU model with a specific ambiguity set that leads to the CEU model with neo-additive capacities. They find that the subjects in their experiment are largely heterogeneous in their ambiguity aversion degree and on average, the population exhibits ambiguity aversion, i.e., the estimate of $\alpha$ across all subjects is larger than $1/2$. Using a different measure of ambiguity aversion, \citet{DimmockEtal2016:AmbiguityAversionAndHousehold} also find in their experimental study that most subjects are ambiguous averse. The above experimental and empirical studies show that the most relevant case in Proposition \ref{prop:AmImMarketPriceRisk} is when $\gamma>1$ and $\alpha\in[1/2,1]$. In this case, the ambiguity-implied density generator $\theta^*$ is strictly decreasing in $\kappa$, showing that the more ambiguity the agent perceives, the more pessimistic she is under the ambiguity-adjusted probability measure.
	
	\begin{proposition}\label{prop:ComparativeStat}
		Let the same assumptions as in Theorem \ref{thm:main} hold.
		\begin{itemize}
			\item[(i)] The consumption propensity $\delta$ is strictly increasing in $\theta^*$ (constant in $\theta^*$ and strictly decreasing in $\theta^*$, respectively) when $\gamma >1$ (when $\gamma=1$ and when $\gamma\in(0,1)$, respectively).
			\item[(ii)] The risk-free rate $r_f$ is strictly increasing in $\theta^*$.
			\item[(iii)] The solution to \eqref{eq:ODEPriceEndwomentRatio} is strictly increasing in $\omega$, depends on $\gamma$, $\rho$, and $\sigma_C$ through $(1-\gamma)\rho\sigma_C$, and is strictly increasing in $(1-\gamma)\rho\sigma_C$ and strictly decreasing in $\delta$. The solution to \eqref{eq:ODEPriceEndwomentRatioHC} is strictly decreasing in $\omega$, depends on $\gamma$, $\rho$, and $\sigma_C$ through $(1-\gamma)\rho\sigma_C$, and is strictly decreasing in $(1-\gamma)\rho\sigma_C$ and $\delta$.
		\end{itemize}
	\end{proposition}
	
	In the market equilibrium, the aggregate market capitalization is proportional to the endowment, so the return of the market portfolio held by the representative agent is the same as the growth rate of the endowment. It is known in the literature that the investment return affects the consumption decision through the income effect and substitution effect; see for instance \citet{Epstein1988:RiskAversionAssetPrices}. When $\gamma$, which is also the reciprocal of the elasticity of intertemporal substitution, is larger than 1, the income effect dominates, so the agent consumes more today when the investment return becomes higher. This explains why when $\gamma>1$, the consumption propensity $\delta$ is increasing in $\theta^*$, which is positively related to the mean growth rate of the endowment process perceived by the agent and thus positively related to the mean investment return. When $\gamma<1$, the substitution effect dominates. In consequence, the consumption propensity $\delta$ is decreasing in $\theta^*$. When $\gamma=1$, the income and substitution effects cancel out and, consequently, $\delta$ does not depend on $\theta^*$.
	
	The risk-free rate does not depend on the dividend-endowment ratio $\omega$. This is because the risk-free rate is determined by comparing the investment in the risk-free asset and investment in the market portfolio. The return of the latter is the same as the endowment growth rate, which does not depend on the dividend-endowment ratio $\omega$, so the risk-free rate is independent of $\omega$. For the same reason, the risk-free rate does not depend on the correlation $\rho$ between the aggregate endowment and the dividend-endowment ratio. On the other hand, the higher $\theta^*$ is, the higher the endowment growth rate perceived by the agent and thus the higher the market portfolio return is. Consequently, the risk-free rate must become higher in the market equilibrium.

	The stock price-endowment ratio, $\varphi_S(\omega)$, is increasing in $\omega$. This is because the higher $\omega$ is, the higher the current dividend amount is and, consequently, the higher the stock value is. On the contrary, the human capital price endowment ratio is decreasing in $\omega$. Note that the increasing stock price endowment ratio $\varphi_S(\omega)$ in $\omega$ does not imply an increasing stock price dividend ratio $\varphi_S(\omega)/\omega$ in $\omega$. Indeed, our numerical study shows that the stock price-dividend ratio is decreasing in $\omega$; see Figure \ref{fig:PDRPVol} below.
	
	In market equilibrium, the agent's wealth is the total market capitalization of the risky assets and her consumption amount is the aggregate endowment. As a result, the market capitalization-endowment ratio is the reciprocal of the agent's consumption propensity $\delta$ and thus is strictly decreasing in $\delta$. Proposition \ref{prop:ComparativeStat}-(iii) shows that both the stock and human capital price-endowment ratios are also strictly decreasing in $\delta$.
	
	By Proposition \ref{prop:ComparativeStat}-(iii), the stock price-endowment ratio is strictly decreasing in (independent of and strictly decreasing in, respectively) $\rho$ when $\gamma>1$ (when $\gamma=1$ and when $\gamma\in (0,1)$, respectively). This can be explained as follows: By \eqref{eq:equilibriumStockEQ}, we conclude that
	\begin{align}
		S_t/C_t = \expect^{\theta^*}_t\left[\int_t^{\infty}e^{-\phi(s-t)}\left(\frac{\bar C_{s}}{\bar C_t} \right)^{1-\gamma}\frac{D_s}{C_s}ds\right].\label{eq:equilibriumStockEndowRatioEQ}
	\end{align}
	It is then straightforward to see that when $\gamma=1$, the correlation between the endowment growth $\bar C_{s}/\bar C_t$ and the dividend-endowment ratio $D_s/C_s$ does not have any effect on the stock price-endowment ratio $S_t/C_t$. When $\gamma>1$, the higher the correlation $\rho$ between $\bar C_{s}/\bar C_t$ and $D_s/C_s$ is, the less correlation between $\left(\bar C_{s}/\bar C_t \right)^{1-\gamma}$ and $D_s/C_s$ is and, consequently, the lower $S_t/C_t$ is. The case of $\gamma\in (0,1)$ can be discussed similarly.
	
	By Proposition \ref{prop:ComparativeStat}-(iii) and (i), $\varphi_S$ and $\varphi_H$ are strictly decreasing (strictly increasing and constant, respectively) in $\theta^*$ when $\gamma>1$ ($\gamma<1$ and $\gamma=1$, respectively). This can  be seen from \eqref{eq:equilibriumStockEndowRatioEQ}. When $\gamma>1$, the risk-adjusted consumption growth rate $\left(\bar C_{s}/\bar C_t \right)^{1-\gamma}$ is strictly decreasing in $\theta^*$, so the stock price-endowment ratio is strictly decreasing in $\theta^*$. The cases when $\gamma<1$ and when $\gamma=1$ can be explained similarly.

	Combining Propositions \ref{prop:AmImMarketPriceRisk} and \ref{prop:ComparativeStat}, we can derive the effect of the agent's ambiguity aversion degree $\alpha$ and perceived ambiguity level $\kappa$ on the risk-free rate and the stock and human-capital prices. In particular, with $\gamma>1$ and $\alpha\in[1/2,1]$, which is the empirically relevant case, when $\alpha$ becomes larger or $\kappa$ becomes larger, the risk-free rate and the agent's consumption propensity become lower and the stock and human capital prices become higher.

Next, we compute the stock risk premium and volatility. We assume that the benchmark probability measure $\P^0$ is the ground truth. Then, straightforward calculation yields the stock risk premium
	\begin{align}
		\mu_S(\omega)-r_f(\omega)
		= \gamma\sigma_C^2 + \gamma\rho\sigma_C\sigma_{\omega}(\omega)\frac{\varphi_S'(\omega)}{\varphi_S(\omega)}
		-\sigma_C \theta^*\label{conditional_equity_premium}
	\end{align}
	and volatility
	\begin{align}
		\begin{split}
			\|\sigma_S(\omega)\|=\sqrt{\sigma_{S,1}^2(\omega)+\sigma_{S,2}^2(\omega)}
			= \sqrt{\sigma_C^2 + 2\rho\sigma_C\sigma_{\omega}(\omega)\frac{\varphi_S'(\omega)}{\varphi_S(\omega)}+ \sigma_{\omega}^2(\omega) \lf(\frac{\varphi_S'(\omega)}{\varphi_S(\omega)} \rh)^2
			}.\label{conditional_volatility}
		\end{split}
	\end{align}
	The following proposition shows the dependence of the stock risk premium and volatility on $\theta^*$ in two special cases.
		\begin{proposition}\label{prop:RiskPremiumVol}
		Let the same assumptions as in Theorem \ref{thm:main} hold.
		\begin{enumerate}
			\item[(i)] Suppose that $\rho(1-\gamma)=0$. Then, the equity premium $\mu_S(\omega)-r_f(\omega)$ is strictly decreasing in $\theta^*$.
			\item[(ii)] Suppose that $\gamma=1$. Then, the stock volatility does not depend on $\theta^*$.
		\end{enumerate}
	\end{proposition}
	
	Combining Propositions \ref{prop:RiskPremiumVol} and \ref{prop:AmImMarketPriceRisk}, we can derive the dependence of the stock risk premium on the agent's perceived ambiguity level $\kappa$ and her ambiguity aversion degree $\alpha$ when $(1-\gamma)\rho=1$. In particular, when $\rho=0$, $\gamma>1$, and $\alpha\in[1/2,1]$, the risk premium is strictly increasing in $\alpha$ and $\kappa$.

	In general, the stock risk premium and volatility depend on $\varphi_S'(\omega)/\varphi_S(\omega)$. Following \citet{LongstaffPiazzesi2004:CorporateEarnings}, we call $\varphi_S'(\omega)/\varphi_S(\omega)$ the {\em elasticity} of the stock price with respect to the dividend-endowment ratio. In a special case, we have a closed-form formula of $\varphi_S'(\omega)/\varphi_S(\omega)$; see the corollary below:
	
	\begin{corollary}\label{Coro:SpecialCase}
		Let the same assumptions as in Theorem \ref{thm:main} hold. Suppose that $(1-\gamma)\rho=0$ and $\mu_\omega$ and $\sigma_\omega$ are given by \eqref{eq:QudraticDiffusion}. Then,
		\begin{align}
			\varphi_S(\omega) = \frac{\omega}{\lambda+\delta} + \frac{\lambda \overline{\omega}}{\delta(\lambda+\delta)},\quad \varphi_H(\omega) = \frac{1-\omega}{\lambda+\delta} + \frac{\lambda (1- \overline{\omega})}{\delta(\lambda+\delta)}\quad \omega \in (0,1). \label{eq:PriceEndowmentRatioClosedForm}
		\end{align}
		In consequence, the stock price-dividend ratio
		\begin{align}
			S_t/D_t= \varphi_S(\omega_t)/\omega_t = \frac{1}{\lambda+\delta} + \frac{\lambda \overline{\omega}}{\delta(\lambda+\delta)\omega_t}\label{eq:PDRatioSpecialCase}
		\end{align}
		is strictly decreasing in $\omega_t$, and the elasticity of the stock price
		\begin{align}
			\varphi_S'(\omega)/\varphi_S(\omega) = \left(\omega + \overline{\omega}\lambda/\delta\right)^{-1}\label{eq:ElasticitySpecialCase}
		\end{align}
		is strictly decreasing in $\omega$ and strictly increasing in $\delta$. The stock risk premium and volatility are
		\begin{align}
			&  \mu_S(\omega)-r_f(\omega)
			= \gamma\sigma_C^2 + \gamma\rho\sigma_C\nu\omega(1-\omega)\left(\omega + \overline{\omega}\lambda/\delta\right)^{-1}
			-\sigma_C \theta^*,\label{eq:RPSpecialCase}\\
			&\|\sigma_S(\omega)\|= \sqrt{\sigma_C^2 + 2\rho\sigma_C\nu \omega(1-\omega)\left(\omega + \overline{\omega}\lambda/\delta\right)^{-1}+ \lf(\nu \omega(1-\omega)\left(\omega + \overline{\omega}\lambda/\delta\right)^{-1} \rh)^2}.\label{eq:VolSpecialCase}
		\end{align}
		Moreover, when $\rho=0$, $\mu_S(\omega)-r_f(\omega)$ is independent of $\omega$ and $\|\sigma_S(\omega)\|$ is first strictly increasing and then strictly decreasing in $\omega$. When $\gamma=1$ and $\rho>0$, both $\mu_S(\omega)-r_f(\omega)$ and $\|\sigma_S(\omega)\|$ are first strictly increasing and then strictly decreasing in $\omega$. When $\gamma=1$ and $\rho<0$, $\mu_S(\omega)-r_f(\omega)$ is first strictly decreasing and then strictly increasing in $\omega$.
	\end{corollary}

	\section{Numerical Studies}\label{se:Numerical}
	
	In this section, we conduct a numerical experiment to complement the comparative statics studied in Section \ref{se:ComparativeStatics} and to quantitatively compute the equilibrium asset returns in our model.
	
	\subsection{Parameter Calibration}\label{subse:Calibration}
	
	To calibrate the model parameters, we collect the annual data on the personal consumption expenditures (PCE) price index and the aggregated corporate earnings and consumption from the National Income and Product Account documented by the Bureau of Economic Analysis in the period from 1933 to 2022.\footnote{The annual consumption data was obtained from the U.S. Bureau of Economic Analysis (BEA) National Income and Product Accounts (NIPA) \href{\beaurl}{Table 2.3.5}. The price index for personal consumption expenditures, used to calibrate real consumption growth rates in this study, was obtained from the U.S. Bureau of Economic Analysis \href{\beaurll}{NIPA Table 2.3.4}, with all values expressed in constant 2017 dollars. Corporate earnings(after tax) data were compiled from the U.S. Bureau of Economic Analysis \href{\corp}{NIPA Tables 6.19A through 6.19D}, which provide comprehensive industry-level data for the period 1929-2022.} Following \citet{SantosVeronesi2006:LaborIncome}, we define the aggregate consumption as the consumption of non-durables plus services. We follow the asset pricing literature, e.g., \citet{MehraRPrescottE:85ep} and \citet{CampbellCochrane:1999ByForceOfHabit}, to calibrate $\mu_C$ and $\sigma_C$ to the log growth rate of the real consumption growth per capita, which is computed by adjusting the aggregate consumption using the PCE price index and then dividing it by the population size.\footnote{The population data is obtained from the Census Bureau (\href{\cens}{U.S. Census Bureau data portal}.)} The calibrated parameter values are reported in Table \ref{ta:ParameterValues}.\footnote{Our estimates are consistent with those in the literature using different data periods, methods of calculating aggregate consumption, and methods of adjusting for inflation; see for instance \cite{CampbellCochrane:1999ByForceOfHabit}, \cite{LongstaffPiazzesi2004:CorporateEarnings}, and \citet{SantosVeronesi2006:LaborIncome}.}

\begin{figure}
		\centering
 		\includegraphics[width=0.6\linewidth]{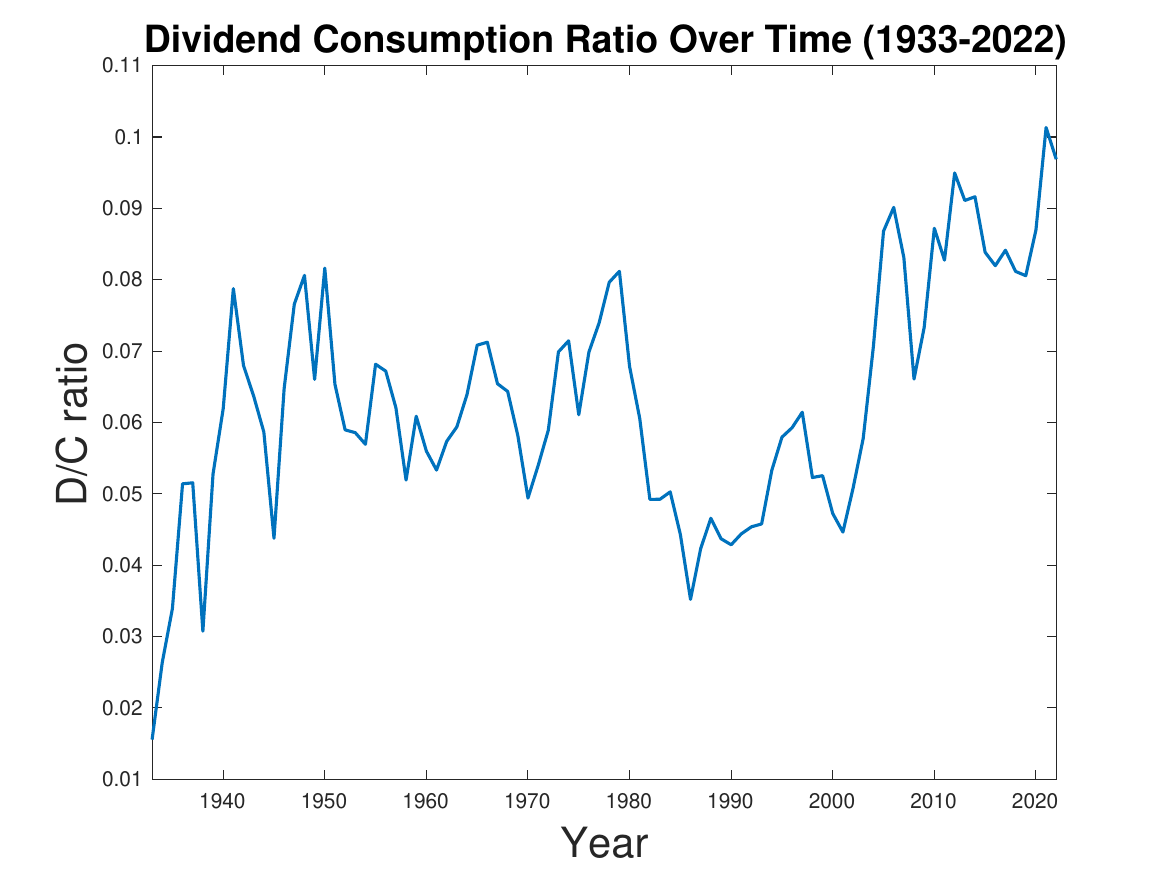}
		\caption{Dividend-endowment ratio in the period from 1933 to 2022.}
		\label{fig:dvidend_consumption_ratio}
	\end{figure}

	We follow \citet{LongstaffPiazzesi2004:CorporateEarnings} to estimate the stock dividend amount to be a fixed dividend payout ratio, assumed to be 50\%, multiplied by the aggregate corporate earnings. The dividend-endowment ratio is then computed to be the ratio of the estimated dividend amount and the consumption amount. We plot the historical dividend-endowment ratio in Figure \ref{fig:dvidend_consumption_ratio}, and we can observe that this ratio is not constant and is in the range [0,0.11] historically.
	
	We use the parametric form \eqref{eq:QudraticDiffusion} for $\mu_\omega$ and $\sigma_\omega$ and estimate $\lambda$, $\overline{\omega}$, $\nu$, and $\rho$ using the maximum likelihood estimation. The estimates are shown in Table \ref{ta:ParameterValues}. In particular, $\rho$ is positive, showing that the aggregate endowment and the dividend-endowment ratio are positively correlated. This is consistent with the estimates in the literature.\footnote{\citet{SantosVeronesi2006:LaborIncome} estimate the correlation between the labor income-endowment ratio and the aggregate endowment to be negative, which implies that the correlation between the dividend-endowment ratio and aggregate endowment is positive; see Section 3.1 therein. \citet{LongstaffPiazzesi2004:CorporateEarnings} estimate that the correlation between the dividend-endowment ratio, referred to as corporate fraction, and the aggregate endowment is positive; see Section 4 therein.}

	\begin{table}
		\centering
		\caption{Parameter values.}\label{ta:ParameterValues}
		\begin{tabular}{c|c|c|c|c|c|c|c}
			\hline
			$\phi$ & $\kappa$ &  $\mu_C$ & $\sigma_C$ & $\lambda$ & $\overline{\omega} $ &  $\rho$&$\nu$\\
			\hline
			0.0984 & 0.2 & 0.0231 & 0.0286 & 0.2232 & 0.0662 &  0.4637&0.1546\\
			\hline
		\end{tabular}
	\end{table}
	
	We set the discount rate $\phi$ to be 0.0984, the same value as in \citet{SantosVeronesi2006:LaborIncome} and also close to the discount rate used in \citet{CampbellCochrane:1999ByForceOfHabit}. In the estimation of the mean growth rate of the endowment, it is known that the standard error is $\sigma_C/\sqrt{T}$, where $T$ is the number of years of available data. Thus, a 95\% confidence interval can be constructed as $[\mu_C-(1.96/\sqrt{T})\sigma_C,\mu_C+(1.96/\sqrt{T})\sigma_C]$, where $1.96$ is the 97.5\% quantile of the standard normal distribution. With the data from 1933 to 2022, we have $T=90$ and $1.96/\sqrt{T}=0.207$. Thus, we set the agent's perceived level of ambiguity $\kappa=0.2$.
	
	For the risk aversion degree $\gamma$ and ambiguity aversion degree $\alpha$, we set them to be various values in the following study. Note that the reasonable value range of $\gamma$ is [1,10]; see for instance \citet{BlissPanigirtzoglou2004:OptionImpliedRiskAversion}. Thus, we set $\gamma \in \{0.5,1,5,10\}$ in the following study, where the choice of $\gamma=0.5$ is of theoretical interest because it is different from the regime that $\gamma>1$. On the other hand, the perceived level of ambiguity $\kappa$ and ambiguity aversion degree $\alpha$ are combined into a single parameter $\theta^*$ to affect asset prices. Because $\theta^*$ is a weighted average of $-\kappa$ and $\kappa$, it can take values in the range $[-\kappa,\kappa]$ as we vary the value of $\alpha$. Given that we set $\kappa=0.2$, we choose different values of $\theta^*$ in $[-0.2,0.2]$ in the following. For all the parameter values used in our numerical study, the assumptions in Theorem \ref{thm:main} hold.
	
	\subsection{Conditional Moments of Asset Returns}
	
\begin{figure}
		\centering
		
		\begin{subfigure}[b]{0.32\textwidth}
			\centering
			\includegraphics[width=\textwidth]{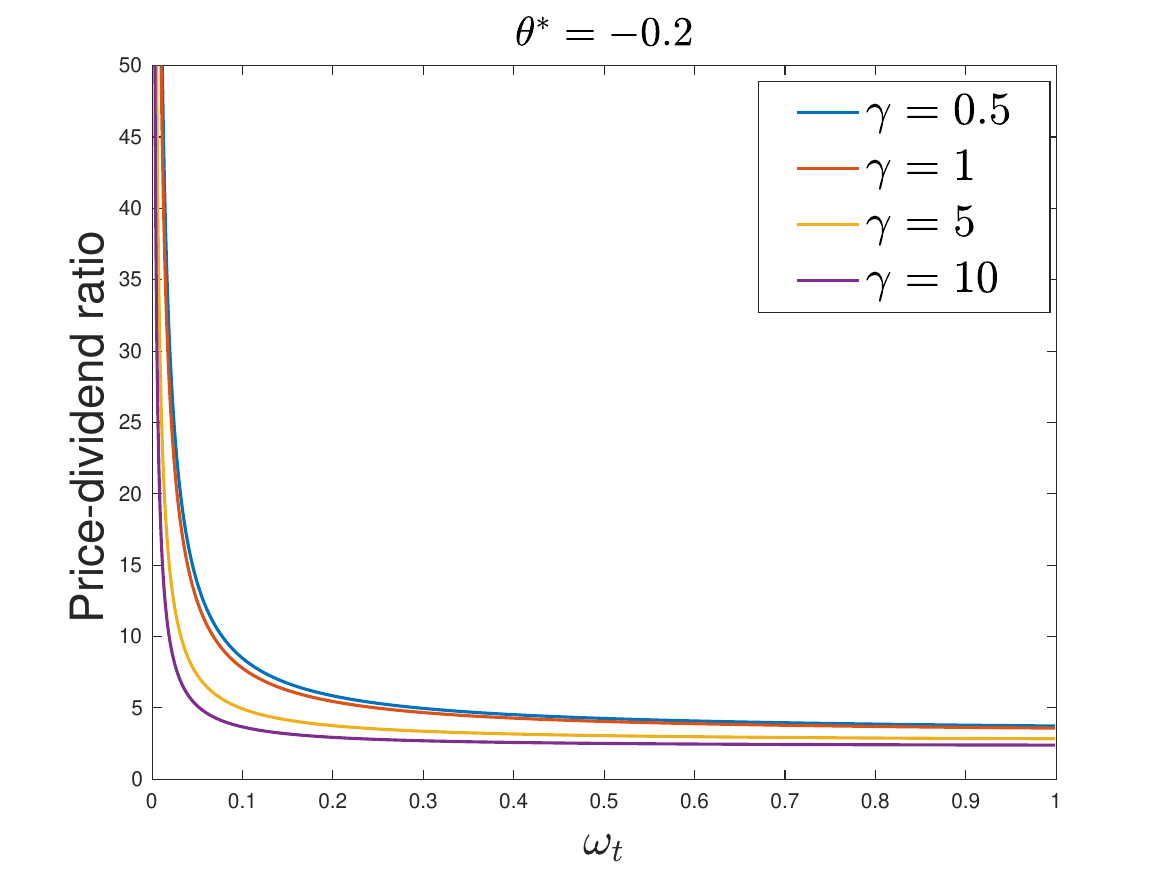}
			%	\caption{$\theta^*=-0.2$}
			%		\label{fig:left}
		\end{subfigure}
		\hfill
		\begin{subfigure}[b]{0.32\textwidth}
			\centering
			\includegraphics[width=\textwidth]{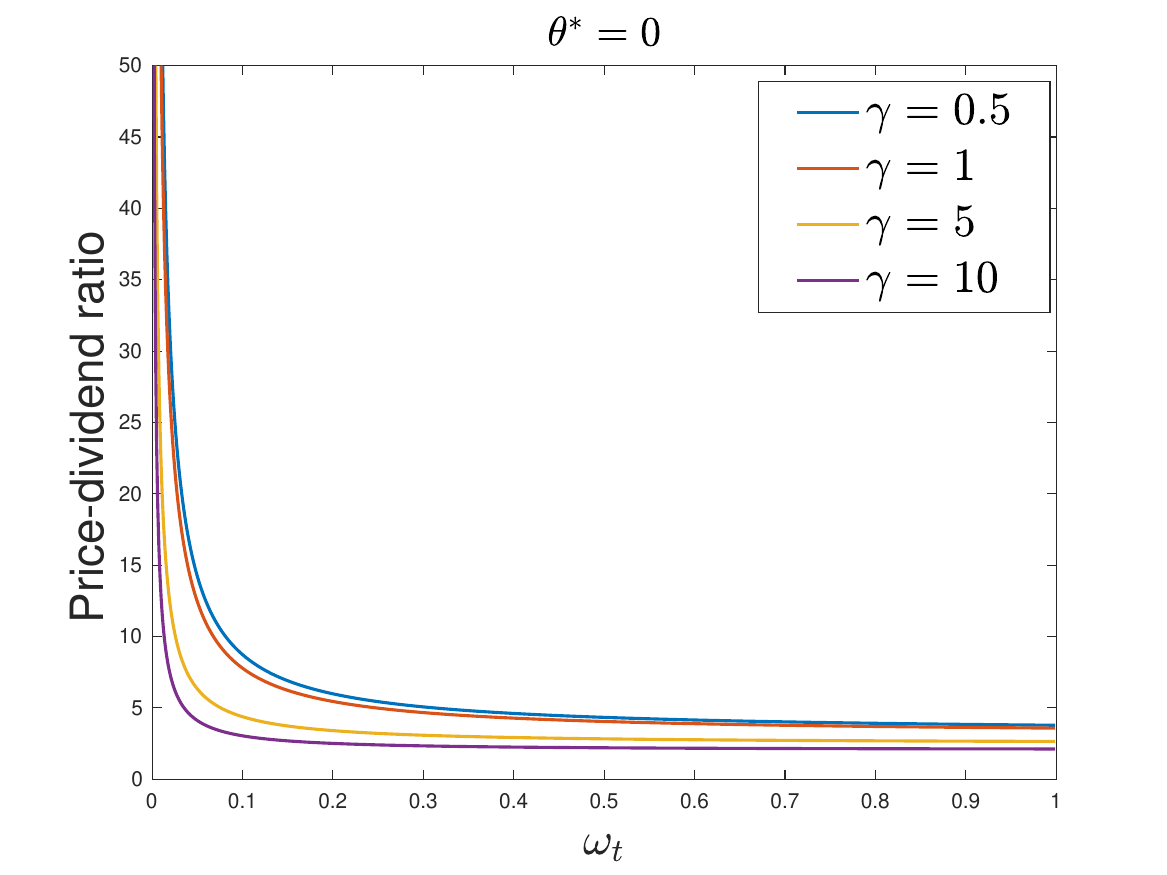}
			%	\caption{$\theta^*=0$}
			%		\label{fig:center}
		\end{subfigure}
		\hfill
		\begin{subfigure}[b]{0.32\textwidth}
			\centering
			\includegraphics[width=\textwidth]{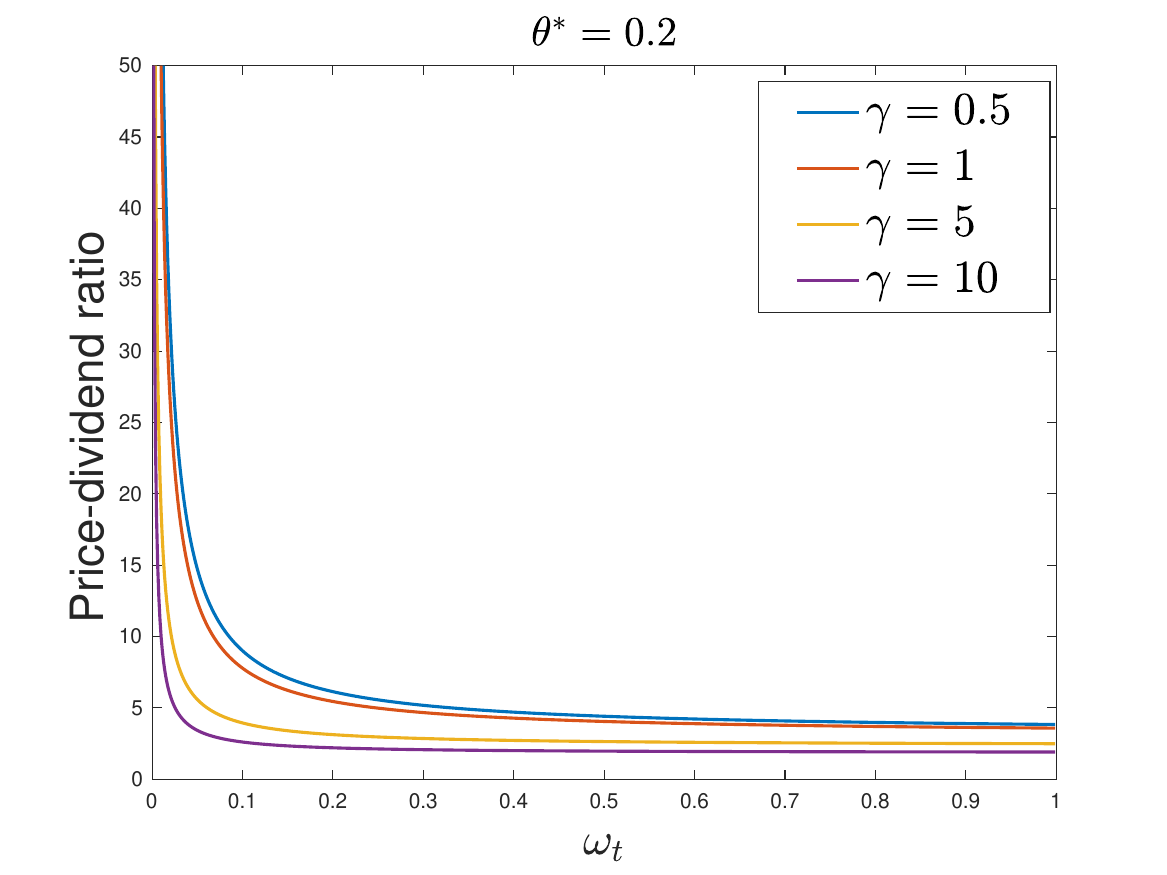}
			%	\caption{$\theta^*=0.2$}
			%		\label{fig:right}
		\end{subfigure}
		
		\begin{subfigure}[b]{0.32\textwidth}
			\centering
			\includegraphics[width=\textwidth]{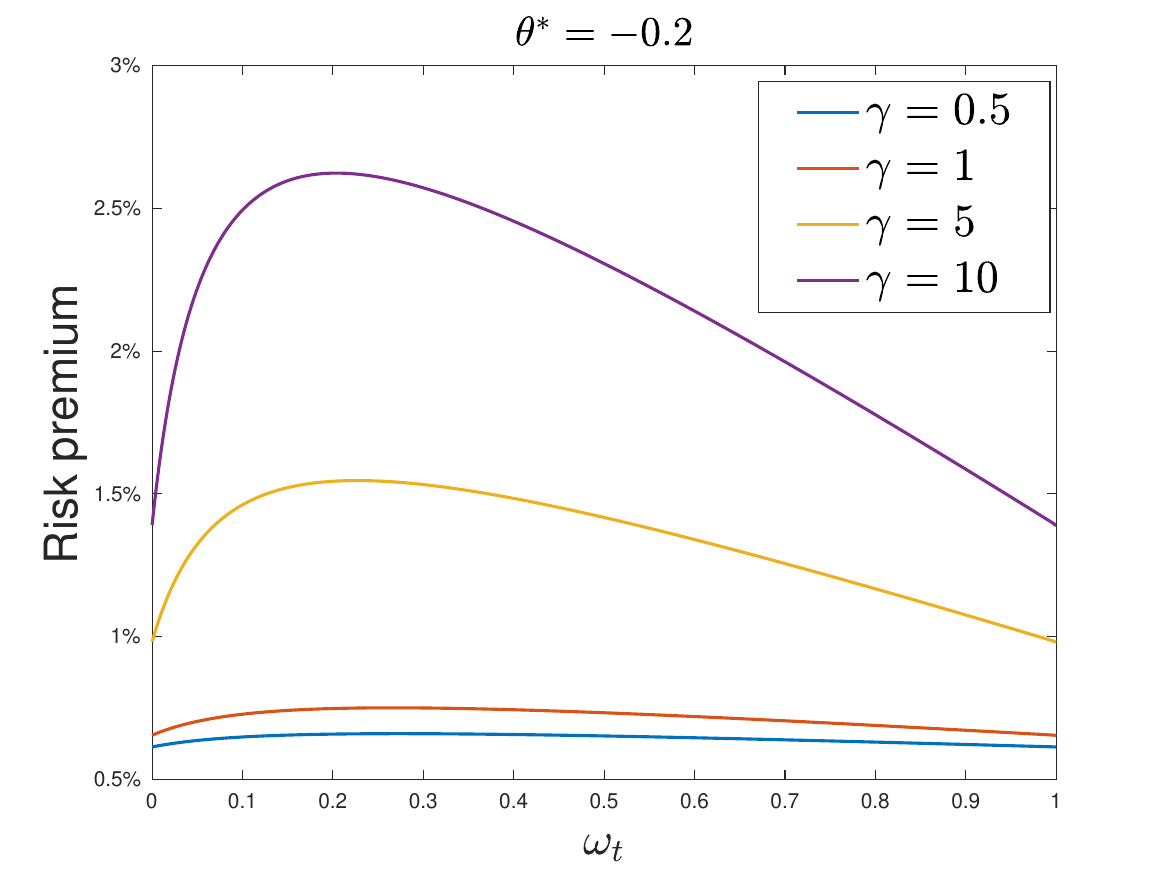}
			%	\caption{$\theta^*=-0.2$}
			%		\label{fig:left_p}
		\end{subfigure}
		\hfill
		\begin{subfigure}[b]{0.32\textwidth}
			\centering
			\includegraphics[width=\textwidth]{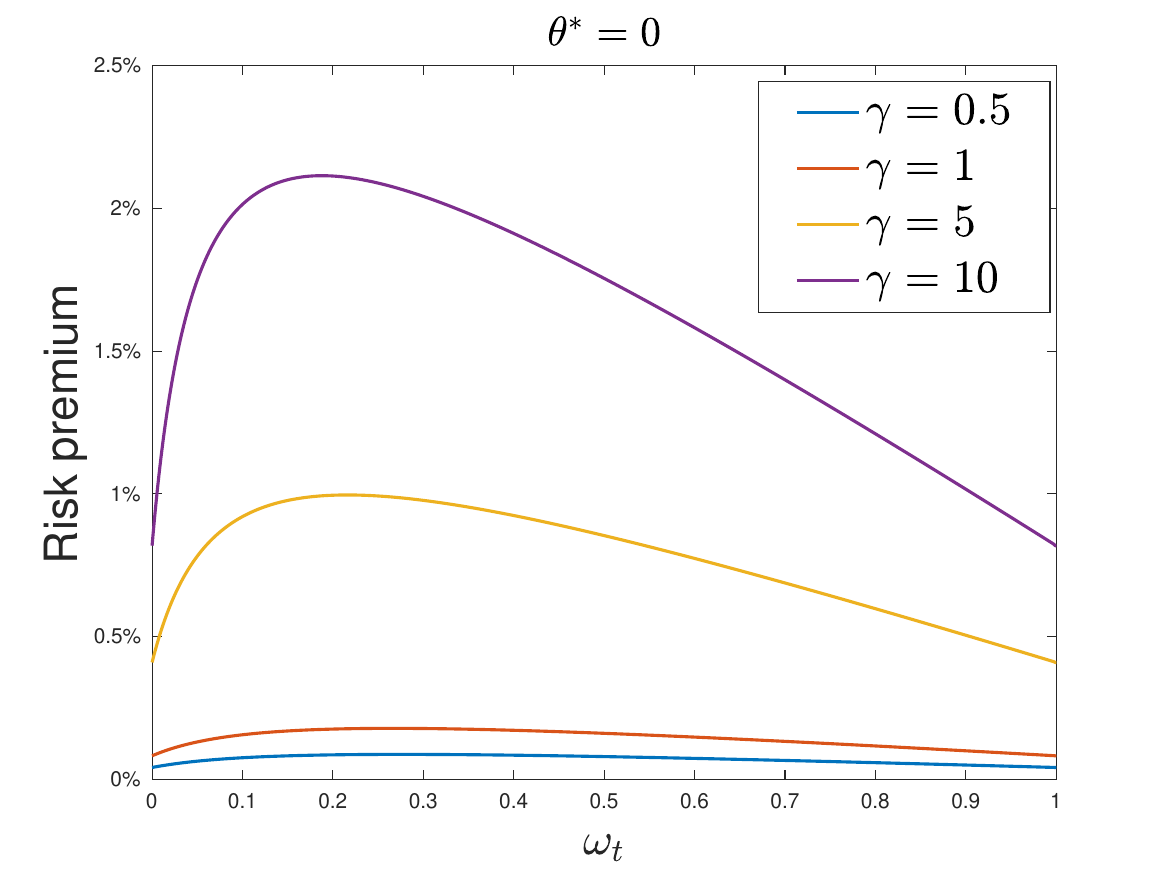}
			%\caption{$\theta^*=0$}
			%		\label{fig:center_p}
		\end{subfigure}
		\hfill
		\begin{subfigure}[b]{0.32\textwidth}
			\centering
			\includegraphics[width=\textwidth]{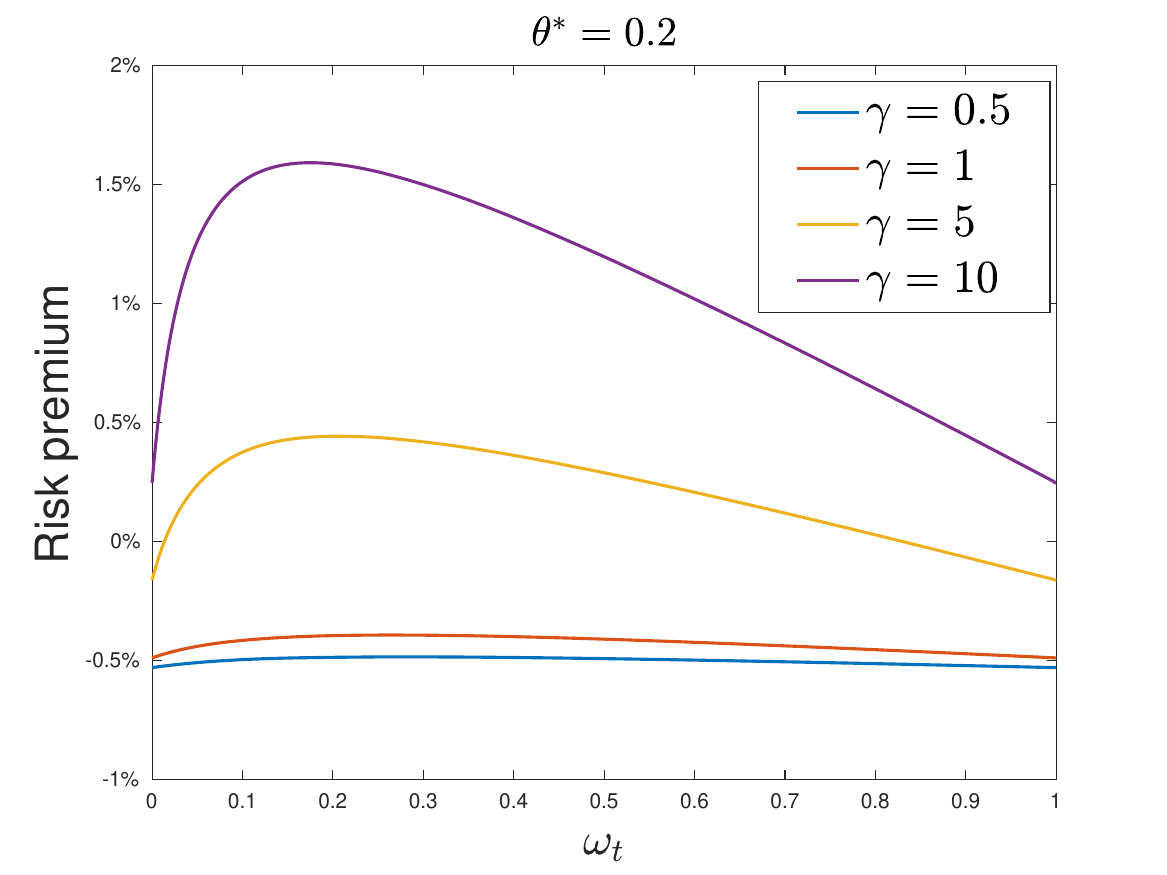}
			%	\caption{$\theta^*=0.2$}
			%		\label{fig:right_p}
		\end{subfigure}
		
		\begin{subfigure}[b]{0.32\textwidth}
			\centering
			\includegraphics[width=\textwidth]{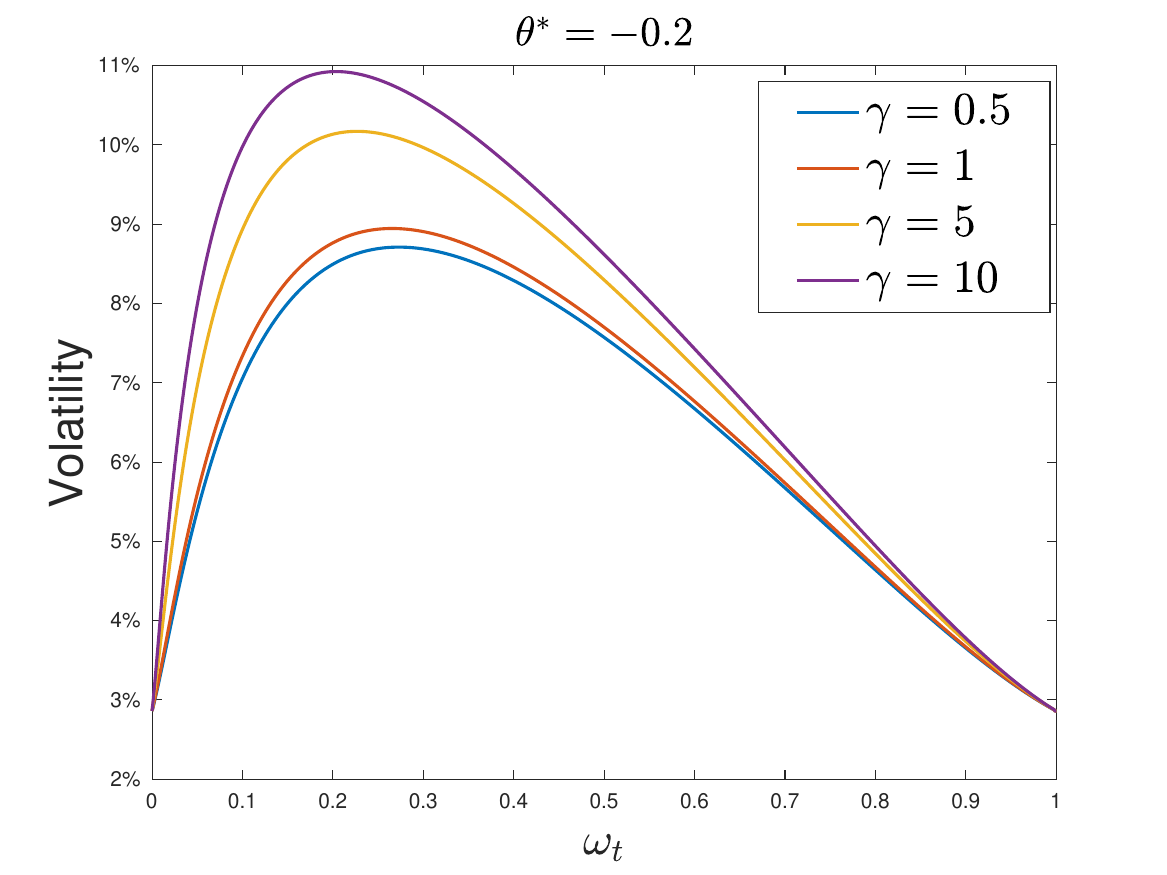}
			%	\caption{$\theta^*=-0.2$}
			%		\label{fig:left_v}
		\end{subfigure}
		\hfill
		\begin{subfigure}[b]{0.32\textwidth}
			\centering
			\includegraphics[width=\textwidth]{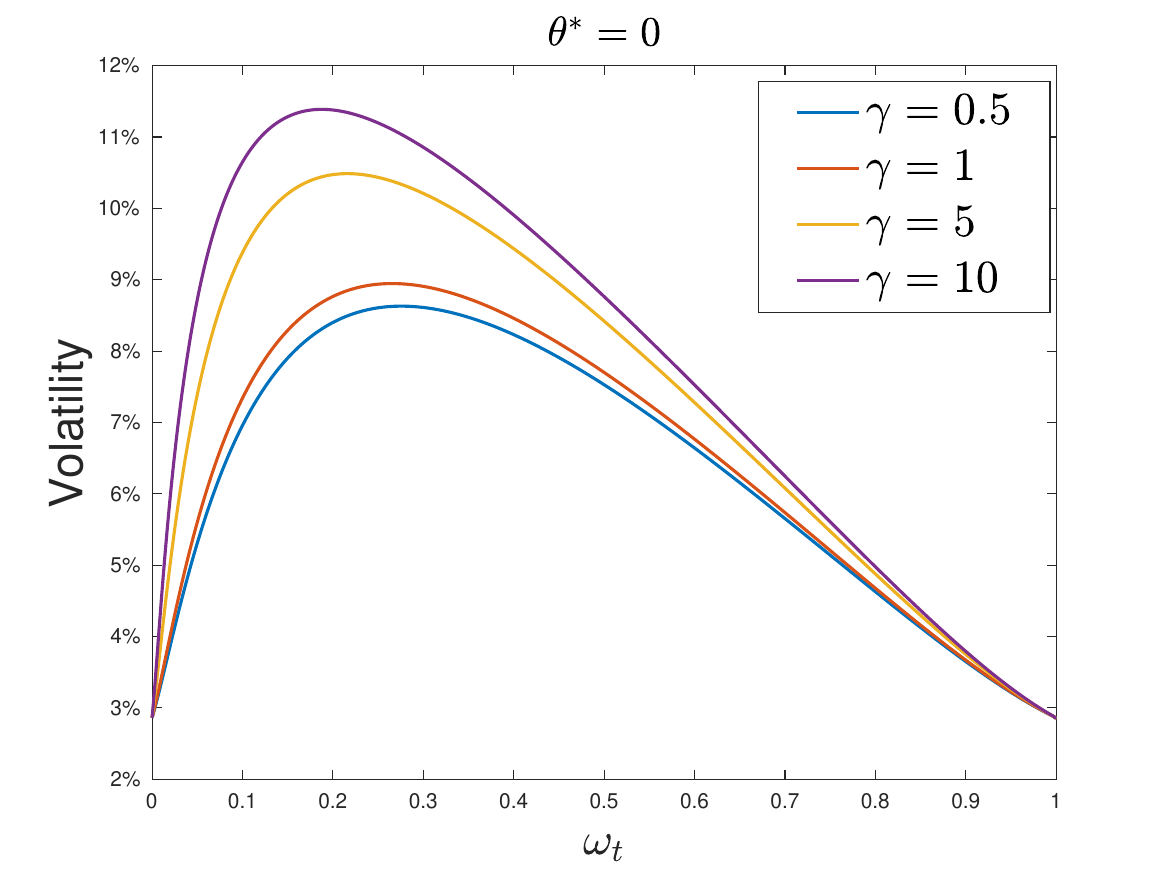}
			%	\caption{$\theta^*=0$}
			%		\label{fig:center_v}
		\end{subfigure}
		\hfill
		\begin{subfigure}[b]{0.32\textwidth}
			\centering
			\includegraphics[width=\textwidth]{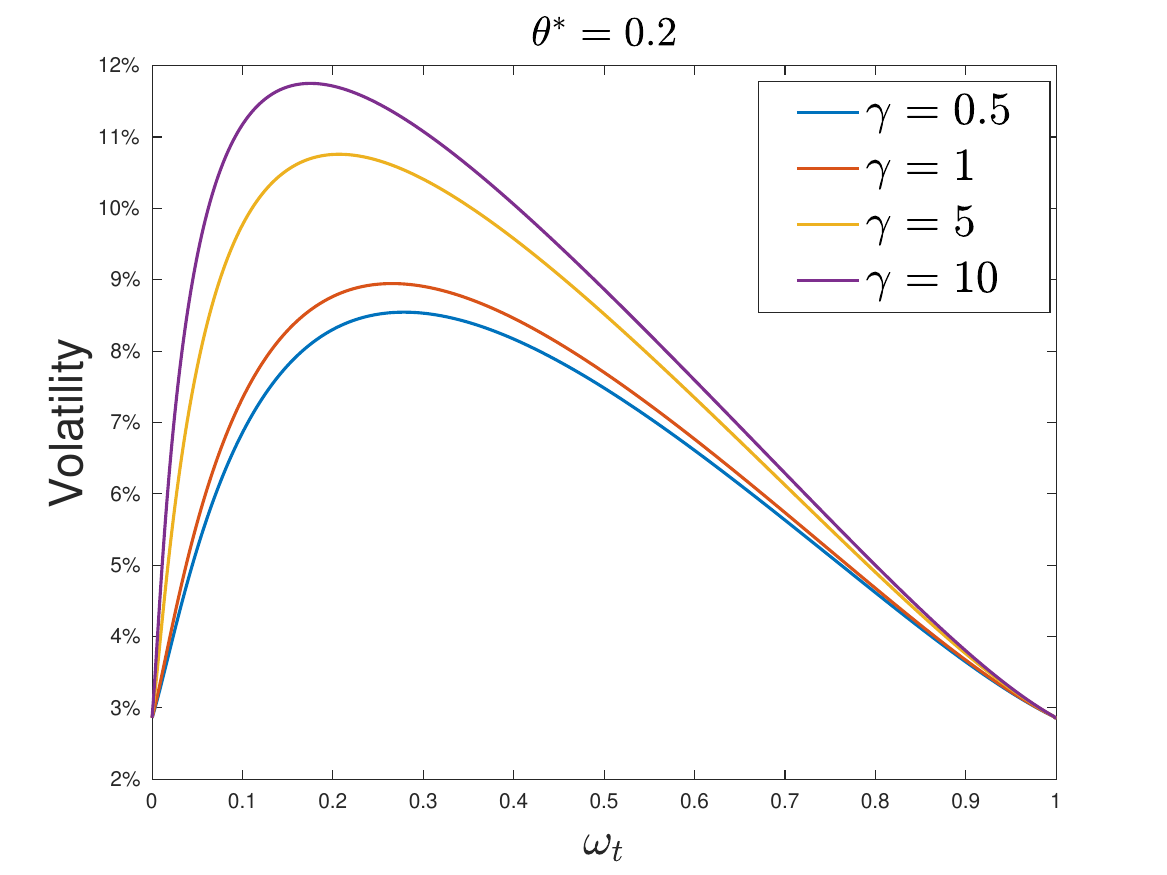}
			%	\caption{$\theta^*=0.2$}
			%		\label{fig:right_v}
		\end{subfigure}
		\caption{Stock price-dividend ratio (top row), risk premium (middle row), and volatility (bottom row) as functions of the dividend-endowment ratio. In each panel, four values of $\gamma$, 0.5, 1, 5, and 10 are used and represented by blue, red, yellow, and purple lines, respectively. $\theta^*$ is set to be $-0.2$ in the left column, 0 in the middle column, and 0.2 in the right column. Other model parameters are given in Table \ref{ta:ParameterValues}.}
		\label{fig:PDRPVol}
	\end{figure}

	We first plot in the top panels of Figure \ref{fig:PDRPVol} the stock price-dividend ratio $S_t/D_t$, which is equal to $\varphi_S(\omega_t)/\omega_t$, as a function of $\omega_t$, where $\varphi_S$ is given in Theorem \ref{thm:main}. The left, middle, and right panels correspond to the cases when $\theta^*$ is set to be $-0.2$, 0, and $0.2$, respectively, and four different values of $\gamma$ are used in each panel. We can observe that the stock price-dividend ratio is strictly decreasing in the dividend-endowment ratio.\footnote{\citet{SantosVeronesi2006:LaborIncome}, \citet{GuasoniWong2020:AssetPrices}, and \citet{GuasoniPiccirilliWang2025:AssetPricing} also find the stock price-dividend ratio to be decreasing in the dividend-endowment ratio in their models; see Proposition 3 and the bottom panel of Figure 3 in \citet{SantosVeronesi2006:LaborIncome}, Figure 1 in \citet{GuasoniWong2020:AssetPrices}, and Figure 1 in \citet{GuasoniPiccirilliWang2025:AssetPricing}.} %We can also observe that the price-dividend ratio is strictly decreasing in the risk aversion degree $\gamma$, which is consistent with Proposition \ref{prop:ComparativeStat}-(iii) because $\rho$ is set to be positive.

	We then plot in the middle and bottom panels of Figure \ref{fig:PDRPVol} the stock risk premium $\mu_S(\omega)-r_f(\omega)$ and volatility $\|\sigma_S(\omega)\|$, respectively. Again, the left, middle, and right panels correspond to the cases when $\theta^*$ is set to be $-0.2$, 0, and $0.2$, respectively, and four different values of $\gamma$ are used in each panel. We can observe that both the stock risk premium and volatility are first increasing and then decreasing in the dividend-endowment ratio. This observation is consistent with Corollary \ref{Coro:SpecialCase} which holds for the special case of $(1-\gamma)\rho$. Recall that historically, the dividend-endowment ratio lies in the range $[0,0.11]$. In this range, both the stock risk premium and volatility are increasing in the dividend-endowment ratio.\footnote{\citet{SantosVeronesi2006:LaborIncome} also find the stock risk premium to be increasing in the dividend-endowment ratio in their model; see Proposition 4 and Figure 3 therein. The model in \citet{CochraneLongstaffStantaClara2008:TwoTrees} also implies an increasing stock risk premium with respect to the dividend-endowment ratio; see Figure 5 therein.} We also observe from Figure \ref{fig:PDRPVol} that fixing $\theta^*$, the stock price-dividend ratio is decreasing in the risk aversion degree $\gamma$, and both the stock risk premium and volatility are increasing in $\gamma$.

\begin{figure}
		\centering
		\begin{subfigure}[b]{0.48\textwidth}
			\centering
			\includegraphics[width=\textwidth]{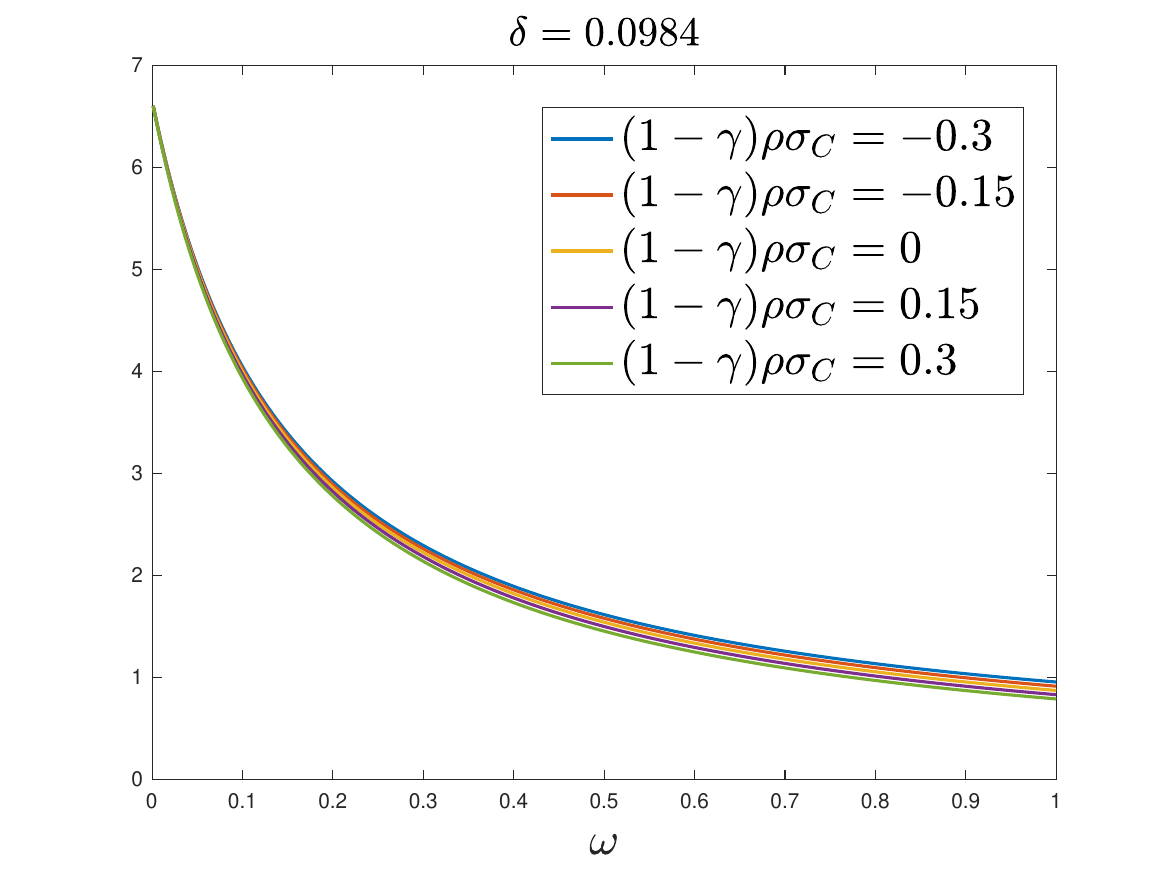}
			%		\caption{$\delta=0.0984$}
			%		\label{fig:grho_delta009}
		\end{subfigure}
		\hspace{0.2cm}
		\begin{subfigure}[b]{0.48\textwidth}
			\centering
			\includegraphics[width=\textwidth]{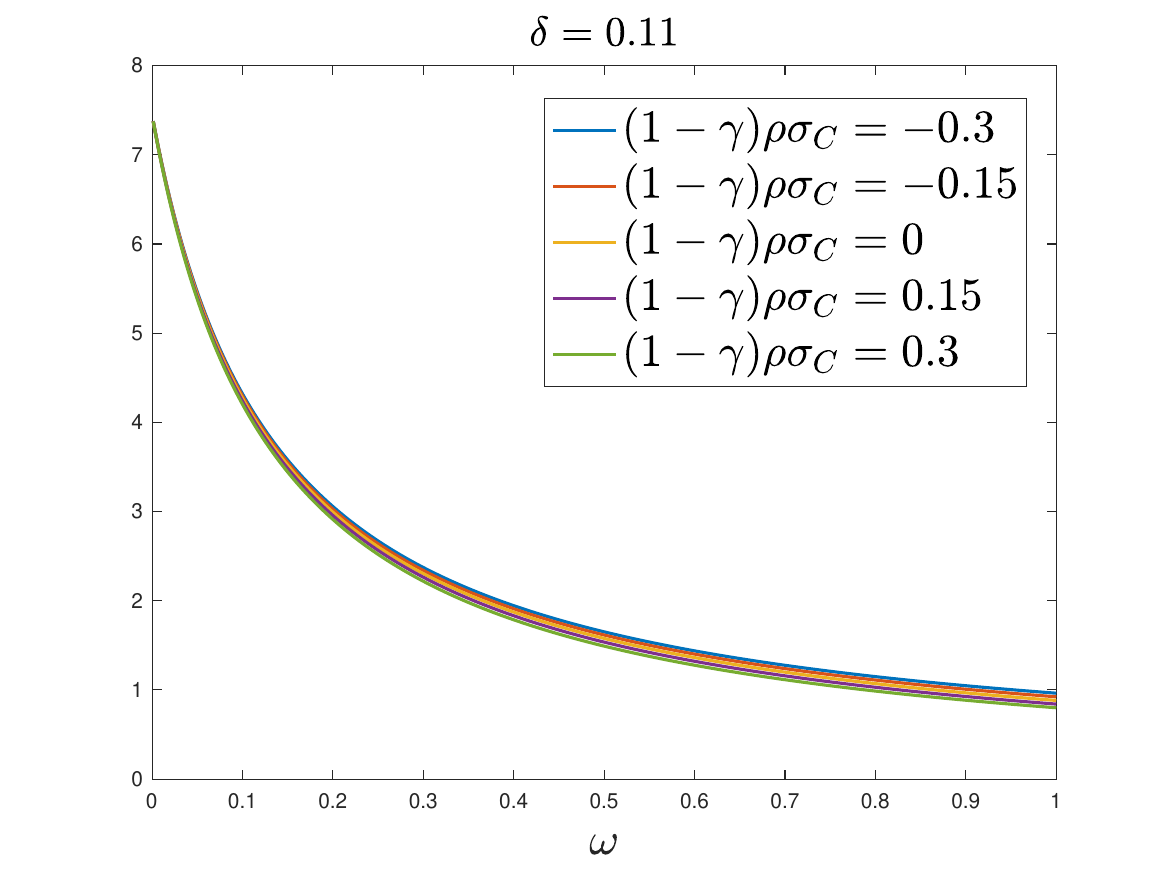}
			%		\caption{$\delta=0.11$}
			%		\label{fig:grho_delta01}
		\end{subfigure}
		
		\begin{subfigure}[b]{0.48\textwidth}
			\centering
			\includegraphics[width=\textwidth]{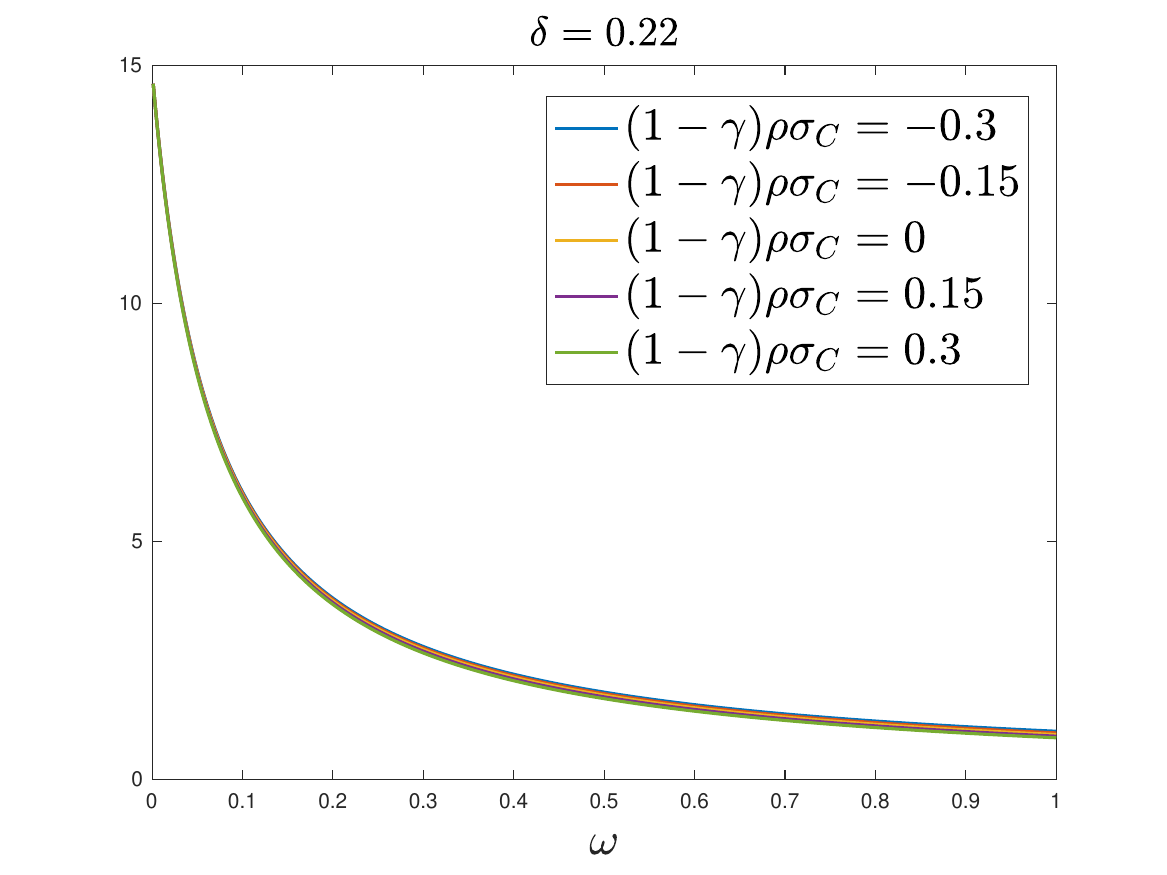}
			%		\caption{$\delta=0.22$}
			%		\label{fig:grho_delta02}
		\end{subfigure}
		\hspace{0.2cm}
		\begin{subfigure}[b]{0.48\textwidth}
			\centering
			\includegraphics[width=\textwidth]{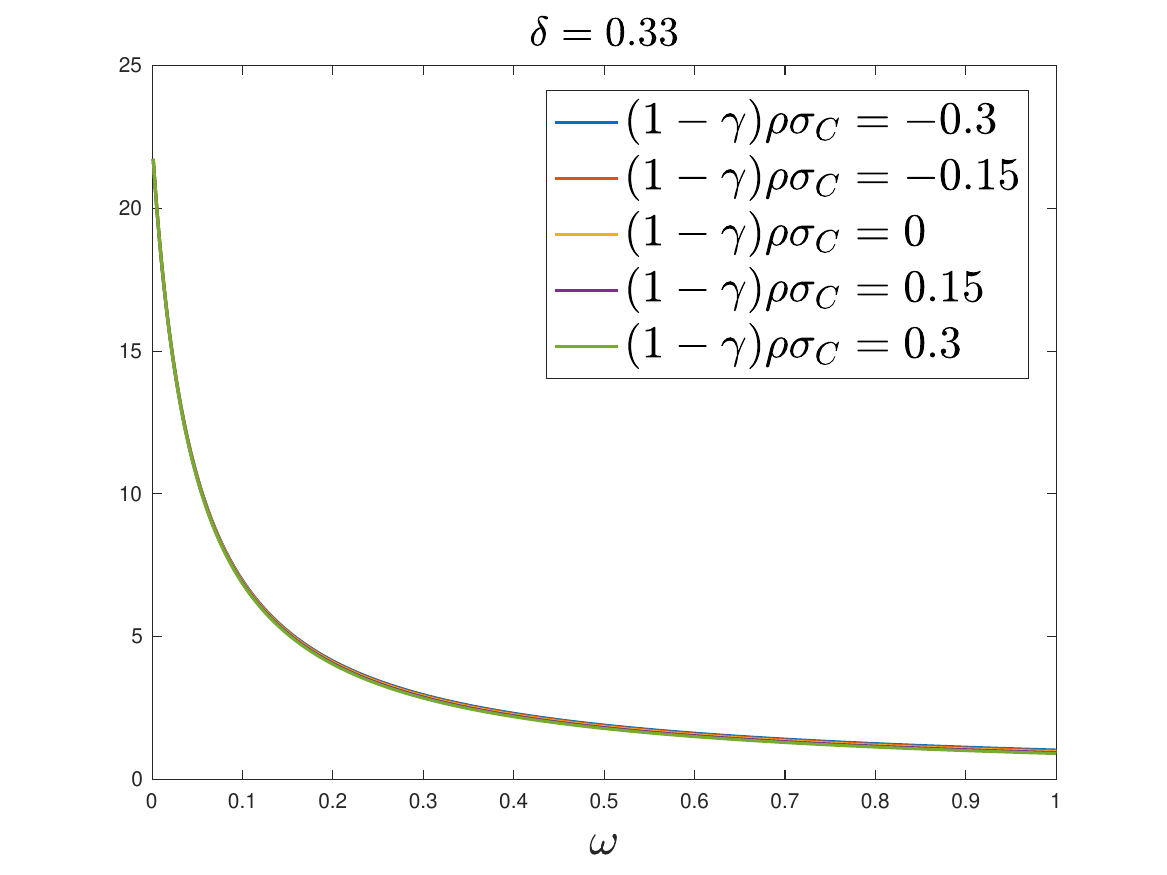}
			%		\caption{$\delta=0.33$}
			%		\label{fig:grho_delta03}
		\end{subfigure}
		\caption{$\varphi_S'/\varphi_S(\omega),\omega\in(0,1)$ with respect to $(1-\gamma)\rho\sigma_C$ with $\delta$ set to be $0.0984$, $0.11$, $0.22$, and $0.33$ in the top left, top right, bottom left, and bottom right panels, respectively. In each panel, five values of $(1-\gamma)\rho\sigma_C$, $-0.3$, $-0.15$, 0, 0.15, and 0.3 are used and represented by blue, red, yellow, purple, and green lines, respectively. Other model parameters are given in Table \ref{ta:ParameterValues}.}
		\label{fig:ElasticityRho}
	\end{figure}

\begin{figure}
		\centering
		\begin{subfigure}[b]{0.48\textwidth}
			\centering
			\includegraphics[width=\textwidth]{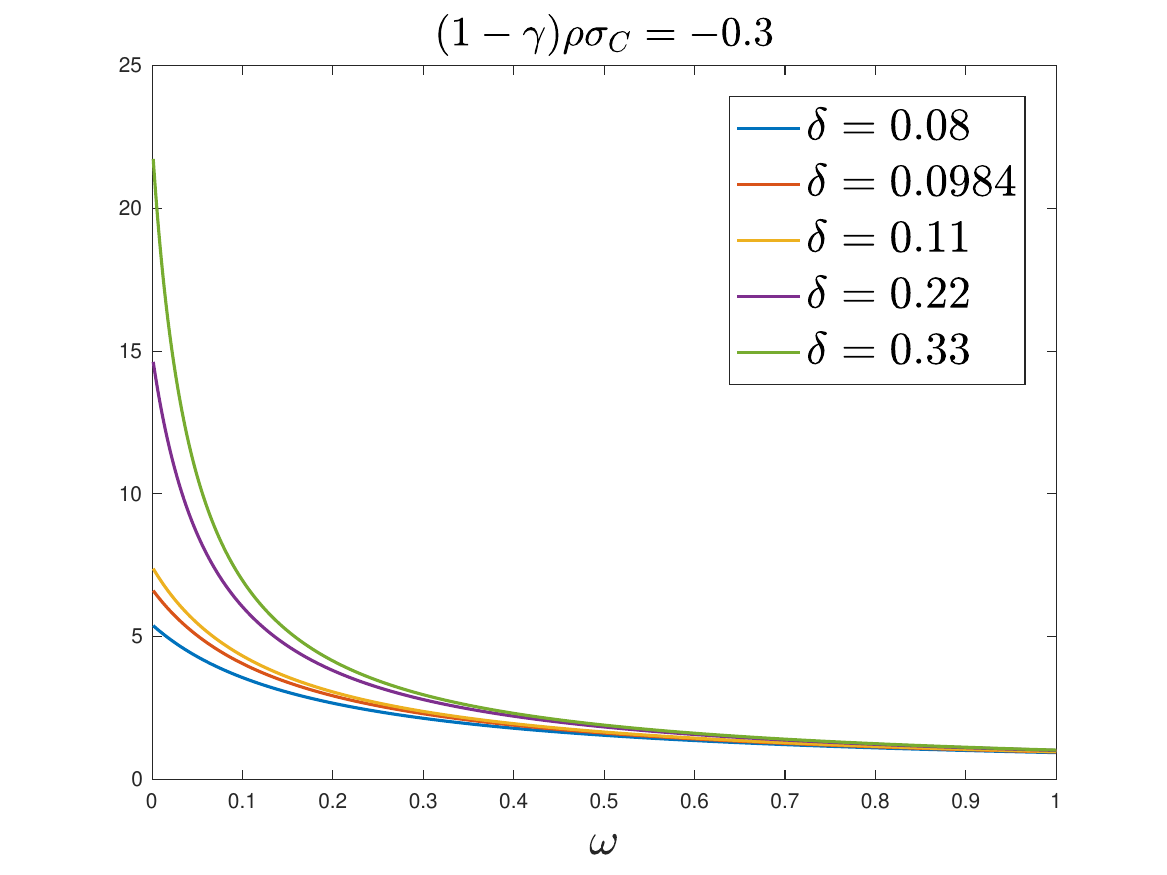}
			%		\caption{$(1-\gamma)\rho\sigma_C=-0.3$}
			%		\label{fig:delta_n03}
		\end{subfigure}
		\hspace{0.2cm}
		\begin{subfigure}[b]{0.48\textwidth}
			\centering
			\includegraphics[width=\textwidth]{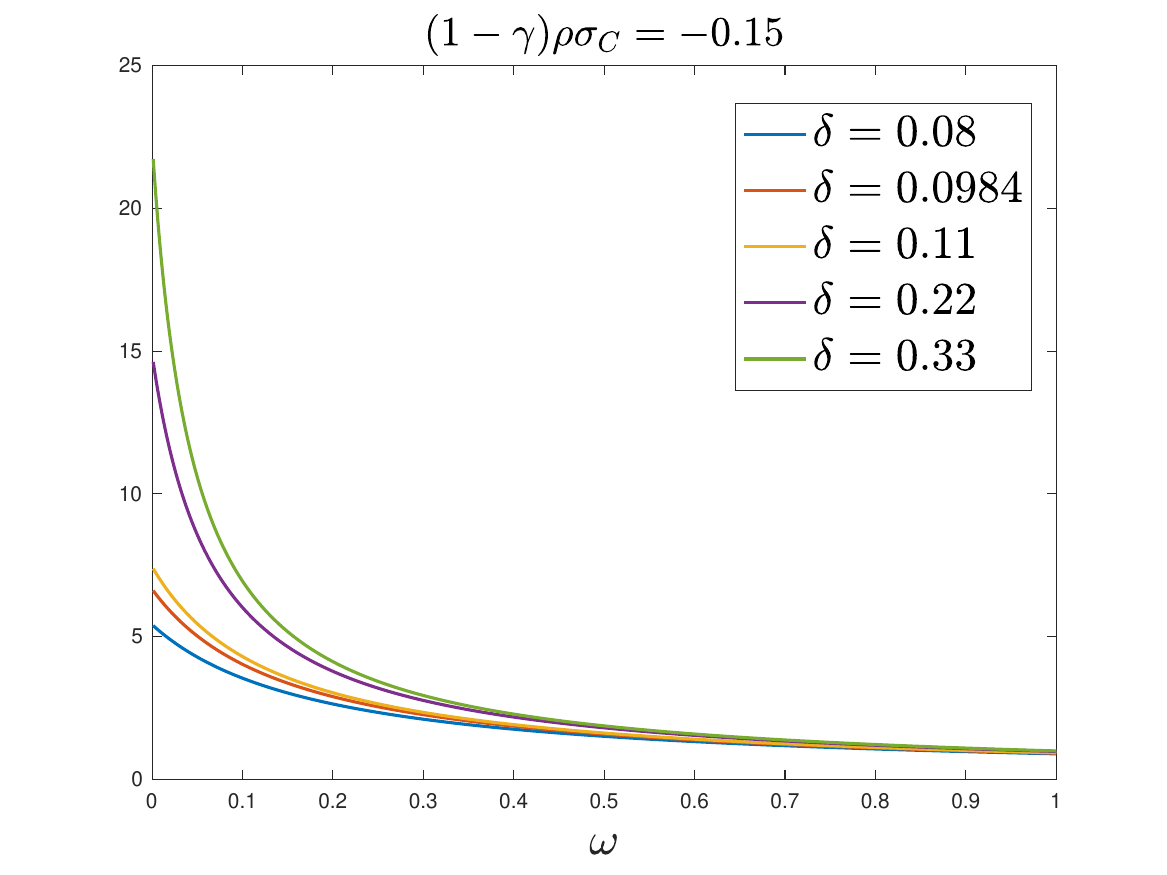}
			%		\caption{$(1-\gamma)\rho\sigma_C=-0.15$}
			%		\label{fig:delta_n015}
		\end{subfigure}
		
		\begin{subfigure}[b]{0.48\textwidth}
			\centering
			\includegraphics[width=\textwidth]{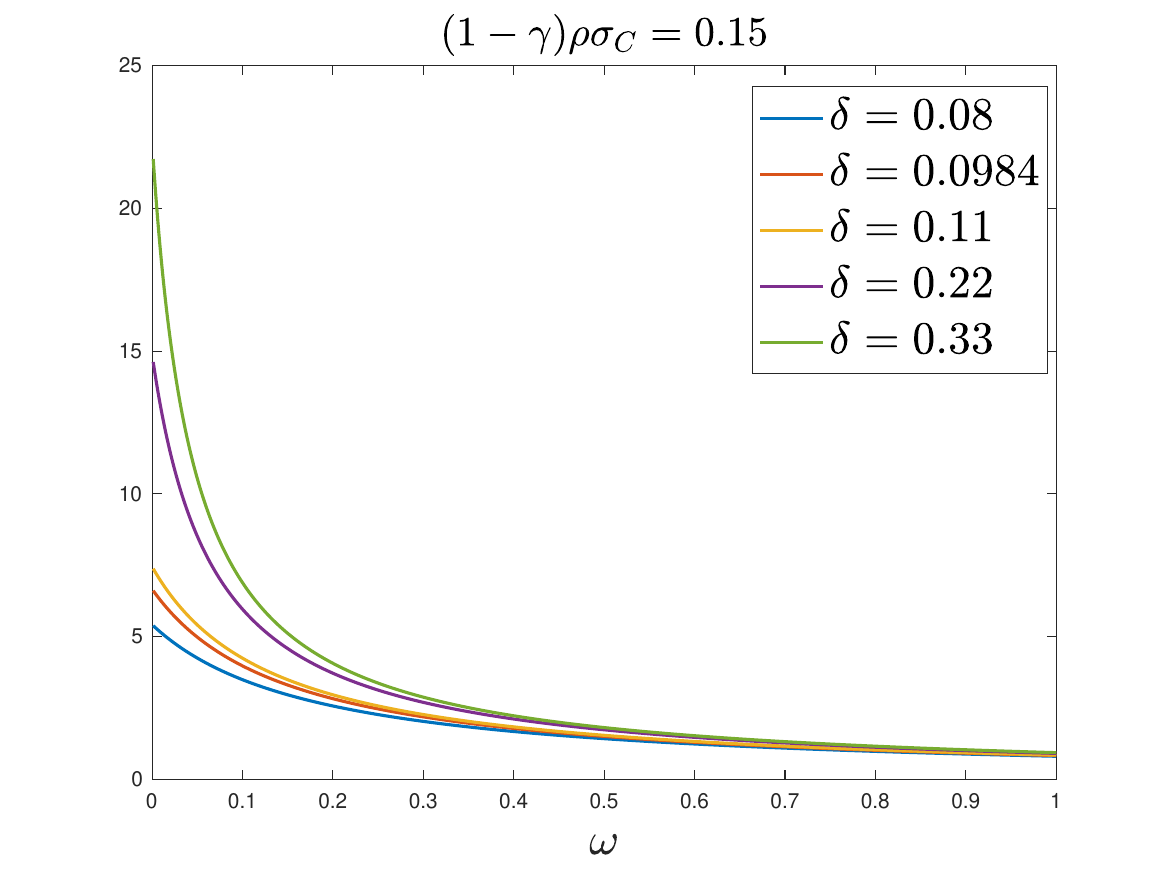}
			%		\caption{$(1-\gamma)\rho\sigma_C=0.15$}
			%		\label{fig:delta_015}
		\end{subfigure}
		\hspace{0.2cm}
		\begin{subfigure}[b]{0.48\textwidth}
			\centering
			\includegraphics[width=\textwidth]{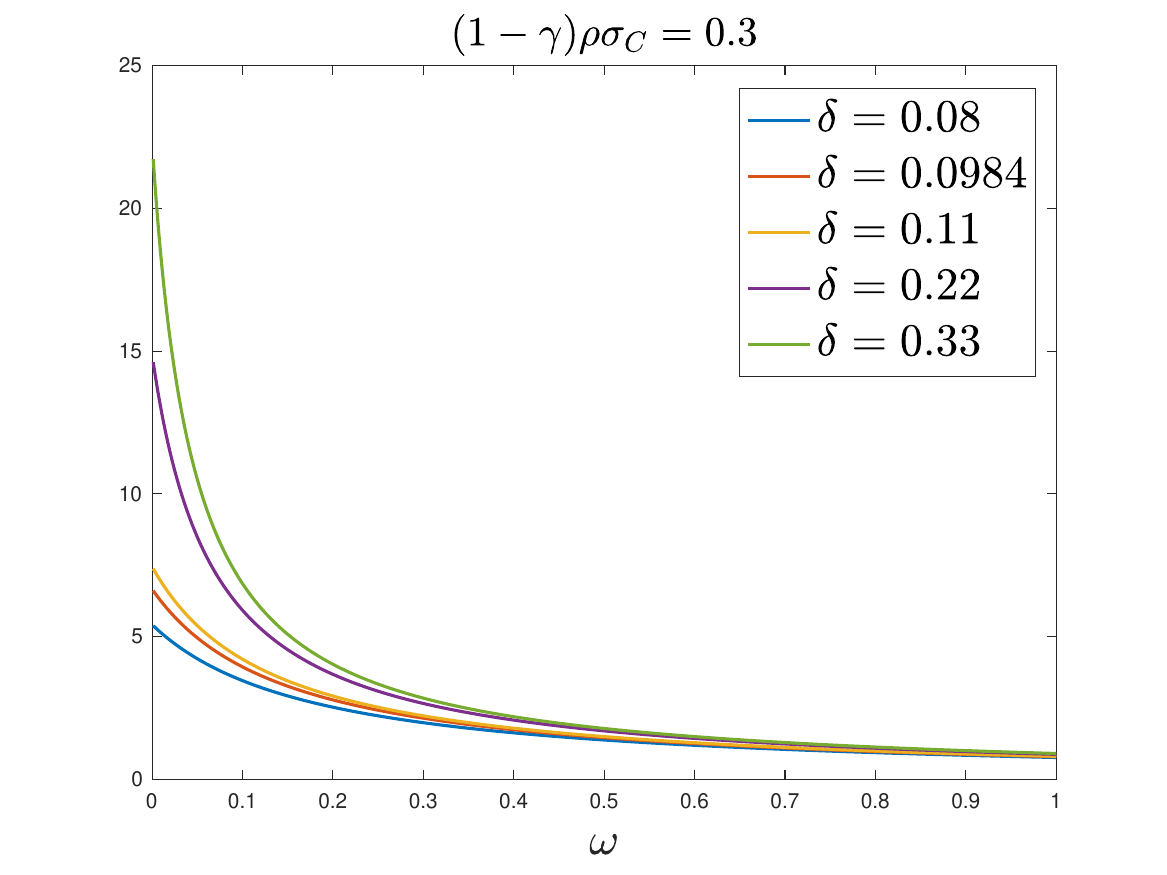}
			%		\caption{$(1-\gamma)\rho\sigma_C=0.3$}
			%		\label{fig:delta_03}
		\end{subfigure}
		\caption{$\varphi_S'/\varphi_S(\omega),\omega\in(0,1)$ with respect to $\delta$ with $(1-\gamma)\rho\sigma_C$ set to be $-0.3$, $-0.15$, $0.15$, and $0.3$ in the top left, top right, bottom left, and bottom right panels, respectively. In each panel, five values of $\delta$, 0.08, 0.0984, 0.11, 0.22, and 0.33, are used and represented by blue, red, yellow, purple, and green lines, respectively. Other model parameters are given in Table \ref{ta:ParameterValues}.}
		\label{fig:ElasticityDelta}
	\end{figure}

	Next, we investigate the impact of $\kappa$ and $\alpha$ on the stock risk premium and volatility. Because $\kappa$ and $\alpha$ affect asset prices only through $\theta^*$, we only need to study how the stock risk premium and volatility depend on $\theta^*$. By \eqref{conditional_equity_premium} and \eqref{conditional_volatility}, both the stock risk premium and volatility depend crucially on the elasticity $\varphi_S'(\omega)/\varphi_S(\omega)$. By Proposition \ref{prop:ComparativeStat}-(iii), $\varphi_S$ and thus $\varphi_S'/\varphi_S$ depend on $\theta^*$ through $\delta$ and depend on $\gamma$, $\rho$, and $\sigma_C$ through $(1-\gamma)\rho\sigma_C$. In Figure \ref{fig:ElasticityRho}, we plot $\varphi'/\varphi$ with respect to $(1-\gamma)\rho\sigma_C\in\{-0.3, -0.15,0,-0.15,0.3\}$ for fixed $\delta$. In Figure \ref{fig:ElasticityDelta}, we plot $\varphi'/\varphi$ with respect to $\delta \in \{0.08, 0.0984, 0.11,0.22, 0.33\}$ for fixed $(1-\gamma)\rho\sigma_C$.\footnote{With the parameter values in Table \ref{ta:ParameterValues}, $\theta^*\in[-0.2,0.2]$, and $\gamma\in[0.5, 10]$, it is straightforward to show that the value range of $\delta$ is $[0.08409,0.3210]$. The five values that we chose for $\delta$ are representative of this range, with the value $0.0984$ corresponding to the case when $\gamma=1$ and thus $\delta=\phi$. With $\gamma$ taking values in $[0.5,10]$, $\rho$ taking values in $[-1,1]$, and $\sigma_C$ as in Table \ref{ta:ParameterValues}, the value range of $(1-\gamma)\rho\sigma_C$ is $[-0.2574,0.2574]$. The five values that we chose for $(1-\gamma)\rho\sigma_C$ are representative of this range, with the value $0$ corresponding to the case when $\gamma=1$ or $\rho=0$.}
	
	We can observe from Figure \ref{fig:ElasticityRho} that $\varphi_S'/\varphi_S$ is decreasing in $(1-\gamma)\rho\sigma_C$. In consequence, when $\gamma>1$, which is empirically relevant, $\varphi_S'/\varphi_S$ is strictly increasing in $\rho$. Then, we can observe from \eqref{conditional_equity_premium} and \eqref{conditional_volatility} that both the stock risk premium and volatility are increasing in $\rho$. Thus, compared to models assuming zero correlation between the endowment and dividend-endowment ratio, such as the model in \citet{GuasoniWong2020:AssetPrices}, accounting for positive correlation as observed in the empirical data can increase both the stock equity premium and volatility and thus help to partially explain the equity premium puzzle \citep{MehraRPrescottE:85ep} and excess volatility puzzle \citep{Shiller1981:DoStockPrices}.

\begin{figure}
		\centering
		\begin{subfigure}[b]{0.32\textwidth}
			\centering
			\includegraphics[width=\textwidth]{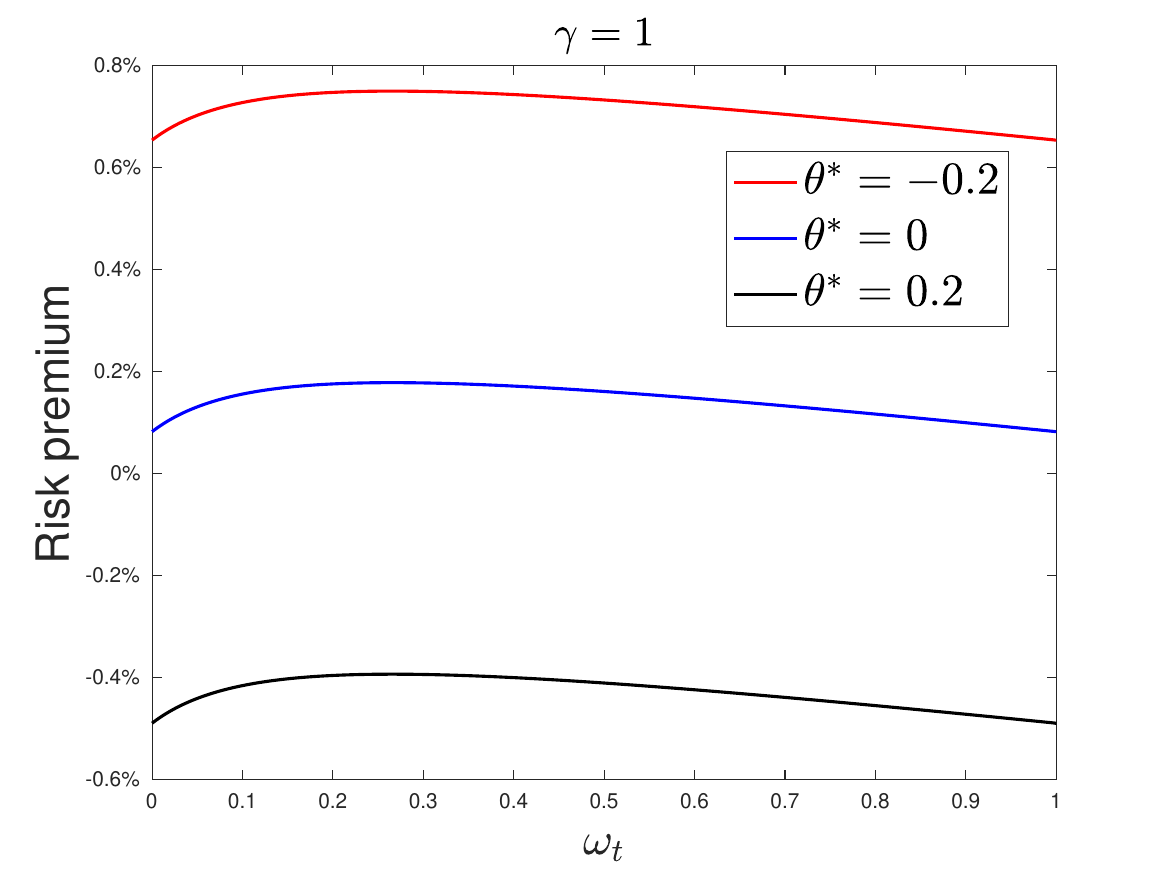}
			%\caption{$\gamma=1$}
			%		\label{fig:left_g}
		\end{subfigure}
		\hfill
		\begin{subfigure}[b]{0.32\textwidth}
			\centering
			\includegraphics[width=\textwidth]{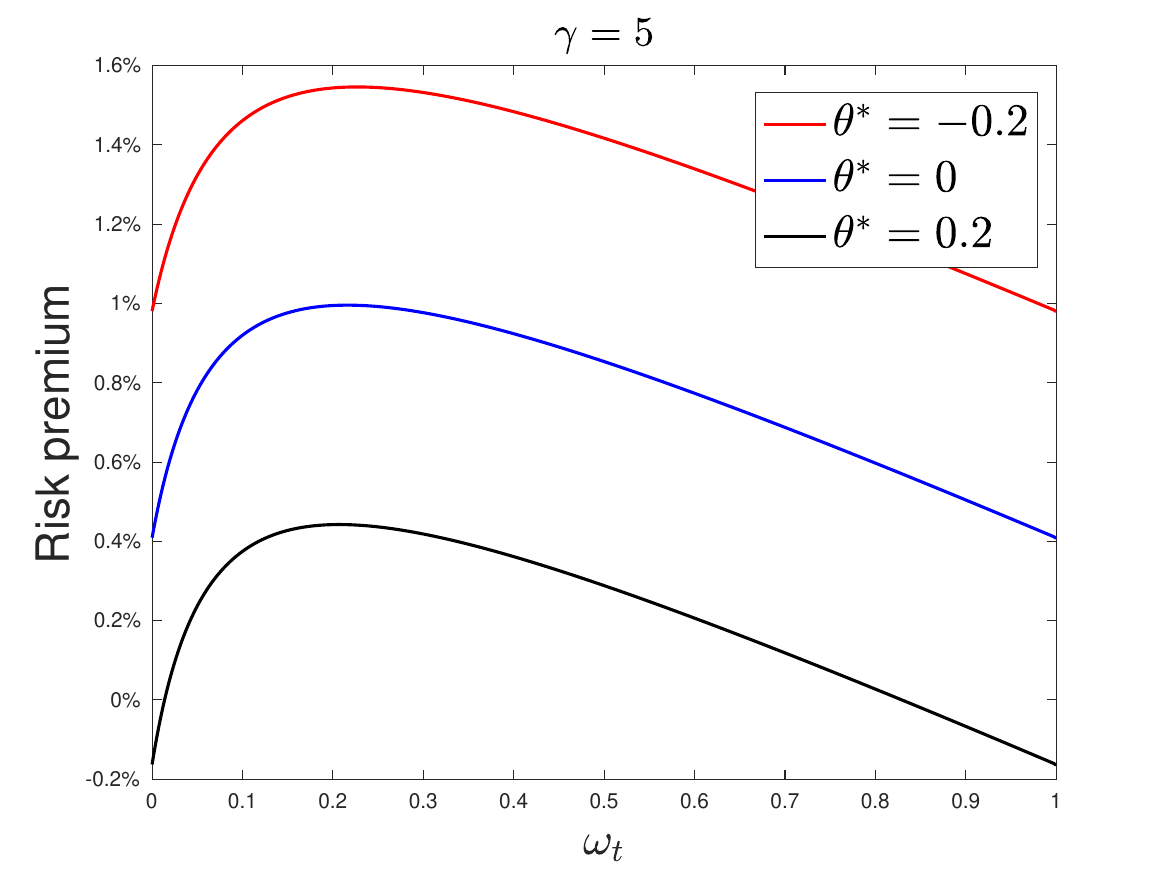}
			%	\caption{$\gamma=5$}
			%		\label{fig:center_g}
		\end{subfigure}
		\hfill
		\begin{subfigure}[b]{0.32\textwidth}
			\centering
			\includegraphics[width=\textwidth]{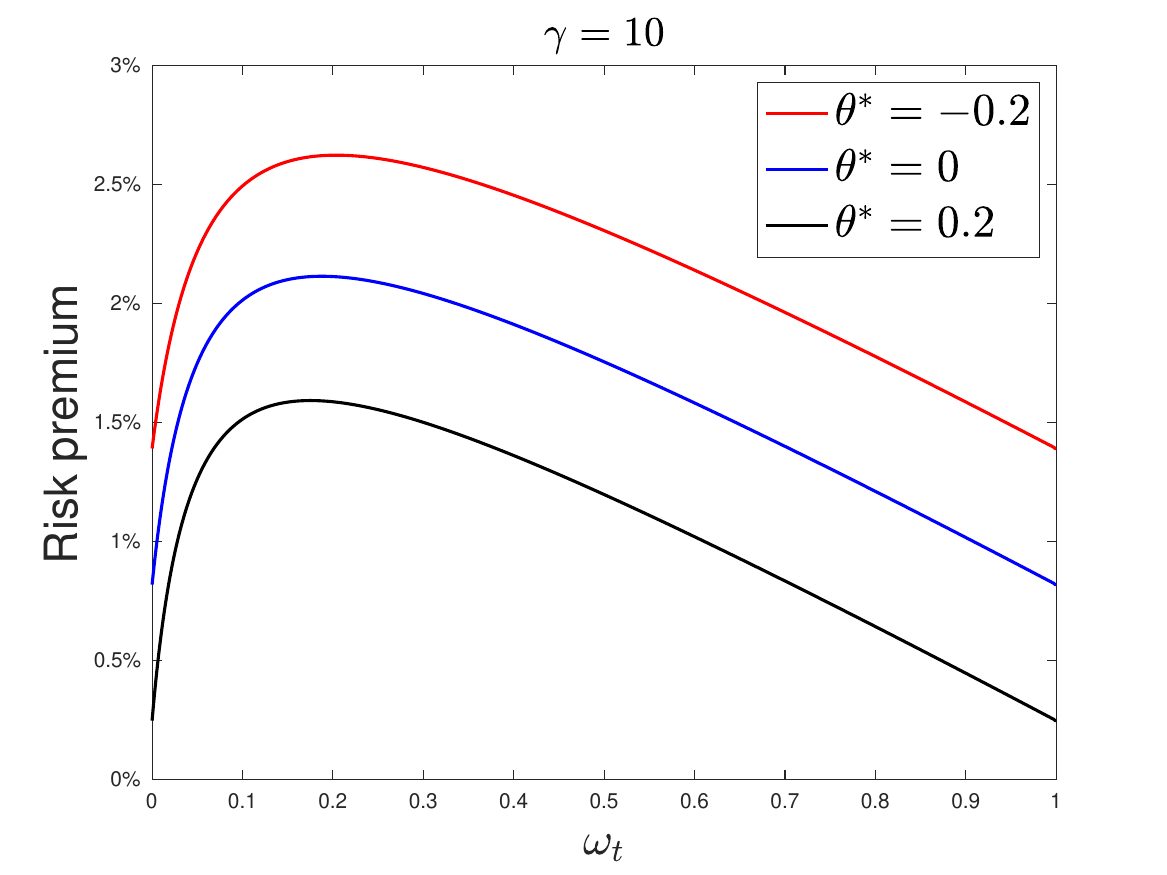}
			%	\caption{$\gamma=10$}
			%		\label{fig:right_g}
		\end{subfigure}
		\caption{Stock risk premium with respect to $\theta^*\in\{-0.2,0,0.2\}$ with $\gamma$ set to be 1, 5, and 10 in the left, middle, and right panels, respectively. Other model parameters are given in Table \ref{ta:ParameterValues}.}
		\label{fig:premium_theta}
	\end{figure}

	We observe from Figure \ref{fig:ElasticityDelta} that $\varphi_S'/\varphi_S$ is increasing in $\delta$. Recall that $\delta$ is decreasing in $\theta^*$ when $\gamma<1$ and increasing in $\theta^*$ when $\gamma>1$. Then, from \eqref{conditional_equity_premium}, we conclude that the stock risk premium is decreasing in $\theta^*$ when $\gamma<1$. When $\gamma>1$, which is empirically more relevant, the second term in \eqref{conditional_equity_premium} is increasing in $\theta^*$ and the third time is decreasing in $\theta^*$. We plot in Figure \ref{fig:premium_theta} the stock risk premium with respect to $\theta^*\in\{-0.2,0,0.2\}$ for each of $\gamma\in\{1,5,10\}$, and find that the stock risk premium is decreasing in $\theta^*$. This shows that for $\gamma\le 10$, the third term dominates the second term in \eqref{conditional_equity_premium}.\footnote{When $\gamma$ is very large, it is possible that the second term dominates. Indeed, when we set $\gamma$ to be 45, an unrealistically high risk aversion degree, the stock risk premium becomes increasing in $\theta^*$ for certain values of the dividend-endowment ratio.} On the other hand, given that $\rho$ is positive, $\delta$ is increasing in $\theta^*$ when $\gamma>1$, and $\varphi_S'/\varphi_S$ is increasing in $\delta$, we conclude from \eqref{conditional_volatility} that the stock volatility is increasing in $\theta^*$ when $\gamma>1$.
	
	Recalling Proposition \ref{prop:AmImMarketPriceRisk}, we conclude that when $\gamma>1$ and $\alpha\in[1/2,1]$, which is the empirically relevant case, the stock risk premium is increasing in the ambiguity aversion degree $\alpha$ and the agent's perceived ambiguity level $\kappa$, but the stock volatility is decreasing in $\alpha$ and increasing in $\kappa$. Therefore, the presence of ambiguity and the agent's ambiguity aversion can help to explain the equity premium but make it more difficult to explain the excess volatility puzzle.
	
	\subsection{Unconditional Moments of Asset Returns}
	
	Finally, we compute the unconditional moments of asset returns. Consider the stock gross return in a small period $[t,t+\Delta t]$, which is approximately equal to
	%\begin{align*}
	%  R_S(t,t+\Delta t):=e^{\int_t^{t+\Delta t}\left(\mu_S(\omega_u) -\frac{1}{2}\|\sigma_S(\omega_u)\|^2\right)du + \int_t^{t+\Delta t}\sigma_S(\omega_u)dB_u^0}.
	%\end{align*}
	%Then, the {\em unconditional} expected gross return is
	%\begin{align*}
	%  \expect[R_S(t,t+\Delta t)]=\expect\left[e^{\int_t^{t+\Delta t}\mu_S(\omega_u) du}\right]
	%\end{align*}
	\begin{align*}
		R_S(t,t+\Delta t):=e^{\left[\mu_S(\omega_t) -\frac{1}{2}\|\sigma_S(\omega_t)\|^2\right]\Delta t + \sigma_S(\omega_t)(B_{t+\Delta t}^0-B_t^0)}.
	\end{align*}
	The annualized unconditional risk premium is then approximately equal to
	\begin{align*}
		\frac{1}{\Delta t}\left(\expect[e^{\mu_S(\omega_t)\Delta t}] - e^{r_f \Delta t}\right)\approx \expect[\mu_S(\omega_t)]-r_f,
	\end{align*}
	where the expectation is taken under the invariant distribution of $\{\omega_t\}_{t\ge 0}$. The unconditional variance of the stock logarithmic return in excess of the risk-free logarithmic return is approximately equal to
	\begin{align*}
		&\mathrm{var}\left(\log R_S(t,t+\Delta t)-r_f\Delta t\right) =\mathrm{var}\left(\left[\mu_S(\omega_t) -\frac{1}{2}\|\sigma_S(\omega_t)\|^2\right]\Delta t +  \sigma_S(\omega_t)(B_{t+\Delta t}^0-B_t^0)\right)\\
		& = \mathrm{var}\left(\left[\mu_S(\omega_t) -\frac{1}{2}\|\sigma_S(\omega_t)\|^2\right]\Delta t\right) + \expect\left[ \sigma_S^2(\omega_t)\Delta t\right] \approx \expect\left[ \sigma_S^2(\omega_t)\right]\Delta t,
	\end{align*}
	where the second equality is due to the conditional variance formula, and the approximation is valid because $\Delta t$ is small. As a result, the annualized unconditional standard deviation of the stock logarithmic excess return, referred to as the unconditional volatility in the following, is approximately equal to $\sqrt{\expect\left[ \sigma_S^2(\omega_t)\right]}$.
	
	To compute the unconditional stock risk premium and volatility, we need to compute the invariant distribution of $\{\omega_t\}_{t\ge 0}$. By Proposition \ref{prop:SDEExistence}, the invariant distribution exists. Moreover, the density function of the invariant distribution, $p(\omega),\omega\in(0,1)$, satisfies the Fokker-Planck equation:
	\begin{align*}
		0 = -\frac{\partial \lf( \mu_{\omega}(\omega) p(\omega) \rh)}{\partial \omega} + \frac{1}{2} \cdot \frac{\partial^2\lf( \sigma_{\omega}^2(\omega) p(\omega) \rh)}{\partial \omega^2},\quad \omega\in(0,1),
	\end{align*}
	which can be solved numerically. On the other hand, the unconditional stock risk premium and volatility can be estimated using the sample mean and sample standard deviation of historical stock returns, respectively.
	
	In the following, we compute the unconditional stock risk premium  $\expect[\mu_S(\omega_t)]-r_f$ and volatility $\sqrt{\expect\left[ \sigma_S^2(\omega_t)\right]}$ in our model and compare them with their empirical estimates in \citet{CampbellCochrane:1999ByForceOfHabit} ($E(r-r^f)$ and $\sigma(r-r^f)$ in the last column of Table 2 therein). In addition, we compute $\exp\big[\expect\big(\ln (S_t/D_t)\big)\big] = \exp\big[\expect\big(\ln(\varphi_S(\omega_t)/\omega_t)\big)\big]$, which is the unconditional price-dividend ratio under the logarithmic transformation, and $\sqrt{\mathrm{var}(\ln (S_t/D_t))}=\sqrt{\mathrm{var}(\ln(\varphi_S(\omega_t)/\omega_t))}$, which is the standard deviation of the log price-dividend ratio, and compare them with the empirical estimates in \citet{CampbellCochrane:1999ByForceOfHabit} ($\exp[E(p-d)]$ and $\sigma(p-d)$ in the last column of Table 2 therein).
	
	We want to investigate whether the unconditional moments of asset returns in our model can match their empirical estimates.  Recall that the agent's risk aversion degree $\gamma$, discount rate $\phi$, perceived ambiguity level $\kappa$, and ambiguity aversion degree $\alpha$ are subjective parameters, with $\kappa$ and $\alpha$ combined into $\theta^*$ to determine the asset returns. In the following, we fix a value of $\theta^*$ and calibrate $\gamma$ and $\phi$ to the estimates of the unconditional stock risk premium and price-dividend ratio in \citet{CampbellCochrane:1999ByForceOfHabit}. Because the asset returns in our model are not in closed form, to simplify the calibration, we use the following approximations. First, recall the stock risk premium \eqref{eq:RPSpecialCase} in the special case when $(1-\gamma)\rho=0$. We approximate $\left(\omega + \overline{\omega}\lambda/\delta\right)^{-1}$ by the first-order Taylor expansion in $\omega$ at $\overline{\omega}$, i.e.,
	\begin{align*}
		\left(\omega + \overline{\omega}\lambda/\delta\right)^{-1}\approx  \big((1+\lambda/\delta)\overline{\omega}\big)^{-1}-\big(\overline{\omega}(1+\lambda /\delta)\big)^{-2}(\omega -\overline{\omega}).%\\
	%	=\overline{\omega}^{-1}(1+\lambda/\delta)^{-2}(2+\lambda/\delta) -\big(\overline{\omega}(1+\lambda /\delta)\big)^{-2}\omega.
	\end{align*}
Plugging the above into \eqref{eq:RPSpecialCase} and taking expectation, we derive
	\begin{align}
		\expect\left[\mu_S(\omega_t)-r_f(\omega_t)\right]%\approx \gamma\sigma_C^2 + \gamma\rho\sigma_C\nu\omega(1-\omega)\left(\omega + \overline{\omega}\lambda/\delta\right)^{-1}					-\sigma_C \theta^*\\
		\approx \gamma\sigma_C^2 + \gamma\rho\sigma_C\nu\expect[\omega_t(1-\omega_t)]\overline{\omega}^{-1}(1+\lambda/\delta)^{-2}(2+\lambda/\delta)\notag\\
		- \gamma\rho\sigma_C\nu \expect[\omega_t^2(1-\omega_t)]\big(\overline{\omega}(1+\lambda /\delta)\big)^{-2}
		-\sigma_C \theta^*.\label{eq:RPApproximate}
	\end{align}
	We use \eqref{eq:RPApproximate} as an approximation of the unconditional stock risk premium for all values of $\gamma$ and $\rho$. Second, using Taylor expansion in $\omega$ at $\overline{\omega}$, we approximate the log stock price dividend ratio in \eqref{eq:PDRatioSpecialCase}, which is valid for the case when $(1-\gamma)\rho$, as
	\begin{align*}
		\ln (S_t/D_t)\approx %-\ln \left(\frac{1}{\lambda+\delta} + \frac{\lambda }{\delta(\lambda+\delta)}\right) + \left(\frac{1}{\lambda+\delta} + \frac{\lambda }{\delta(\lambda+\delta)}\right)^{-1}\left(-\frac{\lambda \overline{\omega}}{\delta(\lambda+\delta)}\right)(\omega_t-\overline{\omega})\\
		%=
		-\ln \delta - \frac{\lambda \overline{\omega}}{\lambda +\delta } (\omega_t-\overline{\omega}).
		%   \frac{1}{\lambda+\delta} + \frac{\lambda }{\delta(\lambda+\delta)}\times \frac{\overline{\omega}}{\omega_t}
	\end{align*}
	Then, we have
	\begin{align}
		\exp\left(\expect[\ln(S_t/D_t)]\right)\approx \delta^{-1}\exp\left(-\frac{\lambda \overline{\omega}}{\lambda +\delta } (\expect[\omega_t]-\overline{\omega})\right).\label{eq:PDRatioApproximate}
	\end{align}
	We use \eqref{eq:PDRatioApproximate} as an approximation of the unconditional stock price-dividend ratio for all values of $\gamma$ and $\rho$.
	
	\iffalse
	Recall that the risk-free rate is given by \eqref{eq:MarketEquiRiskFree}, which can be written as
	\begin{align}
		r_f = \delta+\mu_C+\theta^*\sigma_C -\gamma \sigma_C^2.\label{eq:RiskFreeRateReform}
	\end{align}
	\fi
	
	In the second row of Table \ref{ta:UnconditionalMoments}, we copy the empirical estimates of the unconditional risk premium $\expect[\mu_s(\omega_t) - r_f]$, volatility $\sqrt{\expect[\sigma^2(\omega_t)]}$, price-dividend ratio $\exp\left(\expect[\ln(\frac{S_t}{D_t})]]\right)$, standard deviation of the log price-dividend ratio $\sqrt{\mathrm{var}(\ln(\frac{S_t}{D_t})}$, and the risk-free rate $r_f$ by \citet{CampbellCochrane:1999ByForceOfHabit}. Then, for each $\theta^*\in \{0,-0.05,-0.1,\dots,-0.55,-0.6\}$, we choose $\gamma$ and $\phi$ so that $\expect[\mu_s(\omega_t) - r_f]$ and $\exp\left(\expect[\ln(\frac{S_t}{D_t})]]\right)$ computed from the approximation formula \eqref{eq:RPApproximate} and \eqref{eq:PDRatioApproximate} match the empirical estimates. For $\theta^*\ge -0.35$, the calibrated $\phi$ is negative, showing that our model is not able to match both the risk premium and price-dividend ratio when the degree of ambiguity or the agent's ambiguity aversion degree is low. For $\theta^*\le -0.40$, the calibrated $\gamma$ and $\phi$ are presented in the second and third columns of Table \ref{ta:UnconditionalMoments}. For each calibrated triplet $(\theta^*,\gamma,\phi)$, we compute $\expect[\mu_s(\omega_t) - r_f]$, $\sqrt{\expect[\sigma^2(\omega_t)]}$, $\exp\left(\expect[\ln(\frac{S_t}{D_t})]]\right)$, $\sqrt{\mathrm{var}(\ln(\frac{S_t}{D_t})}$, and $r_f$ in our model exactly, and the results are shown in the fourth to eighth columns of the table. We can observe that even if we assume a very negative $\theta^*$, which represents a high level of perceived ambiguity and a high ambiguity aversion degree, our model is still not able to explain the empirical asset returns satisfactorily. First, the implied risk aversion degree to match the empirical risk premium is still too high to be realistic. Second, the stock volatility in the model is significantly lower than its empirical estimate. Therefore, even with ambiguity, to match the empirical asset returns, we have to go beyond the standard expected utility framework.
	
	\begin{table}
		\centering
		\caption{Empirical estimates by \citet{CampbellCochrane:1999ByForceOfHabit} (2nd row of the table) and model-implied values (4th to 16th rows of the table) of the unconditional stock risk premium $\expect[\mu_s(\omega_t) - r_f]$, volatility $\sqrt{\expect[\sigma^2(\omega_t)]}$, price-dividend ratio $\exp\left(\expect[\ln(\frac{S_t}{D_t})]]\right)$, standard deviation of the log price-dividend ratio $\sqrt{\mathrm{var}(\ln(\frac{S_t}{D_t})}$, and the risk-free rate $r_f$. For the model implied values, we consider $\theta^*\in \{-0.40,-0.45,\dots,-0.6\}$ and choose $\gamma$ and $\phi$ so that the model implied $\expect[\mu_s(\omega_t) - r_f]$ and $\exp\left(\expect[\ln(\frac{S_t}{D_t})]]\right)$, based on the approximation formula \eqref{eq:RPApproximate} and \eqref{eq:PDRatioApproximate}, match their empirical estimates.}\label{ta:UnconditionalMoments}
			\begin{tabular}{c|c|c|c|c|c|c|c}
			\hline
			\multicolumn{3}{c|}{} & {$\expect[\mu_s(\omega_t) - r_f]$} & {$\sqrt{\expect[\sigma^2(\omega_t)]}$} & {$\exp\left(\expect[\ln(\frac{S_t}{D_t})]]\right)$} & { $\sqrt{\mathrm{var}(\ln(\frac{S_t}{D_t})}$}& $r_f$ \\
			\hline
			\multicolumn{3}{c|}{Empirical Estimates} & 3.9\% & 18\% & 21.1 & 0.27 & 2.9\% \\
			\hline
			$\theta^{*}$ & $\gamma$ & $\phi$ \\
			\hline
%			0.00 & 35.53 & -0.25 & 3.92\% & 4.38\% & 19.91 & 0.1792 & 3.55\% \\
%			-0.05 & 34.23 & -0.21 & 3.91\% & 4.38\% & 20.06 & 0.1792 & 3.52\% \\
%			-0.10 & 32.92 & -0.18 & 3.91\% & 4.38\% & 20.21 & 0.1792 & 3.48\% \\
%			-0.15 & 31.62 & -0.14 & 3.91\% & 4.38\% & 20.36 & 0.1792 & 3.44\% \\
%			-0.20 & 30.32 & -0.10 & 3.91\% & 4.38\% & 20.51 & 0.1792 & 3.41\% \\
%			-0.25 & 29.02 & -0.07 & 3.91\% & 4.38\% & 20.67 & 0.1793 & 3.37\% \\
%			-0.30 & 27.71 & -0.04 & 3.91\% & 4.37\% & 20.83 & 0.1793 & 3.33\% \\
%			-0.35 & 26.41 & -0.02 & 3.91\% & 4.37\% & 20.99 & 0.1793 & 3.30\% \\
			-0.40 & 25.11 & 0.01 & 3.91\% & 4.37\% & 21.15 & 0.1793 & 3.26\% \\
			-0.45 & 23.80 & 0.03 & 3.91\% & 4.37\% & 21.31 & 0.1793 & 3.23\% \\
			-0.50 & 22.50 & 0.05 & 3.91\% & 4.37\% & 21.48 & 0.1793 & 3.19\% \\
			-0.55 & 21.20 & 0.07 & 3.91\% & 4.37\% & 21.65 & 0.1793 & 3.15\% \\
			-0.60 & 19.90 & 0.08 & 3.91\% & 4.37\% & 21.82 & 0.1793 & 3.12\% \\
			\hline
		\end{tabular}
	\end{table}

	\section{Conclusions}\label{se:Conclusions}
	In this work, we studied asset pricing in a continuous-time, pure-exchange economy with a representative agent who is ambiguous about the aggregate endowment growth rate in the economy. We assumed multiple cash flows in the economy, one is the dividend distributed by a stock and the other is labor income generated by human capital. To account for mixed attitudes towards ambiguity, we assumed that the representative agent's preferences are represented by the $\alpha$-MEU model. The agent consumes and trades the stock, human capital, and a risk-free asset to maximize the $\alpha$-MEU value of the discounted sum of consumption utility. Given the dynamic inconsistency of the $\alpha$-MEU model, we assumed that the agent takes an intra-personal equilibrium strategy, which can be consistently implemented throughout the entire horizon. We then defined the market equilibrium as the set of asset prices such that the consumption-investment strategy that clears the market is the agent's intra-personal equilibrium strategy.
	
	We proved the existence and uniqueness of the market equilibrium and showed that the asset prices in the equilibrium are the same as those in an economy with a representative agent who does not perceive any ambiguity and believes in a particular probabilistic model of the aggregate endowment. We then showed that with a reasonable risk aversion degree and ambiguity aversion degree, the more ambiguity the agent perceives or the more ambiguity averse she is, the lower the risk-free rate, the higher the stock price, the higher the stock risk premium, and the lower the stock volatility are. We also showed that the stock price-dividend ratio is decreasing, and the stock risk premium and volatility are increasing in the dividend-endowment ratio. Finally, we showed that our model cannot explain the empirical findings of asset returns: Even if we allow for high degrees of  perceived ambiguity and ambiguity aversion and calibrate the risk aversion degree to match the empirical estimate of the stock risk premium, the stock volatility in our model is significantly lower than its empirical estimate. %This negative results suggests that we need to consider non-expected utility frameworks.
	%One future research dIncorporating ambiguity in non-expected utility frameworks

	%%%%%%%%%%%%%%%%%%%%%%%%%%%%%%%%%%%%%%%%%%%%%%%%%%%%%%%%%%%%%%%%%%%%%%%%%%%%
	%%%%%%%%%%%%%%%%%%%%%%%%%%%%%%%%%%%%%%%%%%%%%%%%%%%%%%%%%%%%%%%%%%%%%%%%%%%%
	%%%%%%%%%%%%%%%%%%%%%%%%%%%%%%%%%%%%%%%%%%%%%%%%%%%%%%%%%%%%%%%%%%%%%%%%%%%%
	%%%%%%%%%%%%%%%%%%%%%%%%%%%%%%%%%%%%%%%%%%%%%%%%%%%%%%%%%%%%%%%%%%%%%%%%%%%%
	%%%%%%%%%%%%%%%%%%%%%%%%%%%%%%%%%%%%%%%%%%%%%%%%%%%%%%%%%%%%%%%%%%%%%%%%%%%%

	%%%%%%%%%%%%%%%%%%%
	%%%%%%%%%%%%%%%%%%%%%%%%%%%%%%%%%%%%%%%%%%%%%%%%%%%%%%%%%%%%%%%%%%%%%%%%%%%
	%%%%%%%%%%%%%%%%%%%
	%%%%%%%%%%%%%%%%%%%%%%%%%%%%%%%%%%%%%%%%%%%%%%%%%%%%%%%%%%%%%%%%%%%%%%%%%%%
	%%%%%%%%%%%%%%%%%%%
	%%%%%%%%%%%%%%%%%%%%%%%%%%%%%%%%%%%%%%%%%%%%%%%%%%%%%%%%%%%%%%%%%%%%%%%%%%%

	\appendix
\begin{appendices}
	
	\section{An Ordinary Differential Equation}\label{appx:ODE}
	Consider the following one-dimensional homogeneous stochastic differential equation (SDE)
	\begin{align}
		dX_t = \mu(X_t)dt + \sigma(X_t) dW_t,\label{eq:SDE}
	\end{align}
	where $(W_t)_{t\ge 0}$ is a standard one-dimensional Brownian motion.
	
	\begin{assumption}\label{as:Explosive}
		\begin{enumerate}
			\item[(i)] $\mu$ and $\sigma$ are twice continuously differentiable with bounded derivatives on $(0,1)$.
			\item[(ii)] $\sigma(x)>0,\forall x\in (0,1)$. For an arbitrarily fixed $x_0\in (0,1)$,
			\begin{align*}
				\int_{0}^{x_0}\rho(x)dx=\infty,\quad 	\int_{x_0}^{1}\rho(x)dx=\infty,%\quad \int_0^1 \big(\rho(x)\sigma(x)^2\big)^{-1}dx<\infty,
			\end{align*}
			where $\rho(x):=\exp\left( -\int_{x_0}^x \frac{2\mu(y)}{\sigma^2(y)} dy\right)$, $x\in (0,1)$.
		\end{enumerate}
	\end{assumption}

	Under Assumption \ref{as:Explosive}, for any $x\in (0,1)$, \eqref{eq:SDE} with $X_0=x$ has a unique strong solution and always stays in $(0,1)$; see for instance, Theorem 5.29 and Corollary 5.16 in Chapter 5 of \citet{KaratzasIShreveS:91bmsc}.
	Denote this solution by $X_t^{0,x},t\ge 0$. For any $\delta>0$, consider
	\begin{align}
		f(x):=\expect\left[\int_0^\infty e^{-\delta t}X_t^{0,x}dt\right],\quad x\in (0,1).\label{eq:FKRepresentation}
	\end{align}
	Because $(X_t^{0,x})_{t\ge 0}$ takes values in $(0,1)$, $f$ is a well-defined function on $(0,1)$.

	\begin{theorem}\label{thm:ODE}
		Let Assumption \ref{as:Explosive} hold and fix $\delta>0$.
		\begin{enumerate}
			\item[(i)] Consider $f$ as defined in \eqref{eq:FKRepresentation}. If
			\begin{align}
				\delta >\sup_{y\in (0,1)}\mu'(y),\label{eq:1stDExistCond}
			\end{align}
			then $f$ is continuously differentiable and
			\begin{align*}
				\inf_{x\in(0,1)}f'(x)>0,\quad \sup_{x\in(0,1)}f'(x)<+\infty.
			\end{align*}
			If furthermore,
			\begin{align}
				\delta>\sup_{y\in (0,1)}\left(2\mu'(y)+ (\sigma')^2(y)\right),\label{eq:2ndDExistCond}
			\end{align}
			then $f$ is twice continuously differentiable with bounded derivatives and is a solution to
			\begin{align}
				\frac{1}{2}\sigma^2(x)f''(x)+ \mu(x)f'(x)-\delta f(x)+x=0,\quad x\in (0,1). \label{eq:ODE}
			\end{align}
			\item[(ii)] Let $f$ be a twice continuously differentiable function on $(0,1)$ solving \eqref{eq:ODE} and assume that $f$ is bounded. Then, $f$ must be given by \eqref{eq:FKRepresentation}.
		\end{enumerate}		
	\end{theorem}
	
	\begin{corollary}\label{coro:ODEBoundness}
		Suppose that Assumption \ref{as:Explosive} hold, that \eqref{eq:1stDExistCond} and \eqref{eq:2ndDExistCond} hold, and that
		\begin{align}
			\lim_{x\downarrow 0}\mu(x)>0,\quad \lim_{x\uparrow 1}\mu(x)<0.\label{eq:ODESolutionBoundCondition}%,\quad \lim_{x\downarrow 0}\sigma(x)=\lim_{x\uparrow 1}\sigma(x)=0.
		\end{align}
		Then, the solution $f$ to \eqref{eq:ODE} satisfies
		\begin{align*}
			\inf_{x\in(0,1)}f(x)>0,\quad \sup_{x\in(0,1)}f(x)<1/\delta.
		\end{align*}
	\end{corollary}
	
	\section{Proof of Results in Appendix \ref{appx:ODE}}
	
	%Note that \eqref{eq:1stDExistCond} holds for any positive $G$ as long as $\mu$ is decreasing, i.e., $\mu$ can be represented as the negative of the gradient of a convex function, which is typically formulated in the study of Langevin dynamics.

	\begin{pfof}{Theorem \ref{thm:ODE}}
		We prove part (i) first.
		By Theorem 5.3 in Chapter 5 of \citet{Friedman2012:SDEandApplications}, there exists a version of the derivative of $X_t^{0,x}$ with respect to $x$, denoted as $A_{t}^{0,x}$, such that
		\begin{align*}
			dA_t^{0,x} = \mu'(X_t^{0,x})A_t^{0,x}dt +\sigma'(X_t^{0,x})A_t^{0,x}dW_t,\quad A_{0}^{t,x} = 1.
		\end{align*}
		Moreover, for every $t\ge 0$, defining
		\begin{align*}
			g(t,x):=\expect[X_t^{0,x}],\quad x\in (0,1),
		\end{align*}
		we conclude from Theorem 5.5 in Chapter 5 of \citet{Friedman2012:SDEandApplications} that $g(t,x)$ is continuously differentiable in $x$ and
		\begin{align*}
			g_x(t,x) =\expect[A_t^{0,x}].
		\end{align*}
		Straightforward calculation then yields
		\begin{align*}
			& g_x(t,x)  = \expect\left[\exp\left(\int_0^t\mu'(X_t^{0,x})dt -\frac{1}{2}\int_0^t(\sigma')^2(X_t^{0,x})dt + \int_0^t \sigma'(X_t^{0,x})A_s^{0,x}dW_s\right)\right]\\
			& \le e^{\big(\sup_{y\in(0,1)}\mu'(y)\big)t}.
		\end{align*}
		By Fubini's theorem, we have
		\begin{align*}
			f(x) = \int_0^\infty e^{-\delta t}\expect\left[X_t^{0,x}\right]dt = \int_0^\infty e^{-\delta t}g(t,x)dt.
		\end{align*}
		Therefore, for every $\delta$ satisfying \eqref{eq:1stDExistCond}, we conclude from the dominated convergence theorem that $f$ is continuously differentiable and
		\begin{align*}
			f'(x) = \int_0^\infty e^{-\delta t}g_x(t,x)dt.
		\end{align*}
		Moreover,
		\begin{align*}
			& f'(x)\le \int_0^\infty e^{-\delta t}e^{\big(\sup_{y\in(0,1)}\mu'(y)\big)t}dt=\frac{1}{\delta-\sup_{y\in(0,1)}\mu'(y)}<\infty,\quad \forall x\in (0,1),\\
			& f'(x)\ge \int_0^\infty e^{-\delta t} e^{\big(\inf_{y\in(0,1)}\mu'(y)\big)t}dt = \frac{1}{\delta-\inf_{y\in(0,1)}\mu'(y)}>0,\quad \forall x\in (0,1).
		\end{align*}
		
		By Theorem 5.4 in Chapter 5 of \citet{Friedman2012:SDEandApplications}, there exists a version of the second-order derivative of $X_t^{0,x}$ with respect to $x$, denoted as $B_{t}^{0,x}$, such that
		\begin{align*}
			&dB_t^{0,x}=\mu''(X_t^{0,x})(A_t^{0,x})^2dt + \mu'(X_t^{0,x})B_t^{0,x}dt\\
			&+\sigma''(X_t^{0,x})(A_t^{0,x})^2dW_t + \sigma'(X_t^{0,x})B_t^{0,x}dW_t,\quad B_0^{0,x}=0.
		\end{align*}
		Moreover, by Theorem 5.5 in Chapter 5 of \citet{Friedman2012:SDEandApplications}, $g(t,x)$ is twice continuously differentiable in $x$ and
		\begin{align*}
			g_{xx}(t,x) = \expect[B_t^{0,x}].
		\end{align*}
		Applying It\^o's Lemma, we derive
		\begin{align*}
			d \left(\left(A_t^{0,x}\right)^{-1}B_t^{0,x}\right)=A_t^{0,x}\left[h(X_t^{0,x})dt + \sigma''(X_t^{0,x})dW_t\right],\quad \left(A_0^{0,x}\right)^{-1}B_0^{0,x}=0,
		\end{align*}
		where
		\begin{align*}
			h(y):=\mu''(y) -(\sigma')^2(y)\sigma''(y).
		\end{align*}
		In consequence, for each fixed $t$, we have
		\begin{align*}
			B_t^{0,x} = A_t^{0,x}\int_0^tA_s^{0,x}h(X_s^{0,x}) ds + A_t^{0,x}\int_0^tA_s^{0,x}\sigma''(X_s^{0,x}) dW_s.
		\end{align*}
		Straightforward calculation yields that
		\begin{align}
			&\expect\left[\left|A_t^{0,x}\int_0^tA_s^{0,x}\sigma''(X_s^{0,x}) dW_s\right|\right]\notag\\
			&\le \frac{1}{2}\expect\left[\left(A_t^{0,x}\right)^2\right]+ \frac{1}{2}\expect\left[\left(\int_0^tA_s^{0,x}\sigma''(X_s^{0,x}) dW_s\right)^2\right]\notag\\
			& = \frac{1}{2}\expect\left[\left(A_t^{0,x}\right)^2\right]+ \frac{1}{2}\expect\left[\int_0^t\left(A_s^{0,x}\right)^2\left(\sigma''(X_s^{0,x})\right)^2ds\right]\notag\\
			&\le \frac{1}{2}e^{\big(\sup_{y\in(0,1)}(2\mu'(y)+(\sigma')^2(y))\big)t}+ \frac{1}{2}\sup_{y\in(0,1)}(\sigma''(y))^2\int_0^te^{\big(\sup_{y\in(0,1)}(2\mu'(y)+(\sigma')^2(y))\big)s}ds,\label{eq:PfofODEBound1}
		\end{align}
		where the last inequality is the case due to the dynamics of $(A_t^{0,x})_{t\ge 0}$. Similarly,
		\begin{align}
			&\expect\left[\left|A_t^{0,x}\int_0^tA_s^{0,x}h(X_s^{0,x}) ds\right|\right]\le \sup_{y\in (0,1)}|h(y)|\expect\left[A_t^{0,x}\int_0^tA_s^{0,x} ds\right]\notag\\
			& = \sup_{y\in (0,1)}|h(y)|\int_0^t\expect\left[A_t^{0,x}A_s^{0,x}\right] ds\notag\\
			&\le \sup_{y\in (0,1)}|h(y)|\int_0^t\expect\left[\left(\frac{1}{2}\left(A_t^{0,x}\right)^2+\frac{1}{2}\left(A_s^{0,x}\right)^2\right)\right] ds\notag\\
			& = \frac{1}{2}\sup_{y\in (0,1)}|h(y)|\left(t \expect\left[\left(A_t^{0,x}\right)^2\right] + \int_0^t \expect\left[\left(A_s^{0,x}\right)^2\right]ds\right)\notag\\
			&\le\frac{1}{2}\sup_{y\in (0,1)}|h(y)| \left(t e^{\big(\sup_{y\in(0,1)}(2\mu'(y)+(\sigma')^2(y))\big)t} + \int_0^t e^{\big(\sup_{y\in(0,1)}(2\mu'(y)+(\sigma')^2(y))\big)s}ds\right).\label{eq:PfofODEBound1}
			%  \\
			%    & = \sup_{y\in (0,1)}|h(y)|\int_0^t\expect\left[\expect_s\left[A_t^{0,x}\right]A_s^{0,x}\right] ds\\
			%  & = \sup_{y\in (0,1)}|h(y)|\int_0^t\expect\left[\expect_s\left[A_t^{0,x}/A_s^{0,x}\right]\left(A_s^{0,x}\right)^2\right] ds\\
			%  & \le \sup_{y\in (0,1)}|h(y)|\int_0^te^{\big(\sup_{y\in(0,1)}\mu'(y)\big)(t-s)}\expect\left[\left(A_s^{0,x}\right)^2\right] ds\\
			%  & \le \sup_{y\in (0,1)}|h(y)|\int_0^te^{\big(\sup_{y\in(0,1)}\mu'(y)\big)(t-s)}e^{\big(\sup_{y\in(0,1)}(2\mu'(y)+(\sigma')^2(y))\big)s} ds\\
			%  & = \sup_{y\in (0,1)}|h(y)| e^{\big(\sup_{y\in(0,1)}\mu'(y)\big)t} \frac{1}{a}\left(e^{at}-1\right)\\
			%  &\le \begin{cases}
				%    \sup_{y\in (0,1)}|h(y)|e^{\big(\sup_{y\in(0,1)}\mu'(y)\big)t}\le \sup_{y\in (0,1)}|h(y)|, & \text{if }a\le 0,\\
				%    \frac{1}{a}\sup_{y\in (0,1)}|h(y)|e^{\big(\sup_{y\in(0,1)}(2\mu'(y)+(\sigma')^2(y))\big)t}, &\text{if }a>0.\label{eq:PfofODEBound1}
				%  \end{cases},
		\end{align}
		Combining \eqref{eq:PfofODEBound1} and \eqref{eq:PfofODEBound1} and recalling the dominated convergence theorem, we conclude that for any $\delta$ satisfying \eqref{eq:1stDExistCond} and \eqref{eq:2ndDExistCond}, $f$ is twice continuously differentiable with bounded derivatives and
		\begin{align*}
			f''(x) = \int_0^\infty e^{-\delta t}g_{xx}(t,x)dt.
		\end{align*}
		
		Now, by the law of iterated expectation and the Markovian property of the solution to \eqref{eq:SDE}, we have
		\begin{align*}
			f(x) &= \expect\left[\int_0^t e^{-\delta s}X_s^{0,x}ds + e^{-\delta t} f(X_t^{0,x})\right]\\
			&=  f(x)+ \expect\left[\int_0^te^{-\delta s} \left(X_s^{0,x} - \delta  f(X_s^{0,x}) + \mu(X_s^{0,x}) f'(X_s^{0,x}) + \frac{1}{2}\sigma^2(X_s^{0,x}) f''(X_s^{0,x})\right)ds\right],
		\end{align*}
		where the second equality is the case due to It\^o's lemma and the fact that $f'$ and $\sigma$ are bounded. In consequence,
		\begin{align*}
			\frac{1}{t}\expect\left[\int_0^te^{-\delta s} \left(X_s^{0,x} - \delta  f(X_s^{0,x}) + \mu(X_s^{0,x}) f'(X_s^{0,x}) + \frac{1}{2}\sigma^2(X_s^{0,x}) f''(X_s^{0,x})\right)ds\right]=0.
		\end{align*}
		Sending $t$ to 0 in the above, recalling that $f$, $f'$, $f''$, $\mu$, and $\sigma$ are bounded, and applying the dominated convergence theory, we immediately derive \eqref{eq:ODE}.
		
		Next, we prove part (ii) of the theorem. Let $f$ be a solution to \eqref{eq:ODE} with the assumed regularities. Applying It\^o's lemma to $e^{-\delta t}f(X_t^{0,x}) ,t\ge 0$, we derive, for any stopping time $T$,
		\begin{align*}
			f(x) &= e^{-\delta T}f(X_T^{0,x}) - \int_0^Te^{-\delta s} \left(- \delta  f(X_s^{0,x}) + \mu(X_s^{0,x}) f'(X_s^{0,x}) + \frac{1}{2}\sigma^2(X_s^{0,x}) f''(X_s^{0,x})\right)ds\\
			& \quad - \int_0^Te^{-\delta s} \sigma(X_s^{0,x}) f'(X_s^{0,x}) dW_s\\
			& = e^{-\delta T}f(X_T^{0,x})  +\int_0^Te^{-\delta s}X_s^{0,x}ds- \int_0^Te^{-\delta s} \sigma(X_s^{0,x}) f'(X_s^{0,x}) dW_s,
		\end{align*}
		where the second equality is the case because $f$ satisfies \eqref{eq:ODE}. Setting $T$ to be $n\wedge \tau_n$, where $\tau_n$ is the first time $(X_s^{0,x})_{s\ge 0}$ exists $[1/n,1-1/n]$ for some $n>2$, taking expectation on both sides of the equation, and recalling the continuity of $\sigma$ and $f'$, we derive
		\begin{align*}
			f(x) = \expect\left[e^{-\delta (n\wedge \tau_{n})}f(X_{n\wedge \tau_{n}}^{0,x}) \right] + \expect\left[\int_0^{n\wedge \tau_{n}}e^{-\delta s}X_s^{0,x}ds\right].
		\end{align*}
		Sending $n$ to $\infty$ in the above and recalling that $(X_{s}^{0,x})_{s\ge 0}$ and $f$ are bounded, we derive \eqref{eq:FKRepresentation}.
	\end{pfof}			
	
	\begin{pfof}{Corollary \ref{coro:ODEBoundness}}
		Because $\lim_{x\downarrow 0}\mu(x)>0$ and because of Assumption \ref{as:Explosive}-(ii), we conclude that $\liminf_{x\downarrow 0}\sigma(x)=0$. As a result, there exists a decreasing sequence $\{x_n\}$ in $(0,1)$ with $\lim_{n\rightarrow \infty}x_n=0$ such that $\lim_{n\rightarrow \infty}\sigma(x_n)=0$. By \eqref{eq:ODE}, we have
		\begin{align*}
			\frac{1}{2}\sigma^2(x_n)f''(x_n)+ \mu(x_n)f'(x_n)-\delta f(x_n)+x_n=0.
		\end{align*}
		Because $f$ is increasing, we have
		\begin{align*}
			\lim_{x\downarrow 0}f(x) = \lim_{n\rightarrow \infty} f(x_n) =\frac{1}{\delta}\lim_{n\rightarrow \infty}\left[ \frac{1}{2}\sigma^2(x_n)f''(x_n)+ \mu(x_n)f'(x_n)+x_n\right]>0,
		\end{align*}
		where the inequality is the case because (i) $f''$ is bounded on $(0,1)$, (ii) $\lim_{n\rightarrow \infty}x_n=0$, (iii) $\lim_{x\downarrow 0}\mu(x)>0$, and (iv) $\inf_{x\in(0,1)}f'(x)>0$. Similarly, we can prove that $\lim_{x\uparrow 1}f(x)<1/\delta$. The proof then completes.
	\end{pfof}

	\section{Proof of Results in the Main Text} \label{appx:proofs}
	
	\begin{pfof}{Proposition \ref{prop:SDEExistence}}
		For an arbitrarily fixed $x_0\in (0,1)$, define the scale function
		\begin{align*}
			\rho_\omega(x):= \exp\lf( -\int_{x_0}^x \frac{2\mu_{\omega}(y)}{\sigma_{\omega}^2(y)} d y\rh),\quad x\in (0,1).
		\end{align*}
		Because $\sigma_\omega$ is positive and $\sigma_\omega'$ is bounded on $(0,1)$ and because $\lim_{x\downarrow 0} \sigma_\omega(x)=0$, we conclude that $\sigma_\omega(x)/x$ is positive and bounded on $(0,1)$. In consequence,
		\begin{align*}
			x/\sigma_\omega(x)\ge k,\quad x\in(0,1)
		\end{align*}
		for some positive constant $k$. Because $\lim_{x\downarrow 0}\mu_\omega(x)>0$, we conclude that there exists $\delta\in(0,x_0)$ and $C>0$ such that
		\begin{align*}
			\mu_\omega(x)\ge C,\quad \forall x\in (0,\delta).
		\end{align*}
		In consequence, we have
		\begin{align*}
			\frac{2\mu_{\omega}(x)}{\sigma_{\omega}^2(x)}=\frac{2\mu_{\omega}(x)}{x^2}\left(\frac{x}{\sigma_{\omega}(x)}\right)^2\ge 2Ck^2/x^2,\quad \forall x\in (0,\delta).
		\end{align*}
		As a result, we have
		\begin{align*}
			\rho(x)\ge e^{C\left(x^{-1}-x_0^{-1}\right)},\quad \forall x\in (0,\delta),
		\end{align*}
		which implies that $\int_0^{x_0}\rho(x)dx=\infty$. Moreover, we have
		\begin{align*}
			\left(\rho(x)\sigma_\omega(x)\right)^{-1} = \frac{1}{\sigma_\omega^2(x)}e^{-\int_x^{x_0} \frac{2\mu_{\omega}(y)}{\sigma_{\omega}^2(y)} dy}\le \frac{1}{\sigma_\omega^2(x)}e^{-2C\int_x^{x_0} \frac{1}{\sigma_{\omega}^2(y)} dy}=:h(x),\quad x\in (0,\delta).
		\end{align*}
		In consequence,
		\begin{align*}
			&\int_0^\delta \left(\rho(x)\sigma_\omega(x)\right)^{-1}dx\le \int_0^\delta \frac{1}{\sigma_\omega^2(x)}e^{-2C\int_x^{x_0} \frac{1}{\sigma_{\omega}^2(y)} dy}dx\\
			&=\frac{1}{2C}\lim_{x\downarrow 0}\left[e^{-2C\int_\delta^{x_0} \frac{1}{\sigma_{\omega}^2(y)} dy}-e^{-2C\int_x^{x_0} \frac{1}{\sigma_{\omega}^2(y)} dy}\right]\\
			& = \frac{1}{2C}\lim_{x\downarrow 0}e^{-2C\int_\delta^{x_0} \frac{1}{\sigma_{\omega}^2(y)} dy}<\infty,
		\end{align*}
		where the last equality is the case because $\sigma_{\omega}^{-2}(y) =y^2\sigma_{\omega}^{-2}(y)y^{-2}\ge k^2 y^{-2},\forall y\in (0,1)$. Similarly, we can show that $\int_{x_0}^1\rho(x)dx=\infty$ and that $\left(\rho(x)\sigma_\omega(x)\right)^{-1}$ is integrable in $x$ in the neighborhood of $1$. We then recall Theorem 5.29 and Corollary 5.16 in Chapter 5 of \citet{KaratzasIShreveS:91bmsc} for the existence and uniqueness of the strong solution to \eqref{eq:omega} in $(0,1)$ and refer to Theorem 3.1 in \cite{Cherny2004:InvariantDistributions} for the existence and uniqueness of the invariant distribution.
		%
		%
		%  \begin{align*}
			%    \left(\rho(x)\sigma_\omega(x)\right)^{-1}=\left(\rho(x)\sigma_\omega^2(x)\right)^{-1}
			%  \end{align*}
		%
		%\begin{align*}
		%  \lim_{l\rightarrow 0}\int_{l}^{x_0}\rho(x)dx=\infty,\quad 	\lim_{r\rightarrow 1}\int_{x_0}^{r}\rho(x)dx=\infty,\quad \int_0^1 \big(\rho(x)\sigma_\omega(x)^2\big)^{-1}dx<\infty,
		%\end{align*}
		%where $\rho(x):=\exp\lf( -\int_{x_0}^x \frac{2\mu_{\omega}(y)}{\sigma_{\omega}^2(y)} \d y\rh)$, $x\in (0,1)$.
		%
		%   One can refer to Theorem 5.29 and Corollary 5.16 in Chapter 5 of \citet{KaratzasIShreveS:91bmsc} for the existence and uniqueness of the strong solution to \eqref{eq:omega} and refer to Theorem 3.1 in \cite{Cherny2004:InvariantDistributions} or Exercise 5.40 in Chapter 5 of \citet{KaratzasIShreveS:91bmsc} for the existence and uniqueness of the invariant distribution.
	\end{pfof}

	\begin{pfof}{Proposition \ref{prop:limit_labor}}
		 For any $x>0$, $\omega\in (0,1)$, sufficiently small $\eps>0 $, $c>0$, and $u=(u_S,u_H)\in \R^2$, by \eqref{eq:ConvergentTail}, $\uV^{\bc_{t,\eps,c},\bu_{t,\eps,u}}(x,\omega)$, $\oV^{\bc_{t,\eps,c},\bu_{t,\eps,u}}(x,\omega)$, $\uV^{\hat{\bc}, \hat{\bu}}(x,\omega)$ and $\oV^{\hat{\bc}, \hat{\bu}}(x,\omega)$ are all well defined. Following the proof of Theorem 2.2 in \citet{ChenEpstein2002:AmbiguityRiskAssetReturns}, and recalling \eqref{eq:ConvergentTail} and the Markovian property of $X^{\bc_{t,\eps,c},\bu_{t,\eps,u}}$, we can derive the following dynamic programming principle:
		
		%
		%			By Theorem 2.2 in \citet{ChenEpstein2002:AmbiguityRiskAssetReturns}(\textcolor{red}{By Propsition \ref{prop: Infinite_horizon_BSDE}}) and the definition of $(\bc_{t,\eps,c},\bu_{t,\eps,u})$ as in \eqref{eq:PertubatedStrategy}, we have
		\begin{align}
			&\uV^{\bc_{t,\eps,c},\bu_{t,\eps,u}}(x,\omega) = \inf_{\theta\in\Theta} \E_t^{\theta} \lf[\int_t^{t+\eps} e^{-\phi(s-t)} U(c X^{c,u}_s) \d s + e^{-\phi \eps} \uV^{\hat{\bc}, \hat{\bu}}(X^{c,u}_{t+\eps},\omega_{t+\eps}) \rh],\label{eq:PertubatedStrategyWCValue}\\
			&	\oV^{\bc_{t,\eps,c},\bu_{t,\eps,u}}(x,\omega) = \sup_{\theta\in\Theta} \E_t^{\theta} \lf[\int_t^{t+\eps} e^{-\phi(s-t)} U(c X^{c,u}_s) \d s + e^{-\phi \eps} \oV^{\hat{\bc}, \hat{\bu}}(X^{c,u}_{t+\eps},\omega_{t+\eps}) \rh].\label{eq:PertubatedStrategyBCValue}
		\end{align}
		For any strategy $(\bc,\bu):=(\bc(\omega), \bu_S(\omega), \bu_H(\omega)),\omega\in(0,1)$, by \eqref{eq:wealth-labor-income}, \eqref{eq:StockPrice}, and \eqref{eq:HumanCapitalPrice}, and recalling $\left(B^\theta_{1,t},B^\theta_{2,t}\right)=\left(B^0_{1,t} - \int_0^t \theta_{s} \d s,B^0_{2,t}\right)$, we can derive the associated wealth process $X^{\bc, \bu}$ as follows:
		\begin{align*}
			\d X^{\bc, \bu}_t / X^{\bc, \bu}_t %&= \lf[ - \bc(\omega_t)  + r_f(\omega_t) (1-\bu^P(\omega_t) - \bu^H (\omega_t) ) \rh] \d t \\
			%&  + \bu_S(\omega_t)  \lf( \frac{\d S_t + D_t\d t}{S_t} \rh) + \bu_H(\omega_t) \lf( \frac{\d H_t + L_t\d t}{H_t} \rh)  \\
			& = \Big[-\bc(\omega_t) + (1-\bu_S(\omega_t) - \bu_H(\omega_t))r_f(\omega_t) + \bu_S(\omega_t)\big(\mu_S(\omega_t) + \theta_t\sigma_{S,1}(\omega_t)  \big) \\
			& + \bu_H(\omega_t)\big(\mu_H(\omega_t) + \theta_t\sigma_{H,1}(\omega_t)\big)  \Big]\d t
			+ \big( \bu_S(\omega_t) \sigma_{S,1}(\omega_t) + \bu_H(\omega_t) \sigma_{H,1}(\omega_t)  \big) \d B^{\theta}_{1,t}\\
			& + \big(\bu_S(\omega_t)\sigma_{S,2}(\omega_t) + \bu_H(\omega_t)\sigma_{H,2}(\omega_t)  \big)\d B^{\theta}_{2,t}.
		\end{align*}
		Then, by It\^o's lemma, we have
		\begin{align}
			&\E_t^{\theta} \lf[\int_t^{t+\eps} e^{-\phi(s-t)} U(c X^{c,u}_s) \d s + e^{-\phi \eps} \uV^{\hat{\bc}, \hat{\bu}}(X^{c,u}_{t+\eps},\omega_{t+\eps}) \rh]\notag\\
			&=\uV^{\hat{\bc}, \hat{\bu}}(x,\omega) + \E_t^{\theta}\lf[\int_t^{t+\epsilon}  e^{-\phi(s-t)}\Big(U(c X^{c,u}_s) + \CA_{\theta_s}^{c,u} \uV_{t}^{\hat{\bc}, \hat{\bu}}(X^{c,u}_{s},\omega_{s})\Big)ds\rh].\label{eq:PertubatedStrategyWCValueFixedtheta}
		\end{align}
		Combining \eqref{eq:PertubatedStrategyWCValue} and \eqref{eq:PertubatedStrategyWCValueFixedtheta}, we derive
		\begin{align}
			\uV^{\bc_{t,\eps,c},\bu_{t,\eps,u}}(x,\omega) = \uV^{\hat{\bc}, \hat{\bu}}(x,\omega)  + \inf_{\theta\in\Theta}\E_t^{\theta}\lf[\int_t^{t+\epsilon}  e^{-\phi(s-t)}\Big(U(c X^{c,u}_s) + \CA_{\theta_s}^{c,u} \uV_{t}^{\hat{\bc}, \hat{\bu}}(X^{c,u}_{s},\omega_{s})\Big)ds\rh].\label{eq:PertubatedStrategyWCValueTranformed}
		\end{align}
		
		For each fixed $\bar \theta\in[a,b]$, by considering a constant process in $\Theta$ taking value $\bar \theta$, we conclude from \eqref{eq:PertubatedStrategyWCValueTranformed} that
		\begin{align*}
			&\limsup_{\epsilon\downarrow 0}\frac{1}{\epsilon}\left(\uV^{\bc_{t,\eps,c},\bu_{t,\eps,u}}(x,\omega) - \uV^{\hat{\bc}, \hat{\bu}}(x,\omega)\right)\\
			&\le  \limsup_{\epsilon\downarrow 0}\frac{1}{\epsilon}\E_t^{\bar \theta}\lf[\int_t^{t+\epsilon}  e^{-\phi(s-t)}\Big(U(c X^{c,u}_s) + \CA_{\bar \theta}^{c,u} \uV_{t}^{\hat{\bc}, \hat{\bu}}(X^{c,u}_{s},\omega_{s})\Big)ds\rh]\\
			& = U(cx)+\CA_{\bar \theta}^{c,u} \uV_{t}^{\hat{\bc}, \hat{\bu}}(x,\omega),
		\end{align*}
		where the last equality is due to the dominated convergence theorem, whose use is validated by the observation that both $U(x)$ and $\CA_{\bar \theta}^{c,u} \uV_{t}^{\hat{\bc}, \hat{\bu}}(x,\omega)$ are of polynomial growth in $x\in (0,\infty)$, uniformly in $\omega\in (0,1)$ and that $\expect_t^{\bar \theta}\left[\sup_{s\in[t,t+\epsilon]}|X_{s}^{c,u}|^{n}\right]$ is finite for every $n\in \mathbb{R}$. As a result, we derive
		\begin{align}
			\limsup_{\epsilon\downarrow 0}\frac{1}{\epsilon}\left(\uV^{\bc_{t,\eps,c},\bu_{t,\eps,u}}(x,\omega) - \uV^{\hat{\bc}, \hat{\bu}}(x,\omega)\right)\le U(cx)+\inf_{\bar \theta\in[-\kappa,\kappa]}\CA_{\bar \theta}^{c,u} \uV_{t}^{\hat{\bc}, \hat{\bu}}(x,\omega).\label{eq:pfofCalcFOCEq1}
		\end{align}
		
		On the other hand, we derive from \eqref{eq:PertubatedStrategyWCValueTranformed} that
		\begin{align}
			&\limsup_{\epsilon\downarrow 0}\frac{1}{\epsilon}\left(\uV^{\bc_{t,\eps,c},\bu_{t,\eps,u}}(x,\omega) - \uV^{\hat{\bc}, \hat{\bu}}(x,\omega)\right)\\
			&\ge \limsup_{\epsilon\downarrow 0}\inf_{\theta\in\Theta}\frac{1}{\epsilon}\E_t^{\theta}\lf[\int_t^{t+\epsilon}  e^{-\phi(s-t)}\Big(U(c X^{c,u}_s) + \inf_{\bar \theta\in[-\kappa,\kappa]}\CA_{\bar \theta}^{c,u} \uV_{t}^{\hat{\bc}, \hat{\bu}}(X^{c,u}_{s},\omega_{s})\Big)ds\rh]\\
			& = U(cx)+\inf_{\bar \theta\in[-\kappa,\kappa]}\CA_{\bar \theta}^{c,u} \uV_{t}^{\hat{\bc}, \hat{\bu}}(x,\omega)+\limsup_{\epsilon\downarrow 0}\inf_{\theta\in\Theta}\frac{1}{\epsilon}\E_t^{\theta}\lf[\int_t^{t+\epsilon}  e^{-\phi(s-t)} \left[h\left(X^{c,u}_{s},\omega_{s}\right) -  h\left(X^{c,u}_{t},\omega_{t}\right)\right]ds\rh],\label{eq:pfofCalcFOCEq2}
		\end{align}
		where $h(x,\omega):=U(cx)+\inf_{\bar \theta\in[-\kappa,\kappa]}\CA_{\bar \theta}^{c,u} \uV_{t}^{\hat{\bc}, \hat{\bu}}(x,\omega)$. Because $h(x,\omega)$ is of polynomial growth in $x\in (0,\infty)$, uniformly in $\omega\in (0,1)$, and because $\expect_t^{0}\left[\sup_{s\in[t,t+\epsilon]}|X_{s}^{c,u}|^{n}\right]$ is finite for every $n\in \mathbb{R}$, we can apply the dominated convergence theorem to derive
		\begin{align*}
			\lim_{\epsilon\downarrow 0}\expect_t^0\left[\left(\frac{1}{\epsilon}\int_t^{t+\epsilon}  e^{-\phi(s-t)} \left[h\left(X^{c,u}_{s},\omega_{s}\right) -  h\left(X^{c,u}_{t},\omega_{t}\right)\right]\right)^2\right]=0.
		\end{align*}
		On the other hand, denoting $Z_t:=\frac{d\P^\theta}{d\P^0}\bigg|_{\mathcal{F}_t}$, we have
		\begin{align*}
			\sup_{\theta \in \Theta}\expect^0_t\left[\left(Z_{t+1}^\theta/Z_t^\theta\right)^2\right]% = \sup_{\theta \in \Theta}\expect^0_t\left[e^{-\int_t^{t+1}\theta_s^2ds +2\int_t^{t+1}\theta_s dB^0_{1,s}}\right]
			= \sup_{\theta \in \Theta}\expect^0_t\left[e^{\int_t^{t+1}\theta_s^2ds}e^{-2\int_t^{t+1}\theta_s^2ds +2\int_t^{t+1}\theta_s dB^0_{1,s}}\right]\le e^{\kappa^2},
		\end{align*}
		where the inequality is the case because $\theta_s$ takes values in $[-\kappa,\kappa]$. As a result,
		\begin{align*}
			&\limsup_{\epsilon\downarrow 0}\sup_{\theta\in\Theta}\E_t^{\theta}\lf[\left|\frac{1}{\epsilon}\int_t^{t+\epsilon}  e^{-\phi(s-t)} \left[h\left(X^{c,u}_{s},\omega_{s}\right) -  h\left(X^{c,u}_{t},\omega_{t}\right)\right]ds\right|\rh]\\
			&=\limsup_{\epsilon\downarrow 0}\sup_{\theta\in\Theta}\E_t^{0}\lf[\left|\frac{1}{\epsilon}\int_t^{t+\epsilon}  e^{-\phi(s-t)} \left[h\left(X^{c,u}_{s},\omega_{s}\right) -  h\left(X^{c,u}_{t},\omega_{t}\right)\right]ds\right|\left(Z_{t+1}^\theta/Z_t^\theta\right)\rh]\\
			&\le \left(\sup_{\theta \in \Theta}\expect^0_t\left[\left(Z_{t+1}^\theta/Z_t^\theta\right)^2\right]\right)^{1/2}\limsup_{\epsilon\downarrow 0}\left(\expect_t^0\left[\left(\frac{1}{\epsilon}\int_t^{t+\epsilon}  e^{-\phi(s-t)} \left[h\left(X^{c,u}_{s},\omega_{s}\right) -  h\left(X^{c,u}_{t},\omega_{t}\right)\right]\right)^2\right]\right)^{1/2}\\
			&=0,
		\end{align*}
		where the inequality is due to H\"older's inequality. As a result, we conclude from \eqref{eq:pfofCalcFOCEq2} that
		\begin{align}
			\limsup_{\epsilon\downarrow 0}\frac{1}{\epsilon}\left(\uV^{\bc_{t,\eps,c},\bu_{t,\eps,u}}(x,\omega) - \uV^{\hat{\bc}, \hat{\bu}}(x,\omega)\right)\ge U(cx)+\inf_{\bar \theta\in[-\kappa,\kappa]}\CA_{\bar \theta}^{c,u} \uV_{t}^{\hat{\bc}, \hat{\bu}}(x,\omega).
		\end{align}
		Combining the above with \eqref{eq:pfofCalcFOCEq1}, we complete the proof of \eqref{eq:PertubatedValueRateWC}. \eqref{eq:PertubatedValueRateBC} can be proved similarly and then \eqref{eq:Gamma} follows immediately.
	\end{pfof}
	
	\begin{pfof}{Proposition~\ref{prop:ov_uv_detailformula}}
		In the following, we denote by $\E_{t,x,\omega}^\theta[\cdot]$ the expectation under $\P^\theta$ conditional on the information at time $t$ with wealth level and market state being $(x,\omega)$ at that time.
		
		First, we prove \eqref{eq:ConvergentTail} so that the preference value of the consumption under $(\bc_{t,\eps,c},\bu_{t,\eps,u})$ is well defined. Given the consumption strategy \eqref{eq:ConsumptioninEquilibriumConstant} and wealth process \eqref{eq:WealthInEquilibriumGM}, we have
			\begin{align*}
				&f^\theta(x,\omega):=\E_{t,x,\omega}^\theta\left[\int_t^\infty e^{-\phi(s-t)}\left|U(\hat{\bc}(\omega_s) X^{\hat{\bc}, \hat{\bu}}_s)\right|ds\right]\\
				& = \E_{t,x,\omega}^\theta\left[\int_t^\infty e^{-\phi(s-t)}\left|U\left(\delta x e^{(\mu_C-\frac{1}{2}\sigma_C^2)s + \sigma_C B_{1,s}^0 }\right)\right|ds\right]\\
				& \le (\delta x)^{1-\gamma}\E_t^\theta\left[\int_t^\infty e^{-\phi(s-t)}\left| U\left(e^{(\mu_C-\frac{1}{2}\sigma_C^2)s + \sigma_C B_{1,s}^0 }\right)\right|ds\right] + |U(\delta x)|/\phi
			\end{align*}
			By Assumption \ref{as:GrowthCondition}, it is straightforward to verify that
			\begin{align*}
				K_1:=\sup_{\theta\in\Theta}\E_t^\theta\left[\int_t^\infty e^{-\phi(s-t)}\left| U\left(e^{(\mu_C-\frac{1}{2}\sigma_C^2)s + \sigma_C B_{1,s}^0 }\right)\right|ds\right]<\infty.
			\end{align*}
			In consequence, we have
			\begin{align*}
				&  f^\theta(x,\omega)\le K_1 (\delta x)^{1-\gamma} + |U(\delta x)|/\phi\\
				& \le K_1\delta^{1-\gamma}x^{1-\gamma} + \delta^{1-\gamma}|U(x)|/\phi + |U(\delta)|/\phi,\quad \forall x>0,\omega\in(0,1),\theta\in \Theta.
			\end{align*}
			
			Now, for any $x>0$, $\omega\in (0,1)$, sufficiently small $\eps\ge 0$, $c>0$, and $u=(u_S,u_H)\in \R^2$, we have \begin{align*}
				& \sup_{\theta\in\Theta}\E_{t,x,\omega}^\theta\left[\int_T^\infty e^{-\phi(s-t)}\left|U(\bc_{t,\eps,c}(\omega_s) X^{\bc_{t,\eps,c},\bu_{t,\eps,u}}_s)\right|ds\right]\\
				&  = e^{-\phi(T-t)}\sup_{\theta\in\Theta}\E_{t,x,\omega}^\theta\left[f^\theta(X^{\bc_{t,\eps,c},\bu_{t,\eps,u}}_T,\omega_T)\right]\\
				&\le  e^{-\phi(T-t)}\sup_{\theta\in\Theta}\left(K_1\delta^{1-\gamma}\E_{t,x,\omega}^\theta\left[\left|X^{\bc_{t,\eps,c},\bu_{t,\eps,u}}_T\right|^{1-\gamma}\right]
				+ (\delta^{1-\gamma}/\phi)\E_{t,x,\omega}^\theta\left[\left|U(X^{\bc_{t,\eps,c},\bu_{t,\eps,u}}_T)\right|\right]+ |U(\delta)|/\phi\right)\\
				& = e^{-\phi(T-t)}\sup_{\theta\in\Theta}\bigg(
				K_1\delta^{1-\gamma}\E_{t,x,\omega}^\theta\left[g_1^\theta(T-t-\epsilon,X^{c,u}_{t+\epsilon},\omega_{t+\epsilon})\right]\\
				&\quad
				+ (\delta^{1-\gamma}/\phi)\E_{t,x,\omega}^\theta\left[g_2^\theta(T-t-\epsilon,X^{c,u}_{t+\epsilon},\omega_{t+\epsilon})\right] + |U(\delta)|/\phi\bigg),
			\end{align*}
			where
			\begin{align*}
				g_1^\theta(\tau,x,\omega):=\E_{t,x,\omega}^\theta\left[\left|X^{\hat{\bc}, \hat{\bu}}_{t+\tau}\right|^{1-\gamma}\right],\quad g_2^\theta(\tau,x,\omega):=\E_{t,x,\omega}^\theta\left[\left|U(X^{\hat{\bc}, \hat{\bu}}_{t+\tau})\right|\right].
			\end{align*}
			By the wealth equation \eqref{eq:WealthInEquilibriumGM}, we can immediately derive
			\begin{align*}
				g_1^\theta(\tau,x,\omega)\le x^{1-\gamma} e^{\beta \tau},\quad \forall x>0,\omega\in(0,1),\theta\in \Theta,
			\end{align*}
			where
			\begin{align*}
				\beta:=\max\left\{(1-\gamma)\left(\mu_C +\kappa\sigma_C - \frac{\gamma}{2}\sigma_C^2\right),(1-\gamma)\left(\mu_C -\kappa\sigma_C - \frac{\gamma}{2}\sigma_C^2\right)\right\}.
			\end{align*}
			Because $r_f$, $\mu_S$, $\mu_H$, $\sigma_S$, and $\sigma_H$ are bounded on $(0,1)$, it is straightforward to see that
			\begin{align*}
				K_2:=\sup_{\theta\in\Theta}\E_{t,x,\omega}^\theta\left[\left|X^{c,u}_{t+\epsilon}\right|^{1-\gamma}\right]<\infty.
			\end{align*}
			In consequence,
			\begin{align*}
				&\limsup_{T\rightarrow \infty} e^{-\phi(T-t)}\sup_{\theta\in\Theta}\bigg(
				K_1\delta^{1-\gamma}\E_{t,x,\omega}^\theta\left[g_1^\theta(T-t-\epsilon,X^{c,u}_{t+\epsilon},\omega_{t+\epsilon})\right]\bigg)\\
				&\le \limsup_{T\rightarrow \infty} e^{-\phi(T-t)}K_1\delta^{1-\gamma}K_2 e^{\beta(T-t-\epsilon)}=0,
			\end{align*}
			where the equality is the case due to Assumption \ref{as:GrowthCondition}. Similarly, we can prove that
			\begin{align*}
				&\limsup_{T\rightarrow \infty} e^{-\phi(T-t)}\sup_{\theta\in\Theta}\bigg(
				(\delta^{1-\gamma}/\phi)\E_{t,x,\omega}^\theta\left[g_2^\theta(T-t-\epsilon,X^{c,u}_{t+\epsilon},\omega_{t+\epsilon})\right]\bigg)=0.
			\end{align*}
			Consequently, we derive \eqref{eq:ConvergentTail}.

		Next, we can follow the calculation in Section 3.4 of \citet{ChenEpstein2002:AmbiguityRiskAssetReturns} to derive \eqref{eq:uvGMFormula}. Straightforward calculation then yields \eqref{eq:GeneratorOnContinuitionValueLow}. Similarly, we can derive \eqref{eq:ovGMFormula} and calculate $\CA_{\bar \theta}^{c,u} \oV^{\hat{\bc}, \hat{\bu}}(x,\omega)$.
		
		Finally, straightforward calculation yields that
		\begin{align*}
			\left((1-\gamma)\uv + \phi^{-1}\right)\delta^{1-\gamma} x^{1-\gamma}  = x\times \frac{\partial}{\partial x}\uV^{\hat{\bc}, \hat{\bu}}(x,\omega)>0,
		\end{align*}
		where the inequality is the case because $\uV^{\hat{\bc}, \hat{\bu}}(x,\omega)$ is strictly increasing in $x$. As a result, for $u_S$ and $u_H$ with $u_S+u_H\ge 0$, $\CA_{\bar \theta}^{c,u} \uV^{\hat{\bc}, \hat{\bu}}(x,\omega)$ is increasing in $\bar \theta$ and, consequently,
		\begin{align*}
			\inf_{\bar \theta\in[-\kappa,\kappa]}\CA_{\bar \theta}^{c,u} \uV^{\hat{\bc}, \hat{\bu}}(x,\omega) = \CA_{-\kappa}^{c,u} \uV^{\hat{\bc}, \hat{\bu}}(x,\omega).
		\end{align*}
		Similarly, we can show that
		\begin{align*}
			\sup_{\bar \theta\in[-\kappa,\kappa]}\CA_{\bar \theta}^{c,u} \oV^{\hat{\bc}, \hat{\bu}}(x,\omega) = \CA_{\kappa}^{c,u} \oV^{\hat{\bc}, \hat{\bu}}(x,\omega).
		\end{align*}
		It is then straightforward to derive \eqref{eq:GammaClosedForm} from \eqref{eq:Gamma} and the closed-form formulae of $\CA_{-\kappa}^{c,u} \uV^{\hat{\bc}, \hat{\bu}}(x,\omega)$ and $\CA_{\kappa}^{c,u} \oV^{\hat{\bc}, \hat{\bu}}(x,\omega)$. For $u_S$ and $u_H$ with $u_S+u_H< 0$, $\CA_{\bar \theta}^{c,u} \uV^{\hat{\bc}, \hat{\bu}}(x,\omega)$ and $\CA_{\bar \theta}^{c,u} \oV^{\hat{\bc}, \hat{\bu}}(x,\omega)$ are decreasing in $\bar \theta$ and, consequently,
			\begin{align*}
				\inf_{\bar \theta\in[-\kappa,\kappa]}\CA_{\bar \theta}^{c,u} \uV^{\hat{\bc}, \hat{\bu}}(x,\omega) = \CA_{\kappa}^{c,u} \uV^{\hat{\bc}, \hat{\bu}}(x,\omega),\quad \sup_{\bar \theta\in[-\kappa,\kappa]}\CA_{\bar \theta}^{c,u} \oV^{\hat{\bc}, \hat{\bu}}(x,\omega) = \CA_{-\kappa}^{c,u} \oV^{\hat{\bc}, \hat{\bu}}(x,\omega).
			\end{align*}
			Combining \eqref{eq:Gamma} and the closed-form formulae of $\CA_{\kappa}^{c,u} \uV^{\hat{\bc}, \hat{\bu}}(x,\omega)$ and $\CA_{-\kappa}^{c,u} \oV^{\hat{\bc}, \hat{\bu}}(x,\omega)$, we can then derive \eqref{eq:GammaClosedForm}.
	\end{pfof}

	\begin{pfof}{Theorem \ref{thm:main}}
		Consider a triplet $(r_f,\varphi_S,\varphi_H)$ such that $r_f$ is continuous and bounded, $\varphi_i,i\in\{S,H\}$ are twice continuously differentiable with bounded derivatives, $\varphi_i,i\in\{S,H\}$ are bounded from below by a positive constant, and
		\eqref{eq:ConstantTotalPriceConsumptionRatio} holds with $\delta$ to be determined in the market equilibrium. Then, by \eqref{eq:StockDrift}--\eqref{eq:HCVol}, the corresponding asset mean return rates and volatility are continuous and bounded functions of $\omega\in (0,1)$. By Proposition \ref{prop:ov_uv_detailformula}, we have the form of $\Gamma^{\hat{\bc}, \hat{\bu}}(x,\omega;c,u_S,u_H)$ as in \eqref{eq:GammaClosedForm}. It is straightforward to see that $\Gamma^{\hat{\bc}, \hat{\bu}}(x,\omega;c,u_S,u_H)$ is strictly concave in $(c,u_S,u_H)$ in the convex region
		\begin{align*}
			\Delta:=\{(c,u_S,u_H)\in\mathbb{R}^3:c\in(0,1), u_S+u_H\ge 0\}.
		\end{align*}
		If furthermore, $- \alpha\kappa \big((1-\gamma)\uv + \phi^{-1}\big) + (1-\alpha) \kappa \big((1-\gamma)\ov + \phi^{-1}\big)\le 0$, $\Gamma^{\hat{\bc}, \hat{\bu}}(x,\omega;c,u_S,u_H)$ is strictly concave in $(0,1)\times \R^2$. According to Proposition \ref{prop:limit_labor} and by the definition of intra-personal equilibrium, the market is in equilibrium if and only if $(\hat{\bc}(\omega),\hat{\bu}_S(\omega),\hat{\bu}_H(\omega))$ as given by \eqref{eq:IntraPersonalEquiInMarketEquil} is a maximizer of $\Gamma^{\hat{\bc}, \hat{\bu}}(x,\omega;c,u_S,u_H)$ in $(c,u_S,u_H)$ for each fixed $x>0$ and $\omega\in (0,1)$. Because $(\hat{\bc}(\omega),\hat{\bu}_S(\omega),\hat{\bu}_H(\omega))$ is in the interior of $\Delta$ and thus in the interior of $(0,1)\times \R^2$, and because $\Gamma^{\hat{\bc}, \hat{\bu}}(x,\omega;c,u_S,u_H)$ is strictly concave in $(c,u_S,u_H)$, we conclude that the market is in equilibrium if and only if the following first-order condition holds:
		\begin{align*}
			\frac{\partial \Gamma^{\hat{\bc}, \hat{\bu}}}{\partial c}(x,\omega;c,u_S,u_H)\Big|_{(c,u_S,u_H) = (\hat{\bc}(\omega),\hat{\bu}_S(\omega),\hat{\bu}_H(\omega))} = 0,\\
			\frac{\partial \Gamma^{\hat{\bc}, \hat{\bu}}}{\partial u_S}(x,\omega;c,u_S,u_H)\Big|_{(c,u_S,u_H) = (\hat{\bc}(\omega),\hat{\bu}_S(\omega),\hat{\bu}_H(\omega))} = 0,\\
			\frac{\partial \Gamma^{\hat{\bc}, \hat{\bu}}}{\partial u_H}(x,\omega;c,u_S,u_H)\Big|_{(c,u_S,u_H) = (\hat{\bc}(\omega),\hat{\bu}_S(\omega),\hat{\bu}_H(\omega))} = 0.
		\end{align*}
		Using \eqref{eq:GammaClosedForm} and \eqref{eq:IntraPersonalEquiInMarketEquil}, we conclude that the above first-order condition is equivalent to the following equations:
		\begin{align}
			& 0 = U'\left(\big(\varphi_S(\omega)+\varphi_H(\omega)\big)^{-1}\right)- \big((1-\gamma)(\alpha\uv+(1-\alpha)\ov) + \phi^{-1} \big)\delta^{1-\gamma} ,\label{eq:FirstOrderCondition_c}\\
			& 0= \big(-r_f(\omega) +\mu_S(\omega) \big)\big((1-\gamma)(\alpha\uv+(1-\alpha)\ov) + \phi^{-1} \big) \notag\\
			&\quad + \sigma_C \left[-\alpha \kappa\big( (1-\gamma)\uv+\phi^{-1}\big) + (1-\alpha)\kappa\big((1-\gamma)\ov+\phi^{-1}) \big) \right]\notag\\
			&\quad - \gamma \big((1-\gamma)(\alpha\uv+(1-\alpha)\ov) + \phi^{-1} \big) \bigg[\sigma_C^2 + \rho\sigma_C\sigma_{S,2}(\omega)
			\notag\\
			& \quad + (\sigma_{S,2}(\omega) +\rho\sigma_C )\left(\frac{\varphi_S(\omega)}{\varphi_S(\omega)+\varphi_H(\omega)}\sigma_{S,2}(\omega) + \frac{\varphi_H(\omega)}{\varphi_S(\omega)+\varphi_H(\omega)}\sigma_{H,2}(\omega) \right) \bigg] ,\label{eq:FirstOrderCondition_uS}\\
			& 0= \big(-r_f(\omega) +\mu_H(\omega) \big)\big((1-\gamma)(\alpha\uv+(1-\alpha)\ov) + \phi^{-1} \big) \notag\\
			&\quad + \sigma_C \left[-\alpha \kappa\big( (1-\gamma)\uv+\phi^{-1}\big) + (1-\alpha)\kappa\big((1-\gamma)\ov+\phi^{-1}) \big) \right]\notag\\
			&\quad -  \gamma \big((1-\gamma)(\alpha\uv+(1-\alpha)\ov) + \phi^{-1} \big) \bigg[\sigma_C^2 + \rho\sigma_C\sigma_{H,2}(\omega)
			\notag\\
			&\quad  + (\sigma_{H,2}(\omega) +\rho\sigma_C )\left(\frac{\varphi_S(\omega)}{\varphi_S(\omega)+\varphi_H(\omega)}\sigma_{S,2}(\omega) + \frac{\varphi_H(\omega)}{\varphi_S(\omega)+\varphi_H(\omega)}\sigma_{H,2}(\omega) \right) \bigg] . \label{eq:FirstOrderCondition_uH}
		\end{align}
		Recalling \eqref{eq:ConstantTotalPriceConsumptionRatio}, we derive from \eqref{eq:FirstOrderCondition_c} that
		\begin{align}
			\delta &= \left((1-\gamma)\left(\alpha \uv+ (1-\alpha)\ov\right) + \phi^{-1}\right)^{-1} = \left(\alpha \delta_-^{-1}+(1-\alpha)\delta_+^{-1}\right)^{-1},\label{eq:PfofMainDelta}%\\
			%  & = \delta_-\delta_+\left(\phi-(1-\gamma)\left(\mu_C-\frac{\gamma}{2}\sigma_C^2\right) -(1-\gamma)\sigma_C\left(-\alpha\kappa + (1-\alpha)\kappa\right)\right)^{-1}\\
			%  & = \delta_-\delta_+\left(\phi-(1-\gamma)\left(\mu_C-\frac{\gamma}{2}\sigma_C^2\right) -(1-\gamma)\sigma_C\left(-\alpha\kappa + (1-\alpha)\kappa\right)\right)^{-1},
		\end{align}
		where the second equality is the case due to \eqref{eq:uvconstant} and \eqref{eq:ovconstant}. It is straightforward to see that when $\kappa=0$, the above formula of $\delta$ coincides with the formula in \eqref{eq:MarketEquiG}. When $\kappa>0$, with $\theta^*$ as given by \eqref{eq:MarketEquiModel}, we have
		\begin{align*}
			&\phi-(1-\gamma)(\mu_C-\frac{\gamma}{2}\sigma_C^2)-(1-\gamma)\sigma_C \theta^*= \frac{1}{2}(\delta_++\delta_-) + \frac{1}{2\kappa }(\delta_+-\delta_-)\theta^*\\
			& = \frac{1}{2}(\delta_++\delta_-) + \frac{1}{2}(\delta_+-\delta_-)\times \frac{-\alpha \delta_-^{-1}+(1-\alpha)\delta_+^{-1}}{\alpha \delta_-^{-1}+(1-\alpha)\delta_+^{-1}}\\
			& = \left(\alpha \delta_-^{-1}+(1-\alpha)\delta_+^{-1}\right)^{-1} \left[\frac{1}{2}(\delta_++\delta_-)\left(\alpha \delta_-^{-1}+(1-\alpha)\delta_+^{-1}\right)+\frac{1}{2}(\delta_+-\delta_-)\left(-\alpha \delta_-^{-1}+(1-\alpha)\delta_+^{-1}\right)\right]\\
			& = \left(\alpha \delta_-^{-1}+(1-\alpha)\delta_+^{-1}\right)^{-1}.
		\end{align*}
		Combining the above with \eqref{eq:PfofMainDelta}, we derive \eqref{eq:MarketEquiG}.
		
		In market equilibrium, the agent's wealth $X_t= S_t+H_t$ and her consumption amount is the aggregate endowment $\bar C_t$. We then conclude that her percentage of wealth consumed is
		\begin{align*}
			\frac{\bar C_t}{S_t+H_t} = \frac{1}{S_t/\bar C_t+H_t/\bar C_t} = \frac{1}{\varphi_S(\omega_t)+\varphi_H(\omega_t)} =\delta.
		\end{align*}
		Subtracting \eqref{eq:FirstOrderCondition_uH} from \eqref{eq:FirstOrderCondition_uS} and recalling \eqref{eq:ConstantTotalPriceConsumptionRatio}, we derive
		\begin{align} \label{eq:subtract}
			& \mu_S(\omega) -\gamma\sigma_{S,2}(\omega) \big[\rho\sigma_C + \delta \big(\varphi_S(\omega)\sigma_{S,2}(\omega) + \varphi_H(\omega)\sigma_{H,2}(\omega) \big)  \big]\no \\
			& = \mu_H(\omega) -\gamma\sigma_{H,2}(\omega) \big[\rho\sigma_C + \delta \big(\varphi_S(\omega)\sigma_{S,2}(\omega) + \varphi_H\sigma_{H,2}(\omega) \big)  \big].
		\end{align}
		By \eqref{eq:ConstantTotalPriceConsumptionRatio}, we derive
		\begin{align}
			\varphi_S'(\omega)=-\varphi_H'(\omega),\quad \varphi_S''(\omega)=-\varphi_H''(\omega). \label{eq:vaphiD1D2Eq}
		\end{align}
		Plugging the above together with \eqref{eq:ConstantTotalPriceConsumptionRatio} and \eqref{eq:StockDrift}--\eqref{eq:HCVol} into \eqref{eq:subtract}, we derive \eqref{eq:ODEPriceEndwomentRatio}. Plugging \eqref{eq:ODEPriceEndwomentRatio} into \eqref{eq:FirstOrderCondition_uS} and recalling \eqref{eq:ConstantTotalPriceConsumptionRatio}, \eqref{eq:vaphiD1D2Eq}, and \eqref{eq:StockDrift}--\eqref{eq:HCVol}, we derive \eqref{eq:MarketEquiRiskFree}. Therefore, the first-order conditions \eqref{eq:FirstOrderCondition_c}--\eqref{eq:FirstOrderCondition_uH} are equivalent to  \eqref{eq:MarketEquiG}, \eqref{eq:MarketEquiRiskFree}, and \eqref{eq:ODEPriceEndwomentRatio}. Moreover, $\varphi_S$ is the solution to \eqref{eq:ODEPriceEndwomentRatio} if and only if $\varphi_H=1/\delta-\varphi_S$ is the solution to \eqref{eq:ODEPriceEndwomentRatioHC}. In consequence, to  study the existence and uniqueness of the market equilibrium, we only need to study the existence and uniqueness of the solution to \eqref{eq:ODEPriceEndwomentRatio}.

		Denote
		\begin{align*}
			\mu(\omega):=\mu_{\omega}(\omega) + (1-\gamma)\rho\sigma_C\sigma_{\omega}(\omega),\quad \sigma(\omega):=\sigma_\omega(\omega),\quad \omega\in (0,1).
		\end{align*}
		Then, \eqref{eq:ODEPriceEndwomentRatio} becomes \eqref{eq:ODE}. Using the same proof as for Proposition \ref{prop:SDEExistence}, we can show that Assumption \ref{as:Explosive} holds. Moreover, it is straightforward to see that $\theta^*\in[-\kappa,\kappa]$ and, consequently, $\delta \ge \min(\delta_+,\delta_-)$.
		Therefore, by \eqref{eq:GrowthCondition2}, we conclude that conditions \eqref{eq:1stDExistCond} and \eqref{eq:2ndDExistCond} hold. In consequence, we can apply Theorem \ref{thm:ODE} to conclude that \eqref{eq:ODEPriceEndwomentRatio} admits a unique solution that is $C^2$ on $(0,1)$ and has bounded derivatives. Moreover, it is straightforward to verify that \eqref{eq:ODESolutionBoundCondition} holds. In consequence, the solution $\varphi_S$ to \eqref{eq:ODEPriceEndwomentRatio} satisfies $\inf_{\omega\in(0,1)}\varphi_S(\omega)>0$ and $\sup_{\omega\in(0,1)}\varphi_S(\omega)<1/\delta$. As a result, $\inf_{\omega\in(0,1)}\varphi_H(\omega)= 1/\delta-\sup_{\omega\in(0,1)}\varphi_S(\omega)>0$. The proof then completes.
	\end{pfof}
	
	\begin{pfof}{Proposition \ref{prop:SDF}}
		We show that the asset prices derived in Theorem \ref{thm:main} satisfy \eqref{eq:equilibriumRiskFreeEQ}--\eqref{eq:equilibriumHCEQ}.
		
		It is straightforward to show \eqref{eq:equilibriumRiskFreeEQ} holds. Because $\varphi_S$ is a smooth, bounded solution to \eqref{eq:ODEPriceEndwomentRatio} with bounded derivatives, It\^o's Lemma yields that
		\begin{align*}
			e^{-\phi t}C_t^{1-\gamma}\varphi_S(\omega_t) = \E_t^{\theta^*}\left[\int_t^{\tau}e^{-\phi s}C_{s}^{1-\gamma}\omega_{s}ds + e^{-\phi \tau}C_\tau^{1-\gamma}\varphi_S(\omega_\tau)\right]
		\end{align*}
		for any $\tau \ge t$. Sending $\tau$ to $\infty$, noting that $\varphi_S$ is bounded, and recalling Assumption \ref{as:GrowthCondition}, we conclude that
		\begin{align*}
			S_t = \varphi_S(\omega_t)C_t = \E_t^{\theta^*}\left[\int_t^{\infty}e^{-\phi(s-t)}\left(\frac{C_{s}}{C_t}\right)^{-\gamma}D_{s}ds\right].
		\end{align*}
		In consequence, \eqref{eq:equilibriumStockEQ} holds. Finally, \eqref{eq:equilibriumHCEQ} can be proved similarly.
		%Similarly,
		%\begin{align*}
		%  H_t = \varphi_H(\omega_t)C_t = \E_t^{\theta^*}\left[\int_t^{\infty}e^{-\phi(s-t)}\left(\frac{C_{s}}{C_t}\right)^{-\gamma}(C_t-D_{s})ds\right].
		%\end{align*}
		%Thus, the risky assets are evaluated using the stochastic discount factor $e^{-\phi(s-t)}\left(\frac{C_{s}}{C_t}\right)^{-\gamma}$, which is the marginal utility of consumption, under the model $\P^{\theta^*}$.
		%
		%
		%\begin{align*}
		%  S_t = \E_t^{\theta^*}\left[\int_t^{\tau}e^{-\phi(s-t)}\left(\frac{C_{s}}{C_t}\right)^{-\gamma}D_{s}ds + e^{-\phi(\tau-t)}\left(\frac{C_{\tau}}{C_t}\right)^{-\gamma}S_{\tau}\right]
		%\end{align*}
		%
		%\begin{align*}
		%  e^{-\phi t}C_t^{1-\gamma}\left(S_t/C_t\right) = \E_t^{\theta^*}\left[\int_t^{\tau}e^{-\phi s}C_{s}^{1-\gamma}\omega_{s}ds + e^{-\phi \tau}C_\tau^{1-\gamma}(S_{\tau}/C_\tau)\right]
		%\end{align*}
	\end{pfof}

	\begin{pfof}{Proposition \ref{prop:AmImMarketPriceRisk}}
		It is straightforward to see that $\theta^*=0$ when $\kappa=0$.
		
		Next, denote
		\begin{align*}
			A:=\phi-(1-\gamma)\left(\mu_C  - \frac{\gamma}{2}\sigma_C^2\right),\quad  B:=(\gamma-1)\sigma_C.
		\end{align*}
		Then, Assumption \ref{as:GrowthCondition} implies that $A>0$, that $\kappa$ can only take values in $[0,A/|B|)$ when $\gamma \neq 1$, and that $\kappa$ can take any value in $[0,\infty)$ when $\gamma =1$. Note that
		\begin{align*}
			\delta_\pm &= A\pm B\kappa,\\
			\theta^* &= \frac{-\alpha \delta_-^{-1} + (1-\alpha)\delta_+^{-1}}{\alpha\delta_-^{-1} +(1-\alpha)\delta_+^{-1}}\kappa = \frac{-\alpha \delta_+ + (1-\alpha)\delta_-}{\alpha\delta_+ +(1-\alpha)\delta_-}\kappa = \frac{(1-2\alpha)A\kappa -B\kappa^2}{A + (2\alpha-1)B\kappa}.
		\end{align*}
		By Assumption \ref{as:GrowthCondition} and because $\alpha\in [0,1]$, we have $A + (2\alpha-1)B\kappa>0$.
		
		Straightforward calculation yields
		\begin{align*}
			\frac{\partial \theta^*}{\partial \alpha} &= \left(A + (2\alpha-1)B\kappa\right)^{-2}\left[-2A\kappa \left(A + (2\alpha-1)B\kappa\right)-2B\kappa\left((1-2\alpha)A\kappa -B\kappa^2\right)\right]\\
			& = \left(A + (2\alpha-1)B\kappa\right)^{-2}2\kappa\left(-A^2+B^2\kappa^2\right)\\
			& = -\left(A + (2\alpha-1)B\kappa\right)^{-2}2\kappa\left(A-B\kappa\right)\left(A+B\kappa\right)<0,
		\end{align*}
		where the inequality is the case because $A\pm B\kappa=\delta_\pm>0$. In consequence, $\theta^*$ is strictly decreasing in $\alpha$.
		
		On the other hand,
		\begin{align*}
			\frac{\partial \theta^*}{\partial \kappa}&=\left(A + (2\alpha-1)B\kappa\right)^{-2}\Big[((1-2\alpha) A-2B\kappa)\left(A + (2\alpha-1)B\kappa\right)-(2\alpha-1)B\left((1-2\alpha)A\kappa -B\kappa^2\right)\Big]\\
			&= \left(A + (2\alpha-1)B\kappa\right)^{-2}\Big[(1-2\alpha)A^2 - 2 AB\kappa + (1-2\alpha)^2B^2\kappa^2\Big]=\left(A + (2\alpha-1)B\kappa\right)^{-2}f(\kappa),
		\end{align*}
		where
		\begin{align*}
			f(\kappa):=(1-2\alpha)A^2 - 2 AB\kappa + (1-2\alpha)^2B^2\kappa^2.
		\end{align*}
		We first consider the case when $\gamma=1$. In this case, $B=0$ and $f(\kappa)= (1-2\alpha)A^2$, which is positive (equal to 0 and negative, respectively) when $\alpha\in[0,1/2)$ (when $\alpha=1/2$ and when $\alpha\in(1/2,1]$, respectively). The monotonicity of $\theta^*$ in $\kappa$ then follows immediately.
		
		Next, we consider the case when $\gamma>1$. In this case, $B>0$ and $\kappa$ can take values in $[0,A/B)$. If $\alpha=1/2$, then $f(\kappa) = -2AB\kappa<0$ for any $\kappa>0$. In consequence, $\theta^*$ is strictly decreasing in $\kappa$. If $\alpha\neq 1/2$, $f(\kappa)$ is quadratic in $\kappa\in \mathbb{R}$ and has two zeros
		\begin{align*}
			\kappa_\pm:=\frac{1\pm \sqrt{1-(1-2\alpha)^3}}{(1-2\alpha)^2}\times \frac{A}{B}.
		\end{align*}
		Because $1-2\alpha\in[-1,1]$, we have
		\begin{align*}
			\kappa_+>\frac{1}{(1-2\alpha)^2}\times \frac{A}{B}\ge \frac{A}{B}.
		\end{align*}
		On the other hand, when $\alpha\in (1/2,1]$, $\kappa_-<0$. In this case, $f(\kappa)<0$ for $\kappa\in [0,A/B]$. In consequence, $\theta^*$ is strictly decreasing in $\kappa\in[0,A/B)$. When $\alpha=0$, $\kappa_-=A/B$. In this case, $f(\kappa)>0$ for all $\kappa\in [0,A/B)$. In consequence, $\theta^*$ is strictly increasing in $\kappa\in[0,A/B)$. When $\alpha\in (0,1/2)$, $\kappa_-\in (0,A/B)$. In this case, $f(\kappa)>0$ for $\kappa\in [0,\kappa_-)$ and $f(\kappa)<0$ for $\kappa\in (\kappa_-,A/B)$. In consequence, $\theta^*$ is strictly increasing in $\kappa\in[0,\kappa_-]$ and strictly decreasing on $[\kappa_-,A/B)$.
		
		Finally, we consider the case when $\gamma\in (0,1)$. In this case, $B<0$ and $\kappa$ can take values in $[0,A/|B|)$. If $\alpha=1/2$, then $f(\kappa) = -2AB\kappa>0$ for any $\kappa>0$. In consequence, $\theta^*$ is strictly increasing in $\kappa$. If $\alpha\neq 1/2$, $f(\kappa)$ is quadratic in $\kappa\in \mathbb{R}$ and has two zeros
		\begin{align*}
			\bar \kappa_\pm:=\frac{-1\pm \sqrt{1-(1-2\alpha)^3}}{(1-2\alpha)^2}\times \frac{A}{|B|}.
		\end{align*}
		It is obvious that $\bar \kappa_-<0$. On the other hand, when $\alpha\in [0,1/2)$, $\bar \kappa_+<0$. As a result, $f(\kappa)>0$ for all $\kappa\ge 0$. In consequence, $\theta^*$ is strictly increasing in $\kappa\in (0,A/|B|)$. When $\alpha\in (1/2,1]$, it is straightforward to verify that $\bar \kappa_+\in (0,A/|B|)$. As a result, $f(\kappa)<0$ for $\kappa\in[0,\bar \kappa_+)$ and $f(\kappa)<0$ for $\kappa\in (\bar \kappa_+,A/|B|)$. In consequence, $\theta^*$ is strictly decreasing in $\kappa\in [0,\bar \kappa_+]$ and strictly increasing in $\kappa\in [\bar \kappa_+,A/|B|)$.
	\end{pfof}
	
	\begin{pfof}{Proposition \ref{prop:ComparativeStat}}
		Assertions (i) and (ii) are straightforward to prove. By Theorem \ref{thm:ODE}, the solution to \eqref{eq:ODEPriceEndwomentRatio} has the following representation
		\begin{align*}
			\varphi_S(\omega) = \expect\left[\int_0^\infty e^{-\delta t}X^{0,\omega}_t\right],
		\end{align*}
		where $(X^{0,\omega}_t)_{t\ge 0}$, which always lies in $(0,1)$, is the solution to the following ODE:
		\begin{align*}
			dX^{0,\omega}_t = \left[\mu_{\omega}(X^{0,\omega}_t) + (1-\gamma)\rho\sigma_C\sigma_{\omega}(X^{0,\omega}_t)\right]dt + \sigma_\omega(X^{0,\omega}_t)dW_t,\quad X^{0,\omega}_0=\omega.
		\end{align*}
		Then, $\varphi_H$ is represented by
		\begin{align*}
			\varphi_H(\omega) = 1/\delta -\varphi_S(\omega) = \expect\left[\int_0^\infty e^{-\delta t}(1-X^{0,\omega}_t)\right].
		\end{align*}
		It is then straightforward to see that both $\varphi_S(\omega)$ and $\varphi_H(\omega)$ are strictly decreasing in $\delta $ for every $\omega \in (0,1)$. Furthermore, it is obvious that both $\varphi_S(\omega)$ and $\varphi_H(\omega)$ depend on $\gamma$, $\rho$, and $\sigma_C$ through $(1-\gamma)\rho\sigma_C$. By the comparison theorem for one-dimensional SDEs (Theorem 2.18 in Chapter 5 of \citet{KaratzasIShreveS:91bmsc}, the larger $(1-\gamma)\rho\sigma_C$ is, the larger $X^{0,\omega}_t$ is pathwise. In consequence, $\varphi_S(\omega)$ is strictly increasing in $(1-\gamma)\rho\sigma_C$ and $\varphi_H(\omega)$ is strictly decreasing in $(1-\gamma)\rho\sigma_C$ for each $\omega \in (0,1)$.
		
		Finally, by Theorem \ref{thm:ODE}, $\varphi_S'(\omega)>0$, so $\varphi_S$ is strictly increasing on $(0,1)$. It then follows that $\varphi_H$ is strictly decreasing on $(0,1)$.
	\end{pfof}
	
	\begin{pfof}{Proposition \ref{prop:RiskPremiumVol}}
		We first consider the case when $\rho=0$. In this case, we derive from \eqref{conditional_equity_premium} that
		\begin{align*}
			\mu_S(\omega)-r_f(\omega)
			= \gamma\sigma_C^2 -\sigma_C \theta^*,
		\end{align*}
		which is strictly decreasing in $\theta^*$.
		
		Next, we consider the case when $\gamma=1$. In this case, $\delta$ does not depend on $\theta^*$, and thus $\varphi_S$ does not depend on $\theta^*$. In consequence, $\mu_S(\omega)-r_f(\omega)$ is strictly decreasing in $\theta^*$ and $\sigma_S(\omega)$ does not depend on $\theta^*$.
	\end{pfof}
	
	\begin{pfof}{Corollary \ref{Coro:SpecialCase}}
		It is straightforward to verify that $\varphi_S$ as defined in \eqref{eq:PriceEndowmentRatioClosedForm} is a solution to \eqref{eq:ODEPriceEndwomentRatio} when $(1-\gamma)\rho=0$. Straightforward calculation yields \eqref{eq:ElasticitySpecialCase} and \eqref{eq:RPSpecialCase}. Moreover, by taking the first-order derivative, it is straightforward to show that $\omega(1-\omega)\left(\omega + \overline{\omega}\lambda/\delta\right)^{-1}$ is first strictly increasing and then strictly decreasing in $\omega$. The monotonicity of $\mu_S(\omega)-r_f(\omega)$ and $\|\sigma_S(\omega)\|$ in $\omega$ then follows immediately.
	\end{pfof}

\end{appendices}

%			\newpage
%			\nocite{*}
%			\footnotesize
\bibliographystyle{apalike}
\bibliography{LongTitles,BibFile}
	
\end{document}